\documentclass{aa}  
\usepackage{graphicx}
\usepackage{txfonts}
\usepackage[dvipsnames]{xcolor}
\usepackage{booktabs}
\usepackage{placeins}

\usepackage[T1]{fontenc}
\usepackage[utf8]{inputenc}

\newcommand{\numerror}[3]{#1^{+#2}_{-#3}}

\newcommand{\vpeak}{V_{\rm peak}}

\newcommand{\wpp}{w_{\rm p}}
\newcommand{\xilm}{\xi_{\ell = 0}}
\newcommand{\xilq}{\xi_{\ell = 2}}

\newcommand{\sig}{\sigma_{8}}
\newcommand{\OmM}{\Omega_\mathrm{m}}
\newcommand{\Omb}{\Omega_{\rm b}}
\newcommand{\h}{h}
\newcommand{\ns}{{n_{\rm s}}}
\newcommand{\Mnu}{M_{\rm \nu}}

\newcommand{\OmMh}{\Omega_\mathrm{m}h^2}

\newcommand{\Halpha}{\rm{H}\alpha}

\newcommand{\ihMpcC}{ h^{3}{\rm Mpc}^{-3}}
\newcommand{\ihMpc}{ h^{-1}{\rm Mpc} }
\newcommand{\ihGpc}{ h^{-1}{\rm Gpc} }
\newcommand{\hMpc}{ h^{-1}{\rm Mpc} }
\newcommand{\ihGpcC}{ h^{3}{\rm Gpc}^{-3}}
\newcommand{\hMsun}{ h^{-1}{\rm M_{ \odot}}}

\usepackage{textcomp}

\usepackage{datatool, filecontents}

\usepackage[colorlinks=true,allcolors=blue]{hyperref}

\DTLsetseparator{,}% Set the separator between the columns.

\DTLloaddb[noheader, keys={thekey,thevalue}]{cosmopars}{Figures/cosmoparam.txt}
\newcommand{\cosmopar}[1]{\DTLfetch{cosmopars}{thekey}{#1}{thevalue}}

\begin{document}
\begin{filecontents*}{Figures/cosmoparam.txt}
DESI_sigma8_mean,0.81
DESI_sigma8_down,0.06
DESI_sigma8_up,0.05
MTNGDESI_sigma8_mean,0.82
MTNGDESI_sigma8_down,0.05
MTNGDESI_sigma8_up,0.04
MTNGHalpha_sigma8_mean,0.8
MTNGHalpha_sigma8_down,0.03
MTNGHalpha_sigma8_up,0.04
MTNGDESI_sigma8_dmean,0.01
MTNGDESI_sigma8_ddown,-0.04
MTNGDESI_sigma8_dup,0.05
MTNGHalpha_sigma8_dmean,-0.02
MTNGHalpha_sigma8_ddown,-0.05
MTNGHalpha_sigma8_dup,0.02
DESI_OmMh_mean,0.146
DESI_OmMh_down,0.009
DESI_OmMh_up,0.009
MTNGDESI_OmMh_mean,0.146
MTNGDESI_OmMh_down,0.008
MTNGDESI_OmMh_up,0.009
MTNGDESI_OmMh_dmean,0.004
MTNGDESI_OmMh_ddown,-0.004
MTNGDESI_OmMh_dup,0.013
MTNGHalpha_OmMh_mean,0.135
MTNGHalpha_OmMh_down,0.007
MTNGHalpha_OmMh_up,0.008
MTNGHalpha_OmMh_dmean,-0.007
MTNGHalpha_OmMh_ddown,-0.003
MTNGHalpha_OmMh_dup,0.012
DESI_sigma12_mean,0.82
DESI_sigma12_down,0.06
DESI_sigma12_up,0.06
MTNGDESI_sigma12_mean,0.82
MTNGDESI_sigma12_down,0.06
MTNGDESI_sigma12_up,0.05
MTNGDESI_sigma12_dmean,0.02
MTNGDESI_sigma12_ddown,-0.04
MTNGDESI_sigma12_dup,0.07
MTNGHalpha_sigma12_mean,0.8
MTNGHalpha_sigma12_down,0.03
MTNGHalpha_sigma12_up,0.04
MTNGHalpha_sigma12_dmean,-0.0
MTNGHalpha_sigma12_ddown,-0.03
MTNGHalpha_sigma12_dup,0.03
DESI_S8_mean,0.86
DESI_S8_down,0.06
DESI_S8_up,0.06
MTNGDESI_S8_mean,0.86
MTNGDESI_S8_down,0.06
MTNGDESI_S8_up,0.06
MTNGDESI_S8_dmean,0.04
MTNGDESI_S8_ddown,-0.03
MTNGDESI_S8_dup,0.09
MTNGHalpha_S8_mean,0.81
MTNGHalpha_S8_down,0.04
MTNGHalpha_S8_up,0.05
MTNGHalpha_S8_dmean,-0.02
MTNGHalpha_S8_ddown,-0.05
MTNGHalpha_S8_dup,0.03
DESI_OmM_mean,0.34
DESI_OmM_down,0.04
DESI_OmM_up,0.03
MTNGDESI_OmM_mean,0.33
MTNGDESI_OmM_down,0.03
MTNGDESI_OmM_up,0.03
MTNGHalpha_OmM_mean,0.31
MTNGHalpha_OmM_down,0.03
MTNGHalpha_OmM_up,0.02
DESI_S8noh_mean,0.85
DESI_S8noh_down,0.06
DESI_S8noh_up,0.06
MTNGDESI_S8noh_mean,0.84
MTNGDESI_S8noh_down,0.12
MTNGDESI_S8noh_up,0.06
MTNGHalpha_S8noh_mean,0.81
MTNGHalpha_S8noh_down,0.04
MTNGHalpha_S8noh_up,0.04
DESI_OmMnoh_mean,0.33
DESI_OmMnoh_down,0.02
DESI_OmMnoh_up,0.02
MTNGDESI_OmMnoh_mean,0.32
MTNGDESI_OmMnoh_down,0.03
MTNGDESI_OmMnoh_up,0.02
MTNGHalpha_OmMnoh_mean,0.3
MTNGHalpha_OmMnoh_down,0.01
MTNGHalpha_OmMnoh_up,0.01
\end{filecontents*}

   \title{Cosmological constraints from the small scale clustering of Emission Line Galaxies}

   %\subtitle{I. Overviewing the $\kappa$-mechanism}

   \author{Sara Ortega-Martinez           \inst{1,2}\fnmsep\thanks{\email{sara.ortegamtnez@gmail.com}}
          \and
          Raul E. Angulo \inst{1,4}
          \and
          Sergio Contreras \inst{3}
          \and
          Jon\'as Chaves-Montero \inst{5}
          \and
          Matteo Zennaro \inst{6}
          \and
          Sownak Bose \inst{7}
          \and 
          Boryana Hadzhiyska \inst{8,9}
          \and
          C\'esar Hernández-Aguayo \inst{10}
          \and
          Lars Hernquist \inst{11}
          \and
          Volker Springel \inst{12}
          }

   \institute{Donostia International Physics Center (DIPC), Donostia-San Sebastian, Spain
         \and
             University of the Basque Country UPV/EHU, Department of Theoretical Physics, Bilbao, E-48080, Spain
        \and
            Facultad de F\'isica, Universidad de Sevilla, Campus de Reina Mercedes, Av. Reina Mercedes s/n 41012 Seville, Spain
        \and
            IKERBASQUE, Basque Foundation for Science, 48013, Bilbao, Spain
        \and
            Institut de F\'{\i}sica d'Altes Energies (IFAE), The Barcelona Institute of Science and Technology, 08193 Bellaterra (Barcelona), Spain
        \and 
            Institute of Space Sciences (ICE, CSIC), Campus UAB, Carrer de Can Magrans, s/n, 08193 Barcelona, Spain
        \and 
            Institute for Computational Cosmology, Department of Physics, Durham University, South Road, Durham DH1 3LE, UK
        \and
            Institute of Astronomy, Madingley Road, Cambridge, CB3 0HA, UK
        \and
            Kavli Institute for Cosmology Cambridge, Madingley Road, Cambridge, CB3 0HA, UK
        \and
            LadHyX UMR CNRS 7646, École Polytechnique, Institut Polytechnique de Paris, 91128 Palaiseau Cedex, France       
        \and
            Harvard-Smithsonian Center for Astrophysics, 60 Garden St, Cambridge, MA 02138, USA
        \and
            Max-Planck-Institut für Astrophysik, Karl-Schwarzschild-Str. 1, D-85748, Garching, Germany     
             }

   \date{Received April ??, 2026;}

% \abstract{}{}{}{}{} 
% 5 {} token are mandatory

    \abstract{
    Spectroscopic surveys such as the Dark Energy Spectroscopic Instrument (DESI) and Euclid are mapping the spatial distribution of millions of galaxies, with Emission Line Galaxies (ELGs) serving as the dominant tracer in the redshift range $0.8<z<1.6$. Standard approaches for extracting cosmological information from galaxy clustering, however, typically discard highly constraining measurements from the nonlinear regime. We apply SHAMe-SF — a modification of Subhalo Abundance Matching tailored for star-forming galaxy samples — to analyse the three-dimensional clustering of DESI ELGs from the One-Percent data release, extending, for the first time to our knowledge, their cosmological analysis deep into the nonlinear regime. We validate our pipeline using two mock ELG samples drawn from the state-of-the-art cosmological hydrodynamical simulation MillenniumTNG, demonstrating that our model yields unbiased constraints on $\sig$ and $\OmMh$ down to scales of $0.3~\ihMpc$ on both samples. We find that including scales below $r=0.8\ihMpc$ — within the halo — is critical for mitigating projection effects and obtaining unbiased constraints on $\sig$. Applied to the DESI One-Percent measurements, our analysis yields $\sim6$\% constraints on $\sig = 0.81^{+0.05}_{-0.06}$ and $\OmMh=0.146^{+0.009}_{-0.009}$. Remarkably, the accuracy of these constraints is similar to that obtained from the combined full-shape analysis of all DESI DR1 tracers, yet using only 1\% of the survey volume. 
    A naive extrapolation of our results from the One-Percent to the full survey area suggests that the complete survey could deliver roughly an order-of-magnitude improvement in precision — a prospect that, while subject to significant practical challenges, illustrates the cosmological potential encoded in the nonlinear regime. SHAMe-SF is well positioned to exploit this information and disentangle cosmological parameters from the galaxy–halo connection through small-scale ELG clustering.}
 
  %\abstract
   %{Applied to the DESI One-Percent measurements, our analysis yields $\sim6$\% constraints on $\sig = 0.82^{+0.05}_{-0.06}$ and $\OmMh=0.145^{+0.009}_{-0.008}$. Remarkably, the accuracy of these constraints is similar to that obtained from the combined full-shape analysis of all DESI DR1 tracers, yet using only 1\% of the survey volume.}
%{ A naive extrapolation of our results suggests that the complete survey could deliver roughly an order-of-magnitude improvement in precision — a prospect that, while subject to significant practical challenges, illustrates the cosmological potential encoded in the nonlinear regime. SHAMe-SF is well positioned to exploit this information and disentangle cosmological parameters from the galaxy–halo connection through small-scale ELG clustering.}


   \maketitle
%
%-------------------------------------------------------------------
\section{Introduction}
The large-scale structure of the Universe is nowadays the main cosmological target at low redshifts, with Emission Line Galaxies (ELGs) being a key tracer at redshifts 0.8 < $z$ < 2. Surveys like the Dark Energy Spectroscopic Instrument (DESI, \citealt{DESI:2016}), Euclid \citep{Euclid:2025}, or Nancy Grace Roman Space Telescope \citep{Spergel:2015_Roman, Akeson:2019_Roman} will provide unprecedented amounts of measurements of positions and velocities of galaxies. In general, this requires selecting different tracers depending on the redshift interval, with redder, older galaxies as the main targets at lower redshifts and ELGs dominating at intermediate redshifts, given their strong emission lines. These emission lines are generally linked to recent star-formation episodes, thus indicating bluer, younger galaxies. 

State-of-the-art cosmological analyses using ELGs rely on perturbative methods, which are valid for any type of tracer. Some examples are \cite{Ivanov:2021} for ELGs measured in the Extended Baryon Acoustic Oscillation Spectroscopic Survey (eBOSS, \citealt{Dawson:2016eboss}), or the DESI analyses of the Baryonic Accoustic Oscillations \citep{DESI:2024_BAODR1, DESI:2025_BAODR2} and the full shape of the power spectrum \citep{DESI:2025_fullshapeDR1} of ELGs. In these approaches, only quasi-linear scales are used, given the difficulty of incorporating small scales due to non-linear evolution, and the complex connection between galaxies and their host dark matter halos (see, e.g., \citealt{Wechsler:2018}). Finding a way to leverage small scales is crucial to obtaining the maximum information from the available data, especially when linear scales start to be saturated by cosmic variance.

Modelling small scales has been a longstanding problem in cosmological data analysis. Given the availability of observations, the first analyses of small-scale clustering focused on red galaxies and historically relied on the galaxy-halo connection described by the Halo Occupation Distribution (HOD). HODs \citep{Jing:1998a, Benson:2000, Peacock:2000, Berlind:2003, Zheng:2005, Zheng:2007, Guo:2015a,C23_HOD}  parametrise the number of galaxies hosted by a halo of a given mass. HODs have been widely used to derive cosmological constraints using different summary statistics \citep{Tinker:2012,BeyondCollaboration:2024, Yuan:2024_knn}, and also in combination with galaxy-galaxy lensing data \citep{Lange:2023,Lange:2025,Gao:2026}. 

The advent of higher-resolution simulations favoured the development of models which also included subhalos. The most explored was Subhalo abundance matching (SHAM, \citealt{Vale:2006, Shankar:2006, Conroy:2006, Trujillo-Gomez:2011, ChavesMontero:2016}), which associates a subhalo property (typically a mass proxy) with the mass or luminosity of the observed tracer. Regarding cosmological analyses using SHAMs, we highlight the works of \cite{Simha:2013} and \cite{Granett:2019}, using observations from SDSS (\citealt{Zehavi:2011,Abazajian:2009}, and \cite{C22:MTNGCosmology} for the MillenniumTNG hydrodynamical simulation. The model used in the latter (SHAMe, \citealt{C21c}) was subsequently applied by \cite{Mahony:2025}, combining galaxy-galaxy lensing from KiDS-1000 (\citealt{Kuijken:2019}) and clustering measurements from GAMA (\citealt{Driver:2022}), to constrain $\sig$ and $\OmM$.

Models for blue galaxies were developed much later than for red galaxies, given their more complex galaxy-halo connection, the need for higher-resolution simulations, and the limited number of available observations at the time. In recent years, with the prospects of observations from DESI, Roman, and Euclid, important milestones in the development of models of the galaxy-halo connection for ELGs have been reached, now being able to describe the small-scale clustering down to scales of 0.04 $\ihMpc$ \citep{Rocher2023:DESI}. 

In the case of HODs, describing the small-scale clustering of ELGs required modifications of the occupation functional form, satellite prescriptions and assembly bias \citep{Alam:2020,Avila:2020,Hadzhiyska:2022_onehalo,Hadzhiyska2022b,Gao1:2022,Gao2:2023,Reyes-Peraza:2024,Yuan:2024_conformity,VosGines:2024, Rocher2023:DESI}. Using the implementation from \cite{Rocher2023:DESI}, \cite{Paviot:2024} also reproduced the lensing measured in the Year 3 catalogue of the Dark Energy Survey \citep{Sevilla-Noarbe:2021,Gatti:2021} . 

In the case of abundance matching models, \cite{Simha:2012} applied the SHAM formalism to star-forming galaxies from a hydrodynamical simulation, finding it was an accurate description of galaxy clustering for $z>2$. \cite{Granett:2019} obtained similar results for samples from VIPERS \citep{Guzzo:2014,Scodeggio:2018} at $0.5<z<1$ including blue galaxies. Further SHAM modifications can be divided in two approaches: conditional abundance matching, where a secondary selection is performed \citep{Gao1:2022,Gao2:2023,Gao3:2024}, sometimes using an additional (sub)halo property \citealt{Hearin:2013b,Favole:2022,Lin:2023, Favole:2025}; and modifying the sorting criteria before doing the matching, employing  different functional forms, satellite quenching or conformity \citep{Prada:2023,C21c,Yu:2024_DESISHAM,Gao3:2024,SOM2024}. Beyond HODs and abundance-matching-based models, there are also new analytical models that reproduce the luminosity functions of ELGs \citep{Hagen:2025}.

Compared to samples of red galaxies, there are fewer cosmological constraints derived from small scales including blue galaxies. To our knowledge, only \cite{Granett:2019} analysed the small-scale clustering of galaxies from VIPERS (down to 0.1~$\ihMpc$), using SHAM to constrain $\sig$ and the growth index $\gamma$. A key question arises: can we obtain robust inferences from the small scale clustering of ELGs that allow us to maximally exploit upcoming data in modern surveys?

With the advent of these new data, simulations with sufficient resolution and the development of models focused on ELGs, a new path is open to constrain cosmology using small scales. However, to our knowledge, none of these models have been previously applied to spectroscopic ELG data in cosmological analyses. In this work, we provide the first cosmological constraints from small-scale redshift-space clustering using ELGs. We employ the SHAMe-SF model, an extension of SHAM presented in \cite{SOM2024, SOM:2025DESI}. To meet the requirement for high-res simulations across different cosmologies, we use the Bacco simulation suite \citep{Angulo:2021}, combined with cosmological rescaling \citep{Angulo:2010} to span $\sig$, $\OmM$, $\Omb$, $\ns$, $h$, and $\Mnu$. Using scales between 0.3~$\ihMpc$ and 40~$\ihMpc$, the projected correlation function and the monopole and quadrupole of the correlation function, we are able to obtain $\sim$5\% constraints on $\sig$ and $\sim$6\% in $\OmMh$ using ELGs from the DESI One-Percent data: $\sig=\numerror{\cosmopar{DESI_sigma8_mean}}{\cosmopar{DESI_sigma8_up}}{\cosmopar{DESI_sigma8_down}}$ and $\OmMh=\numerror{\cosmopar{DESI_OmMh_mean}}{\cosmopar{DESI_OmMh_up}}{\cosmopar{DESI_OmMh_down}}$. These constraints are statistically consistent with the results of the full-shape analysis combining all DESI tracers and Planck.

We validated our inference pipeline with two samples from the MillenniumTNG hydrodynamical simulation (MTNG, \citealt{Pakmor2022}): one based on the DESI selection criteria and another more similar to an H$\alpha$ selection, and find unbiased constraints in both cases. We further analyse the effect of adding or removing scales to our constraints, finding that scales below $1~\ihMpc$ allow us to break degeneracies between SHAMe-SF and cosmological parameters, and to avoid projection effects, especially for $\sig$. 

The outline of this paper is as follows: in Section~\ref{sec:mocks}, we describe the tools to produce mocks with different cosmologies: the gravity-only simulations used and how to modify them, and the SHAMe-SF model. In Section~\ref{sec:validation}, we describe how we build the validation samples to evaluate the inference pipeline. The target observational data and the details on the inference pipeline are defined in Section~\ref{sec:dataandinferencepipeline}. Afterwards, we present cosmological constraints and robustness tests for the MTNG mock samples (Sect.~\ref{sec:MTNGconstraints}) and DESI (Sect.~\ref{sec:DESIonstraints}). We present our conclusions in Section~\ref{sec:conclusions}. In Appendix~\ref{app:emulator}, we analyse the impact of the emulator and the rescaling on the inferences. We discuss the effect of removing small scales in Appendix~\ref{app:scaleproblems}, and show the posterior distribution of all parameters in Appendices~\ref{app:fullposteriors} and~\ref{app:DESIhigh}.

%--------------------------------------------------------------------
\section{Galaxy clustering with the SHAMe-SF model}
\label{sec:mocks}
In this section, we describe the main elements of our predictions for galaxy clustering: the simulations (Sect.~\ref{sec:baccosim}), the scaling algorithm (Sect.~\ref{sec:scaling}) to generate subhalo catalogues across different cosmologies, and the SHAMe-SF (Sect.~\ref{sec:SHAMeSF}) model to account for how galaxies populate subhalos. Combining them, we describe how we compute galaxy clustering and build an emulator to accelerate the predictions in Sect~\ref{sec:clusteringstatistics}.

\subsection{Bacco simulation suite}
\label{sec:baccosim}

\begin{table}
\caption{Specifications and cosmological parameters of the Bacco simulations.}
\begin{center}
\begin{tabular}{llllll}
 & Vilya & Nenya & Narya & TheOne & Barahir \\ \cline{1-6} \vspace{-0.1cm}\\
$\sig$ & 0.9 & 0.9 & 0.9 & 0.9 & 0.9  \\
$\OmM$ & 0.27 & 0.315 & 0.36 & 0.307 & 0.403 \\
$\Omb$ & 0.06 & 0.05 & 0.05 & 0.048 & 0.05 \\
$\ns$ & 0.92 & 1.01 & 1.01 & 0.96 & 0.965 \\
$\h$ & 0.65 & 0.6 & 0.7 & 0.6777 & 0.7  \\
$\Mnu$ & 0.00 & 0.0 & 0.0 & 0.0 & 0.15  \\
$m_p$ & 2.77 & 3.24 & 3.7 & 3.16 & 4.11  \\
\end{tabular}
\end{center}
\tablefoot{Neutrino masses are expressed in eV, and the particle masses of the simulation ($m_p$) in units of $10^9\hMsun$.}
\label{tab:baccosims}
\end{table}

We employ gravity-only simulations to evaluate SHAMe-SF on different cosmologies and to construct covariance matrices. In both cases, we use simulations from the ``BACCO simulation project'' \citep{Angulo:2021}. For the emulator, we take the five $1024 \, \hMpc$ boxes ($3072^3$ particles) with the cosmologies and specifications described in Table~\ref{tab:baccosims}. We discuss further in Sect.~\ref{sec:scaling} how cosmology is varied. For the covariance matrices, we use the $1.5 \, \ihGpcC$ boxes, which are three times larger than MTNG, and run with 4320$^{3}$ particles. 

The initial conditions of all simulations were computed using 2LPT, fixing the mode amplitudes to the ensemble average to suppress the variance on large scales \citep{Angulo:2016}. Halos and subhalos are identified on-the-fly using the Friends-of-Friends algorithm \citep[\texttt{FoF;}][]{Davis:1985} and a modified version of \texttt{SUBFIND} \citep{Springel:2001} respectively, computing the properties required by SHAMe-SF without needing to reanalyse the merger trees.

\subsection{Cosmological rescaling}
\label{sec:scaling}
Using the scaling technique \citep{Angulo:2010}, we can modify the output of a gravity-only simulation to match it to a simulation run with different cosmological parameters, reproducing its matter distribution. This is achieved by displacing the particles, halos, and subhalos of the simulation, and modifying their properties, such as host halo mass and $\vpeak$, the maximum value over redshift of the maximum circular velocity of a subhalo.

We cover a cosmological parameter space around 10$\sigma$ of the Planck cosmology, similar to \cite{C20, C22:MTNGCosmology, Mahony:2025}. Using the five simulations from Table~\ref{tab:baccosims}, the accuracy of the scaling of the power spectrum is better than 2\% (see also \citealt{Ruiz:2011, Guo:2013a,Zennaro:2019} for further details). Reading and scaling a snapshot with the full particle data to a target cosmology takes $\sim$15 minutes on a single CPU.

\subsection{The SHAMe-SF model}
\label{sec:SHAMeSF}
\begin{figure}
                \centering
                \includegraphics[width=0.40\textwidth]{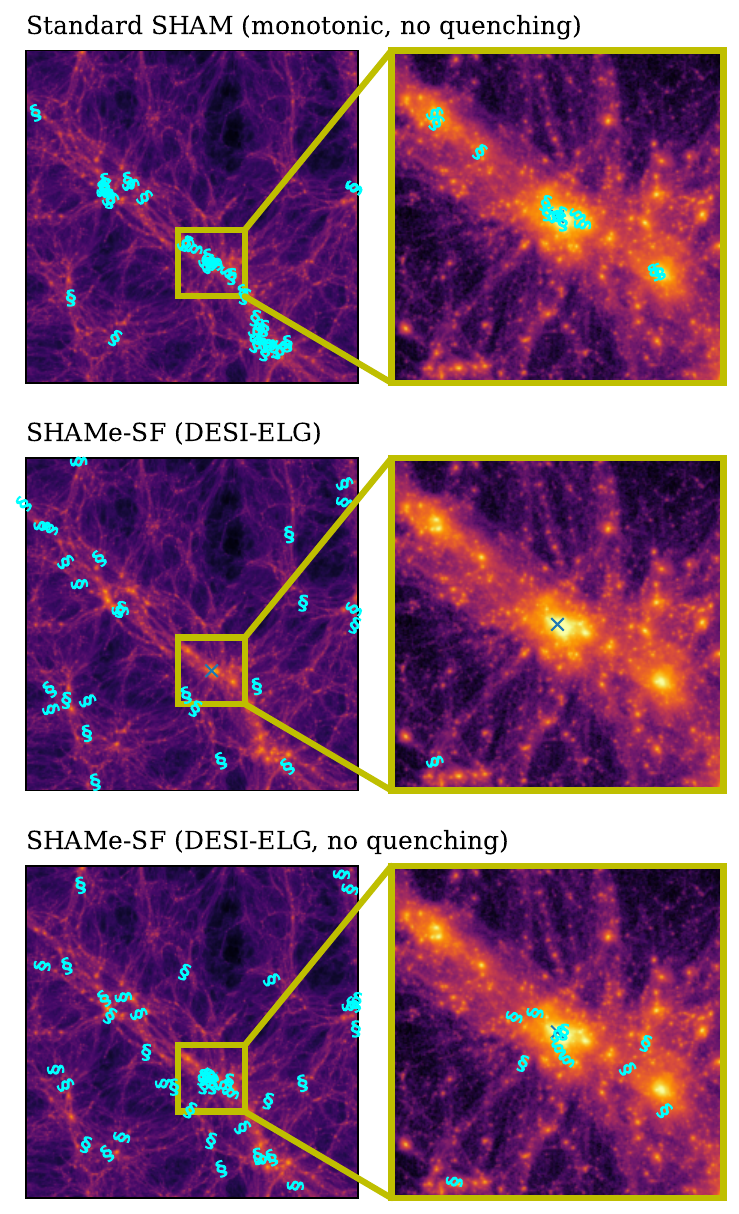}       
        \caption{Visualizations of the subhalo selection for different SHAMe-SF parameter choices. We show, for one slice of one of the Bacco simulations, the dark matter density field, and the subhalos selected by three sets of SHAMe-SF parameters (cyan randomly-aligned section symbols, \S). On the top SHAMe-SF realization, the chosen subhalos have the highest $\vpeak$ values (without sorting or quenching mechanism), similar to the standard SHAM. In the middle and bottom realizations, we choose values of the best fit parameters to DESI-ELGs, and change the quenching factors (note the difference in the presence of satellites between both panels).}
        %\caption{Visualizations of the subhalo selection for different parameter choices of SHAMe-SF. We show, for one slice of one of the Bacco simulations, the dark matter density field (left), the subhalos identified by Subfind (centre-left), and the subhalos selected by three sets of SHAMe-SF parameters (full-field on the centre-right panel, zoom for two halos on the rightmost panel). On the top SHAMe-SF realization, the chosen subhalos have the highest vpeak values. In the middle and bottom realizations, we choose values of vpeak closer to other ELG studies and change the quenching factors (note the difference in the presence of satellites between both panels).}
        \label{fig:lss}
\end{figure}

SHAMe-SF is an extension of SHAM designed to reproduce the clustering of star-forming-like galaxy samples. It was built using physically motivated prescriptions from the TNG simulation in \cite{SOM2024}, and further modified and applied to observational data in \cite{SOM:2025DESI}. We refer the reader to those papers for detailed descriptions and validation; here, we provide an overview of the model for completeness and readability. 

We use SHAMe-SF to populate subhalos of gravity-only simulations based on their properties, rank-ordering them by assigning a score. Subhalos are defined as every structure identified by the \texttt{SUBFIND} algorithm. Thus, each halo would be comprised of a central subhalo (the main object) and potentially several satellite subhalos. 

First, we set an initial score using a functional form depending on $\vpeak$, the peak value over redshift of the maximum circular velocity of the subhalo. We name this initial score ${\rm SFR}|_{({\vpeak})}$ since the model was originally designed for SFR-like samples, and is defined as:

\begin{equation}
  {\rm SFR}|_{({\vpeak})} \propto
    \begin{cases}
       \left[ \left( \frac{\vpeak}{V_{1}}\right)^{-\beta} + \left( \frac{\vpeak}{V_{1}}\right)^{\gamma} \right]^{-1}, & \text{if $\vpeak \leq V_{1}+\Delta V_1$}\\
       & \\
      \text{SFR}_{V_{1}+\Delta V_1} \left/~  \left( \frac{\vpeak}{V_{1}+\Delta V_1}\right)^{\gamma + \Delta\gamma}  \ \right.\ & \text{if $\vpeak$ > $V_{1}+\Delta V_1$}
    \end{cases}       
    \label{eq:model}
\end{equation}
where $\beta$, $\gamma$, $\Delta\gamma$, $V_1$ and $\Delta V_1$ are the first five free parameters. Then, we add a log-normal scatter with width $\sigma$, incorporating partial internal sorting based on a proxy for subhalo concentration. The sorting is done separately for centrals and satellites to keep the satellite fraction fixed, and is regulated by two additional parameters, $f_{k,\rm{cen}}$ and $f_{k,\rm{sat}}$. When $f_{k,\rm{i}}$ = 1, subhalos are fully sorted within the $\vpeak$ interval based on the secondary property, while $f_{k,\rm{i}}$ = 0 keeps the distribution random. Negative values result in anti-correlation; i.e. at fixed $\vpeak$, star forming galaxies tend to live in less concentrated subhaloes.

After this initial assignment, we add a quenching mechanism dependent on the host halo mass and the time since the peak mass of each subhalo. This process mainly tackles the quenching of satellites when falling into a larger halo, but also splashback galaxies (centrals that had a satellite phase through their evolution), which would have lost mass (and gas) due to tidal stripping. The quenching is controlled by three parameters: $M_{\rm crit}$ determines the slope of the quenching function and the minimum halo mass where quenching becomes efficient, and $\alpha_0$ and $\alpha_{\rm exp}$ regulate the amplitude of quenching for each host halo mass. 

We show three examples of galaxy distributions for different combinations of SHAMe-SF parameters in Figure~\ref{fig:lss}. The upper panel is equivalent to traditional abundance matching, which selects subhalos with the highest $\vpeak$ (monotonic power law, no secondary sorting or quenching). Objects are typically located in high-density regions. The middle panel shows the distribution for the best-fit parameters to DESI-ELGs. Galaxies populate less massive subhalos, further from the nodes of the cosmic web. The third panel shows the galaxy distribution keeping the same parameters for the functional form and the semi-sorting, but turning off the quenching mechanism.

To compute the clustering of a sample with number density $\bar{n}$, we assign the SHAMe-SF score to all subhalos and then take $N = \bar{n}V$ subhalos with the highest values, where $V$ is the volume of the gravity-only simulation. Note that we did not include a normalisation factor in the expression. The clustering signal will be independent of this normalisation as long as the same objects are chosen. Similar reasoning applies to the degeneracies between parameters.

\subsection{Galaxy clustering and emulator building}

\label{sec:clusteringstatistics}
\label{sec:emulator}

To obtain the cosmological constraints, we fit the projected correlation function, the monopole and the quadrupole of the correlation function, all of them considering redshift space distortions. The projected correlation function is computed as:
 
\begin{equation}
    w_{\rm p} \equiv 2 \times \int_{0}^{\pi_{\rm max}} d\pi\, \xi(r_{\rm p}, \pi),
\end{equation}
\noindent where $\xi(\rm p, \pi)$ is the correlation function, and using the line-of-sight integration limit $\pi_{\rm max}=40~\ihMpc$, as in \cite{Rocher2023:DESI}. We choose logarithmically spaced bins in $r_{\rm p}$ between 0.1 and 50$~\hMpc$ (given the resolution of the simulations, and how it changes with the cosmological rescaling). 

We compute the multipoles of the correlation function as:
\begin{equation}
    \xi_{\ell} \equiv \frac{2 \ell+1}{2} \int^{1}_{-1} d\mu\, \xi(s,\mu)P_\ell(\mu),
\end{equation}
\noindent where $P_{\ell}$ is the $\ell$-th order Legendre polynomial.

Since we are dealing with different cosmologies, it is necessary to consider the Alcock-Paczynski effects \citep{AlcockPaczynski}. We take the same approach as \cite{Lange:2019}, where they are included at the catalogue level by reshaping the box and changing the coordinates accordingly. We computed all the distances assuming the same cosmological parameters as the DESI One-Percent cosmology (see \citealt{Rocher2023:DESI}), for both the SHAMe-SF realisations and the mock ELGs from MTNG.

Scaling a simulation, creating a mock with SHAMe-SF, and computing clustering statistics takes $\sim$15 minutes, making it infeasible to perform an MCMC-like analysis using direct evaluations of the model. To speed up model evaluations, we train one feedforward neural network emulator per clustering statistic \citep[see ][]{Angulo:2021,Arico:2020,Arico:2021,Pellejero:2023,Zennaro:2023,C2023:lensing,SOM2024}. Each emulator can vary the SHAMe-SF parameters, cosmological parameters ($\sig$, $\OmM$, $\Omb$, $\ns$, $h$ and $\Mnu$), redshift, and number density simultaneously within the following ranges for the SHAMe-SF parameters:
\begin{eqnarray}
        \beta & \in & [0,20]\nonumber \\
        \gamma & \in & [-10,25] \nonumber \\
        \Delta\gamma & \in & [-10,10] \nonumber \\
        V_1 & \in & [10^{1.2},10^{3.5}] \text{ (km/s)} \nonumber \\
        \Delta V_1 & \in & [10^{0.2},10^{1.9}] \text{ (km/s)} \nonumber \\
        \sigma & \in & [0,1.7] \nonumber \\
        f_{k,\rm{(cen+sat)/2}} & \in & [-1,1] \nonumber \\
        f_{k,\rm{(cen-sat)/2}}  & \in & [-1,1] \nonumber \\
        \alpha_0 & \in & [0,8] \nonumber\\
        \alpha_{\rm exp} & \in & [-8,8]  \\ 
        M_{\rm crit} & \in & [9,14.5] \ (\log(\hMsun)) \nonumber ,
        \label{eq:par_range}
\end{eqnarray}
for the cosmological parameters:
\begin{eqnarray}
\nonumber\\
        \sig & \in & [0.65,0.9] \nonumber\\
        \OmM & \in & [0.23,0.4] \nonumber\\
        \Omb & \in & [0.04,0.06]\\
        \ns & \in & [0.92,1.01] \nonumber\\
        \h & \in & [0.6,0.8] \nonumber\\
        \Mnu & \in & [0,0.4] \nonumber      
        ,
        \label{eq:par_range}
\end{eqnarray}
and for sample-specific parameters. 
\begin{eqnarray}
        z & \in & [0,2] \nonumber\\
        \log(\bar{n}) & \in & [-4,-2.25] \ (\log(\ihMpcC))        
        ,
        \label{eq:par_range}
\end{eqnarray}

To create the training set, we use 23200 cosmologies distributed in 15 Latin hypercubes. After choosing the simulation of the suite closest to each target cosmology (see \citealt{C22:MTNGCosmology}), we find the associated redshift for each snapshot of the simulation for the target cosmologies. We assign a random set of SHAMe-SF parameters to each rescaled snapshot and compute clustering for six logarithmically spaced number densities in the range $[3.16,50.62]\times 10^{-4}\ihMpcC$. This leads to 179514 combinations of SHAMe-SF parameters and redshifts, and thus $\sim 10^6$ points are used to train the emulator, given the different number densities.

We train each emulator with the Keras front-end of the TensorFlow library \citep{tensorflow}, using the Adam optimization algorithm with a learning rate of 0.001, and the mean squared error loss function, weighting each scale by the diagonal of the preliminary covariance matrix defined in Section~\ref{sec:covariancematrices}. Each emulator comprises 5 fully connected hidden layers of 400 neurons with GeLU activation functions, and we add dropout layers with a dropout rate of 0.05 during training to prevent over-fitting.

We built two additional test sets: one with the same ranges of cosmological parameters, restricted to scale factors between 0.48 and 0.52, comprising 3508 combinations of cosmologies, SHAMe-SF parameters and number densities; and another test set with the Planck cosmology and 1000 SHAMe-SF parameters. We use these sets to estimate the expected errors on the emulator and the best-fit finder discussed in Section~\ref{sec:PSO}, which are discussed in Appendix~\ref{app:emulator}. When comparing the emulator prediction to the test set, emulator errors are below 3\% of the DESI-ELG signal for all scales considered in the analysis of the projected correlation function and the monopole, and within 7\% for the quadrupole (dominated by points where it changes from positive to negative values). As analysed in Appendix~\ref{app:emulator}, across all scales, the emulator error is smaller than the error from the DESI sample analysed in this work (see Section~\ref{sec:DESIobs}).

\section{Validation mocks in MilleniumTNG}
\label{sec:validation}
Before applying SHAMe-SF to observations, we test its performance using mock samples defined on a hydrodynamical simulation. We begin by introducing the MillenniumTNG hydrodynamical simulation (Sect.~\ref{sec:MTNGsim}), used to define the validation samples. We discuss the criteria for defining ELG samples that mimic different surveys in Section~\ref{sec:mockELG}.

\subsection{MilleniumTNG}
\label{sec:MTNGsim}
The MilleniumTNG simulation suite is a set of simulations designed for analyses of galaxy formation on volumes comparable to current cosmological surveys. The fiducial run has the same size as the Millennium gravity-only simulation and the baryonic physics prescriptions of the TNG simulation suite \citep{Nelson:2017TNGcolors, TNGb, TNGc, TNGd, TNGe}. Presented in \cite{HernandezAguayo2022,Pakmor2022,Barrera2022,Kannan2022,Delgado2022,Ferlito2022,C22a,Hadzhiyska:2022_onehalo,Hadzhiyska:2022,Bose2022}, the fiducial run evolves 4320$^3$ dark matter particles and gas cells, with masses of 1.7$\times 10^8$~M$_\odot$ and 3.1$\times 10^7$~M$_\odot$, respectively, on a 500~$\ihMpc$-side (740 Mpc) box.

MTNG, run with the moving-mesh code AREPO \citep{AREPO}, accounts for radiative cooling and star formation, black hole growth, and their associated feedback processes, as well as supernova feedback among other astrophysical processes. The cosmological parameters adopted for the MTNG simulation are the ones from \cite{Planck2015}, the same as for IllustrisTNG: $\OmM$ = 0.3089, $\Omb$ = 0.0486, $\sig$ = 0.8159, $\ns$ = 0.9667, $h$ = 0.6774, and $\Mnu = 0$ MeV.

\subsection{ELG selection in MTNG}
\label{sec:mockELG}
Accurate modeling of emission lines and colours would require accounting for the emission of all stellar populations and processes such as nebular emission, burstiness (star formation with time scales below the time resolution of the simulation), or dust obscuration (see e.g. \citealt{Knebe:2022,Madar:2024,Rapoport:2025,Osato:2026}). In this work, we take a simplified approach and define the mock ELG samples using properties of galaxies such as the stellar mass and the star formation rate (SFR). In \cite{Hadzhiyska:2021}, the authors showed the equivalence of the HOD and clustering of two samples selected by colours and cuts in stellar mass and sSFR. This eases the modelling effort on defining ELG samples, an approximate option to test different scenarios for the model and the inference pipeline. We build two mock catalogues from MTNG, a DESI-like sample, whose ELGs are selected by colour cuts and spectroscopic confirmation of the [OII] doublet, and a $\Halpha$-like sample, more similar to surveys like Euclid and Nancy Roman.

\subsubsection{MTNG-DESI}
Cross-matching DESI's ELGs with galaxies from the COSMOS survey \citep{Weaver:2022_Cosmos}, \cite{Yuan:2024_conformity} obtained the stellar masses and star formation rates of DESI ELGs in the redshift interval $0.8 < z <1.1$. The authors found that DESI's colour and magnitude selection discards massive galaxies even if they have high SFRs, and DESI ELGs are well localised in a region in SFR, $M_*$ (above the Main Sequence).  Similar to \cite{SOM:2025DESI}, we matched COSMOS' cumulative stellar mass and SFR distributions to the $z = 0.95$ snapshot of MTNG (the effective redshift of the sample), and defined the equivalent thresholds in star formation rate and stellar mass, as well as introducing the main sequence cut. The obtained number density was $\bar{n} = 3.3\times 10^{-3}\,\ihMpcC$. The resultant cuts for stellar mass and SFR were: $10^{8.7}<M_*<10^{10.4}$ $M_\odot$, and SFR$>3$ $M_\odot$yr$^{-1}$.

\subsubsection{MTNG-$H\alpha$}
Following a similar strategy, we define a sample equivalent to an $\Halpha$ selection by taking the galaxies in MTNG with the highest SFR. We also choose $z = 0.95$ to ease the comparison with MTNG-DESI, and select a number density of $\bar{n} = 3.1\times 10^{-3}~\ihMpcC$, in the range of the values from the HOD-populated lightcones of \cite{Euclid51_powerrealspace} ($\bar{n} = 3.7\times 10^{-3}$ and  $2.0\times 10^{-3}~\ihMpcC$), but higher than the estimations from \cite{Pozzetti:2016}: $\bar{n} = 2.5\times 10^{-3}$, $2.8\times 10^{-3}$, and $1.3\times 10^{-3}~\ihMpcC$, for different parametrizations of the redshift evolution of the H$\alpha$ luminosity function in the redshift range $0.9<z<1.1$.

\section{Inference pipeline and observational data}
\label{sec:dataandinferencepipeline}
In this section, we discuss the technical aspects of the inference pipeline and the specifics of the target observational sample (Sect.~\ref{sec:DESIobs}). We describe the sampling algorithm used to explore the parameter space in Section~\ref{sec:NUTS} and find the best fit in Section~\ref{sec:PSO}, where we use the covariance matrices defined in Section~\ref{sec:covariancematrices}. We discuss the priors in Sect.~\ref{sec:priors}.

\begin{figure}
                \centering
                \includegraphics[width=0.48\textwidth]{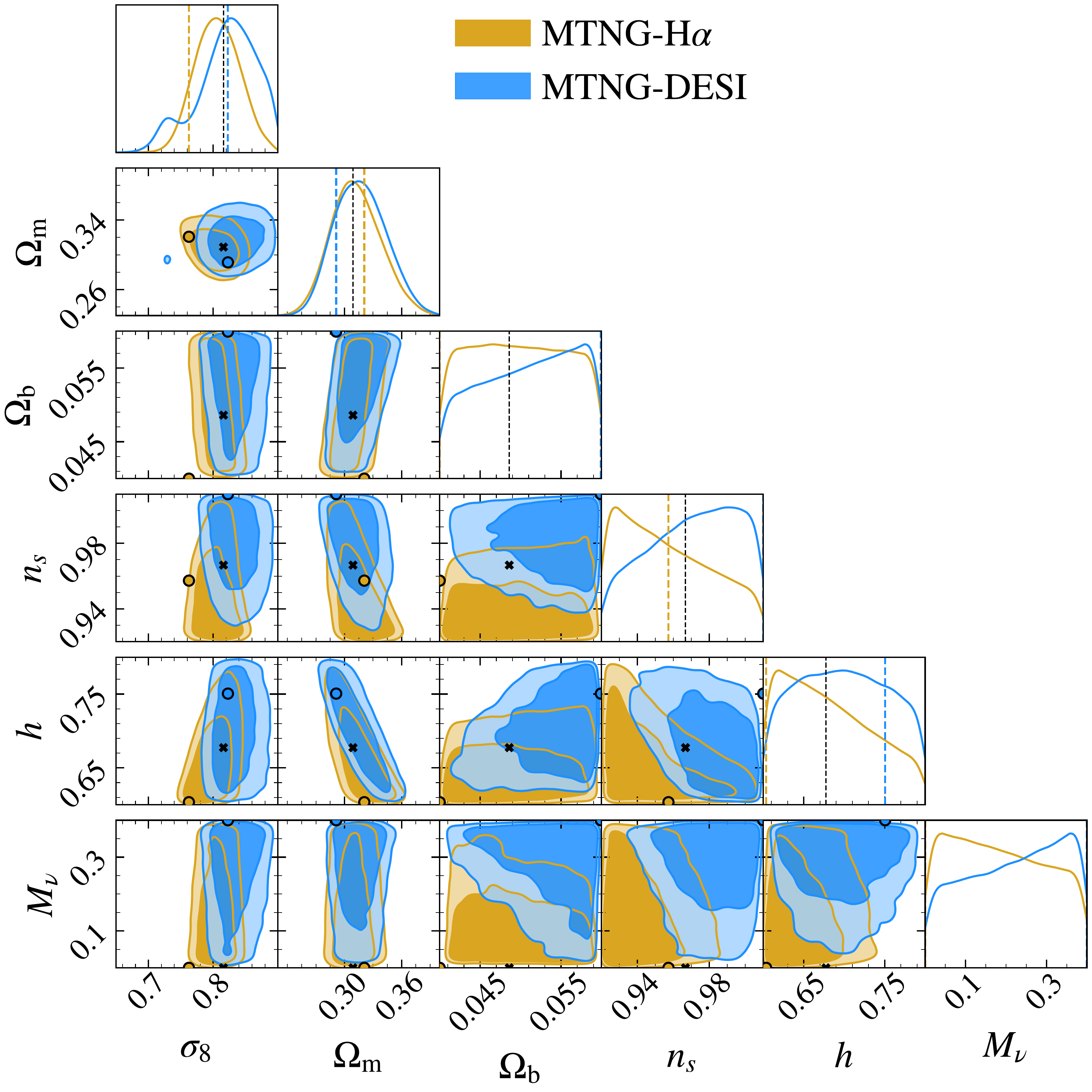}       
        \caption{Posteriors (1 and 2$\sigma$ contours) on the cosmological parameters for a MTNG-DESI (blue) and MTNG-H$\alpha$ (yellow) preliminary fit with all the cosmological parameters free. The circles (2D) and vertical dashed lines (1D) show the best-fit parameters. The black cross and black dashed lines mark the true cosmology of the MTNG simulation.}%\jonas{Explain the meaning of the points, the shaded areas, the vertical lines, etc.}}
        \label{fig:fullposterior}
\end{figure}

\subsection{ELGs in DESI One-Percent data release}
\label{sec:DESIobs}

In June 2023, the DESI Survey publicly released measurements from the first one per cent of its footprint \citep{DESI:2023_EDR}, as part of the Survey Validation program \citep{DESI:2023_validation}, and the spectra from the first year of data collection \citep{DESI:2025_DR1}, used in the BAO and Full-Shape analyses \citep{DESI:2024_BAODR1,DESI:2025_fullshapeDR1}. Galaxies are first identified based on colour selection (for ELGs: \citealt{DESI:2016, Raichoor:2020_DESIELG, Raichoor:2023_DESIELG}) and are targeted afterwards for spectroscopic confirmation.  For a given field of view, photometrically identified targets are given a priority based on their type to use the spectrograph, with ELGs having a relatively low priority. Thus, there are sections of the survey footprint where the spectra of ELGs have not yet been measured. This is not the case for DESI One-Percent, where there is even more than one passage per object for reliable redshift estimations. Even if the sky area scanned consisted of 140 sq deg, since we are interested in smaller scales, this is sufficient for a first test of our inference pipeline on real data without requiring us to model fibre assignment systematics (see e.g. \citealt{Pinon:2025}) due to overlap in the field of view at different redshifts for multiple tracers. 

As in \cite{SOM:2025DESI}, we use the clustering measurements from \cite{Rocher2023:DESI}, for scales $r \in [0.3, 30]\,\hMpc$. Specifically, we take the samples divided into two redshift intervals: a low-$z$ and a high-$z$ bin: $z\in[0.8, 1.1]$ and $ z\in [1.1, 1.6]$. Even if we analysed both bins, we only discuss in the main text the results for the low-z bin (from now on, DESI-ELG) since the high-z sample did not pass the Emulator test described in Appendix~\ref{app:emulator}: populating a scaled simulation with the resultant parameters does not match the prediction of the emulator within the error bars. We show our inferences and tests for the high-$z$ sample anyway in Appendix~\ref{app:DESIhigh}.

\subsection{Exploration of the parameter space}
\label{sec:NUTS}
In previous works, we used {\tt emcee} \citep{emcee} to explore the parameter space and infer the posterior distributions. Since neural network-based emulators are differentiable by construction, it is possible to use alternative sampling strategies that include gradients to inform the proposal of new steps on the chains, such as Hamiltonian Monte Carlo (HMC, \citealt{Duane:1987_HMC}, see also \citealt{Betancourt:2017_HMC}). Conceptually,  the likelihood is used as a potential, and new steps are sampled from trajectories with a momentum drawn from a Gaussian distribution. This approach enables a more efficient exploration of degeneracies between parameters. We use the No-U-Turn Sampler (NUTS, \citealt{Hoffman:2011_NUTS}), an extension of HMC where the step size and mass matrix are self-calibrated on an initial warm-up phase. When simulating the trajectories for the new state proposal, the algorithm includes a stopping criterion when a turning point is reached, avoiding exploring the same region twice on the same step. HMC and NUTS have been applied in cosmological and astrophysical parameter inference setups over the last few years, finding a great improvement in efficiency compared to typical MCMC sampling \citep[see e.g.,][]{Piras:2023,Mootoovaloo:2024,Zacharegkas:2025_sampling}.

We chose the NUTS implementation in NumPyro \citep{NumPyro}, and transformed the weights and biases of the Keras NN-emulators to a JAX-based function. We define the likelihood as a multivariate Gaussian: 
$\log \mathcal{L}(\vec{\theta}) = -\chi^2/2 = -(\vec{d}-\vec{t})^t \mathcal{C}^{-1} (\vec{d}-\vec{t})/2$, where $\vec{d} = \{\wpp, \xi_{\ell=0}, \xi_{\ell=2} \}$ is the data vector and $\mathcal{C}$ is the covariance matrix discussed in Section~\ref{sec:covariancematrices}. We take uniform flat priors on the SHAMe-SF parameters coinciding with the emulator ranges, and fix $\bar{n}$ and $z$ depending on the sample fitted. The priors on the cosmological parameters will be further discussed in Section~\ref{sec:params}, when analysing the constraints and setting the fiducial analysis, but the ranges considered coincide with the emulator intervals for which the scaling technique has been validated. We run 500 chains of 7000 warm-up steps and 2000 sampling steps on one GPU, taking two and a half hours in total per run. To check that chains are converged, we divide them into 7 segments and compare their average likelihoods to the chain's average likelihood. 

In a non-exhaustive comparison with MCMC, we tested how many steps each algorithm required to converge to the final sampled $\log \mathcal{L}$ distribution for the same number of chains (for the setup described in Sect.~\ref{sec:priors}, with all parameters freed). For NUTS, the likelihood distribution was already very similar after roughly 100-200 warm-up steps, while MCMC needed 8000 steps to converge to the $\log \mathcal{L}$ distribution. Comparing running times per step, we find $\sim$1.1s/step for NUTS and $\sim$0.17s/step for the MCMC analyses, returning minimum convergence times of 22 minutes for MCMC and 4 minutes for NUTS. Note that we set a larger number of warm-up steps in NUTS to ensure convergence regardless of the covariance matrix, statistics, or scales used for each sample. 

\subsection{Finding the best fit}
\label{sec:PSO}
To find the best fit for each case, we run a particle swarm optimisation algorithm \citep[PSO,][]{PSO}. It relies on a group of ``particles” (a swarm) that explore the parameter hyperspace, updating their positions at each step using a combination of their individual best positions, the global best position of the ensemble, and an inertia term. Positions are evaluated using the $\chi^2$ distribution defined in the previous section, and adding in the fiducial cosmological run the priors defined in Section~\ref{sec:priors}
. We use the implementation from \cite{PSOBACCO}\footnote{\url{https://github.com/hantke/pso_bacco}}. We run the optimiser five times with 500 walkers for 1000 steps and keep the run with the lowest final $\chi^2$ to mitigate dependencies on the initial positions of the walkers.

\begin{figure*}
                \centering
                \includegraphics[width=0.90\textwidth]{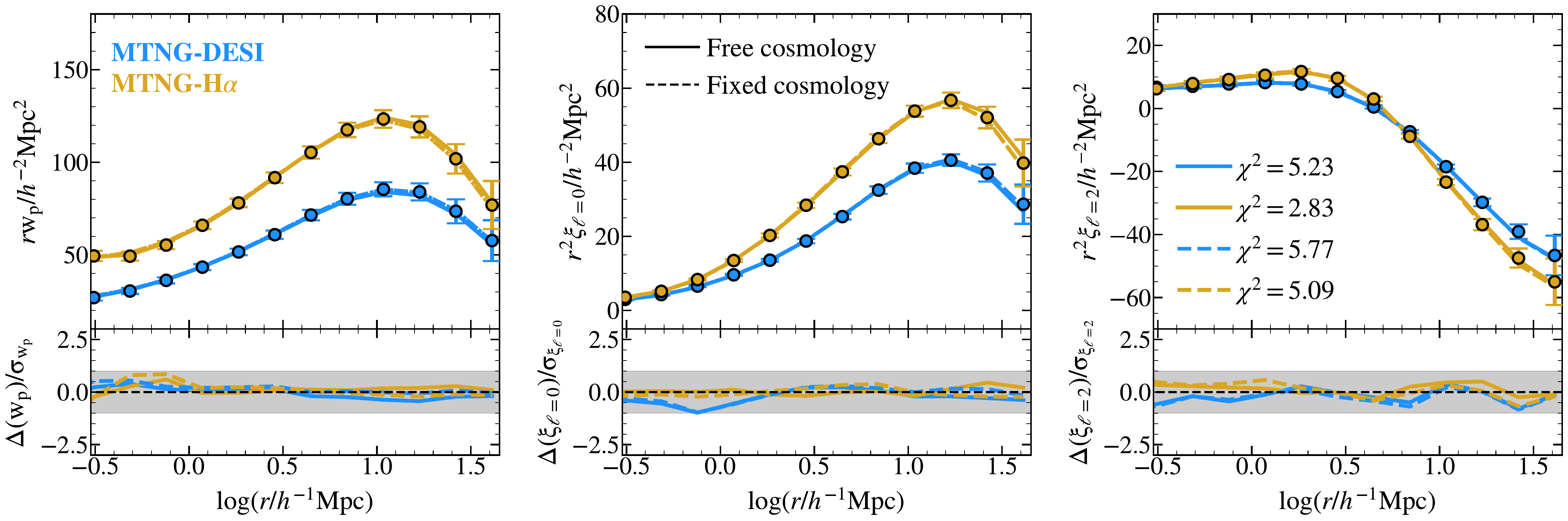}

        \caption{Projected correlation function ($\wpp$) and the monopole ($\xi_{\ell=0}$) and quadrupole ($\xi_{\ell=2}$) of the redshift--space correlation function of ELGs in MTNG-DESI (blue) and MTNG-H$\alpha$ (yellow), together with the corresponding best-fit SHAMe-SF model fixing all the cosmological parameters (dashed line) and with the default configuration described in freeing $\sigma_8$, $\Omega_{m}$, $\Omega_{b}$, $n_s$ and $h$ (solid) with the priors discussed in Section~\ref{sec:priors}. Bottom panels: Difference between the data and the fit with SHAMe-SF in units of the diagonal elements of the respective covariance matrix. }
        \label{fig:clusteringMTNG}
\end{figure*}

\subsection{Covariance matrices}
\label{sec:covariancematrices}

In our covariance matrices, we consider three contributions: from the sample, the emulator, and the scaling algorithm. To estimate the sample covariance matrices of the validation MTNG samples, we compute a preliminary covariance matrix with $3^3$ jackknife (JN, \citealt{Zehavi:2002,Norberg:2009}) samples using the MTNG simulation. Using the diagonal terms, we run the sampler, fixing the cosmological parameters to those of the simulation, to obtain an initial fit with the SHAMe-SF model. We use those parameters to obtain the non-diagonal terms of the covariance matrix by populating one of the largest boxes in the Bacco simulation suite (L = 1.5~$\ihGpc$), scaled to the Planck cosmology, and recomputing the JN covariance matrix with $9^3$ samples. 

To compute the emulator contribution, we quantify the differences between the test set and the emulator prediction using the 100 points whose clustering signal is closest to each sample (weights are computed using the previously computed sample covariance matrix). We normalize the differences in $\wpp$ by the value of $\wpp$ and the differences in the monopole and quadrupole by the monopole, and scale them back again by the same functions evaluated on each MTNG sample, and then compute the covariance of the differences. 

To quantify the error of the scaling technique, we use an additional test set with fixed Planck cosmology and 1000 combinations of SHAMe-SF parameters. For each point, we compute the clustering populating a simulation run with Planck cosmology (target cosmology) and TheOne simulation scaled to the same cosmological parameters (see Table~\ref{tab:baccosims} for the original parameters of the simulation). As in the case of the emulator, we take the 100 points with the closest clustering to each sample, compute and rescale the differences, and build their contribution to the covariance matrix. Ideally, this should be done for a range of cosmologies. To avoid underestimating the error introduced by the scaling, which is only computed for the Planck cosmology (due to the lack of simulations of the same size), we double the contribution to the error bars from this term. The resultant scaling error is roughly 3\% of the signal, compatible with calibrations from \cite{C20} and \cite{C22:MTNGCosmology}. Similar to the latter, the contributions of the scaling and the emulator to the covariance matrices are comparable for all the samples and statistics considered except for the projected correlation function of the MTNG-DESI sample, where the scaling is larger ($\sim$3\% vs. $\sim$2\% of the clustering signal, respectively). For both mock samples the JN error is smaller than the contribution of the emulator and the scaling for scales below 8~$\ihMpc$. In the case of DESI-ELG, the main contribution to the covariance matrix is the data term.

%################

%############

\subsection{Priors on the cosmological parameters}
\label{sec:priors}
\label{sec:params}
Given the large parameter space, we expect to find degeneracies between both SHAMe-SF and cosmological parameters, as well as unconstrained parameters that do not have a significant effect on the clustering signal. We dedicate this section to analysing priors on the cosmological parameters.

To set our fiducial setup, we first analyse the cosmological parameters. We run the sampler, freeing both the SHAMe-SF parameters and the cosmological parameters, and examine the degeneracies and our constraining power on them. Constraints on the cosmological parameters are shown in Figure~\ref{fig:fullposterior}. We have constraining power on $\sig$ and $\OmM$, but not on $\Omb$, $\ns$, $h$ or $\Mnu$. Additionally, there is a degeneracy between $\OmM$ and $h$. For the MTNG samples, we set the neutrino mass value to $\Mnu=0$ MeV, and keep $\Mnu=0.06$ MeV for DESI \citep{Planck:2018}. We take a Gaussian prior for $\ns$ from Planck measurements for the DESI analysis given by $ \ns=\mathcal N(0.965,0.004)$, and the fiducial value of MTNG with the same width for the samples defined in the simulation. We also set a Gaussian prior on $\Omb h^2$ of $\mathcal N(0.0224,0.0001)$ for DESI, and $\mathcal N(0.0223,0.0001)$ for MTNG. We choose a flat prior for $\sig$, $\OmM$, and $h$, but we compute $\OmMh$ when analysing the posteriors given the strong degeneracy between both parameters.

Given the difference in meaning between the elements of the model, and to avoid limiting its performance for other samples, we keep the 11 SHAMe-SF parameters free using the priors set by the emulator limits.

\section{Cosmological constraints from galaxy clustering on MTNG}
\label{sec:MTNGconstraints}

\begin{figure}
                \centering
                \includegraphics[width=0.40\textwidth]{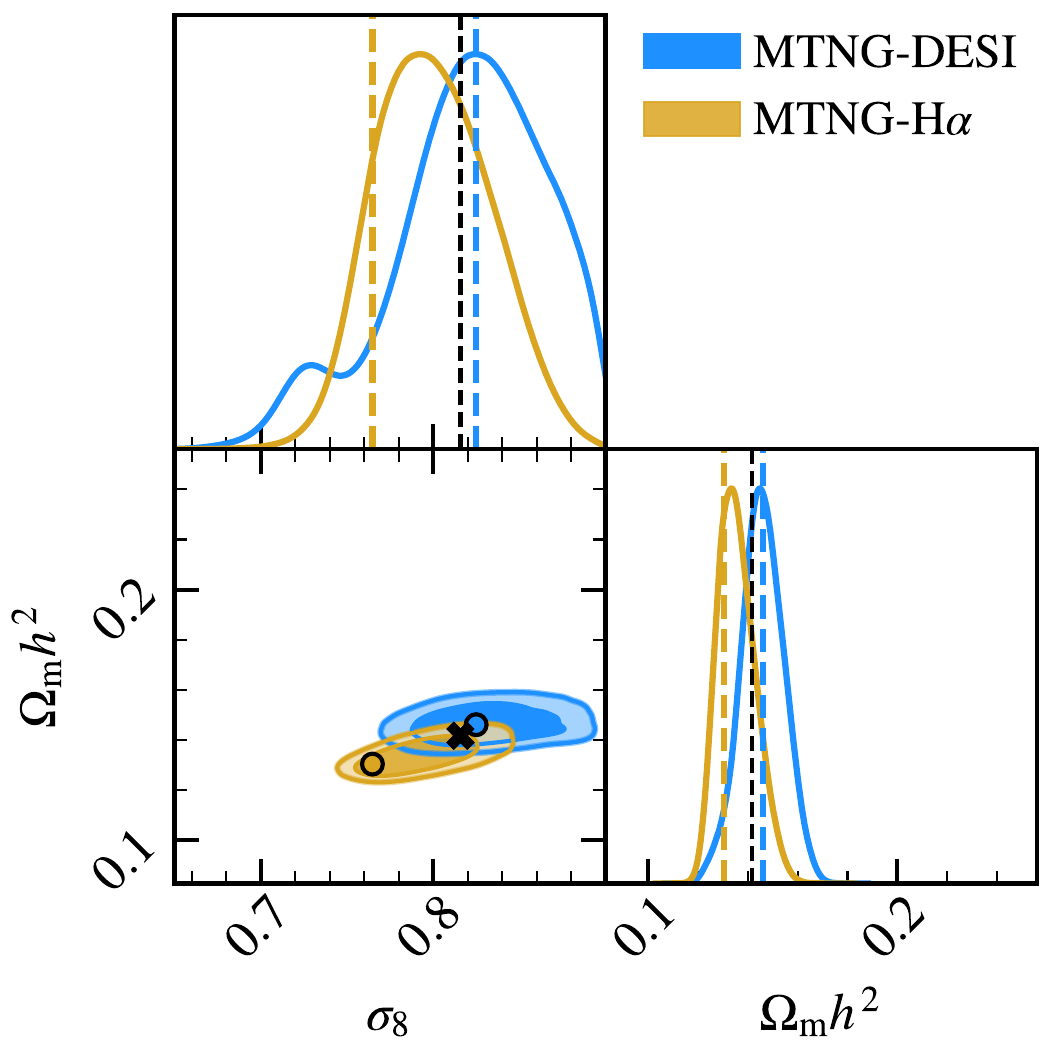}
            
        \caption{Projected constraints on $\sig$, and $\OmMh$ for MTNG-DESI (blue) and MTNG-H$\alpha$ (yellow) mock samples. The black cross and black dashed lines on the histograms mark the true cosmology of the MTNG simulation. The best-fit of each sample (used to compute the clustering statistics in Figure~\ref{fig:clusteringMTNG}) is marked using coloured circles for the 2D distributions and dashed lines on the histograms.}
        \label{fig:constraintsMTNG}
\end{figure}

Combining all the elements of the pipeline from Section~\ref{sec:dataandinferencepipeline} and using the priors from Section~\ref{sec:priors}, we fit the projected correlation function, monopole, and quadrupole of MTNG-DESI and MTNG-$\Halpha$. We run two configurations for each sample, one with the priors on the cosmological parameters and another fixing their value to the true value of the simulation. We set the minimum fitting scale to $r_{\rm min} = 0.3~\hMpc$. Even if in \cite{SOM:2025DESI} we used smaller scales ($r_{\rm min} = 0.1~\hMpc$), this is beyond the resolution limits of some simulations after the cosmological rescaling, so we choose a slightly more conservative minimum scale. We discuss the effect of choosing different combinations of the clustering statistics in Section~\ref{sec:MTNGstatistics}, and the effect of selecting different minimum scales in Section~\ref{sec:MTNGscales}.

 Figure~\ref{fig:clusteringMTNG} shows the measured clustering statistics for MTNG-DESI (blue) and MTNG-$\Halpha$ (yellow) and the best fits with SHAMe-SF (sect.~\ref{sec:PSO}) when fixing the cosmological parameters (dashed) and with the fiducial cosmological priors (solid). In both cases, SHAMe-SF is a good description, with a very similar $\chi^2$ and behaviour for both setups. 

We show the 1-$\sigma$ confidence intervals on $\sig$ and $\OmMh$ for both samples in Figure~\ref{fig:constraintsMTNG}. The best fit is shown with the coloured dashed lines on the 1D histogram and with a point in the 2D contour. The black lines and the cross mark the true cosmology of the MTNG simulation. We are able to constrain both parameters and recover the correct cosmology for both samples within 1-$\sigma$, finding $\Delta\sig=\numerror{\cosmopar{MTNGDESI_sigma8_dmean}}{\cosmopar{MTNGDESI_sigma8_up}}{\cosmopar{MTNGDESI_sigma8_down}}$ (MTNG-DESI) and $\Delta\sig = \numerror{\cosmopar{MTNGHalpha_sigma8_dmean}}{\cosmopar{MTNGHalpha_sigma8_up}}{\cosmopar{MTNGHalpha_sigma8_down}}$ (MTNG-$\Halpha$) and $\Delta\OmMh=\numerror{\cosmopar{MTNGDESI_OmMh_dmean}}{\cosmopar{MTNGDESI_OmMh_up}}{\cosmopar{MTNGDESI_OmMh_down}}$ (MTNG-DESI), and $\Delta\OmMh=\numerror{\cosmopar{MTNGHalpha_OmMh_dmean}}{\cosmopar{MTNGHalpha_OmMh_up}}{\cosmopar{MTNGHalpha_OmMh_down}}$ (MTNG-$\Halpha$).

In galaxy-galaxy lensing analyses, a degeneracy between $\sig$ and $\OmM$ appears (see e.g. \citealt{Abbott:2018}). This degeneracy is condensed in the $S_8$ parameter, defined as:

\begin{equation}
S_8 = \sigma_8 \sqrt{\OmM/0.3}.
\end{equation}

We measure $\Delta S_8 = \numerror{\cosmopar{MTNGDESI_S8_dmean}}{\cosmopar{MTNGDESI_S8_up}}{\cosmopar{MTNGDESI_S8_down}}$ and $\Delta S_8 = \numerror{\cosmopar{MTNGHalpha_S8_dmean}}{\cosmopar{MTNGHalpha_S8_up}}{\cosmopar{MTNGHalpha_S8_down}}$ for MTNG-DESI and MTNG-H$\alpha$, respectively, both in agreement with the value of the MTNG simulation. 

\cite{Sanchez:2020} proposes eliminating the dependence on $h$ of the measurement of the amplitude of matter density fluctuations, computed by averaging on spheres of radius 8~$\ihMpc$ ($\sig$). Different values of $h$ imply comparing different scales, which can be alleviated by setting a characteristic scale in physical units, e.g., 12 Mpc (transforming 8~$\ihMpc$ assuming $h\sim 0.67$ from CMB values). We compute the values of $\sigma_{12}$ for our fiducial analysis using \texttt{baccoemu}\footnote{https://bacco.dipc.org/emulator.html} \citep{Arico:2021} , and show the posteriors on $\sigma_{12}$-$\OmMh$ for the three samples in Figure~\ref{fig:MTNGallapp} (Appendix~\ref{app:fullposteriors}). We obtain $\Delta\sigma_{\rm 12, \  MTNG-DESI}=\numerror{\cosmopar{MTNGDESI_sigma12_dmean}}{\cosmopar{MTNGDESI_sigma12_up}}{\cosmopar{MTNGDESI_sigma12_down}}$, and $\Delta\sigma_{\rm 12, \  MTNG-H\alpha}=\numerror{\cosmopar{MTNGHalpha_sigma12_dmean}}{\cosmopar{MTNGHalpha_sigma12_up}}{\cosmopar{MTNGHalpha_sigma12_down}}$, also in agreement with MTNG.

\subsection{Information on different clustering statistics}
\label{sec:MTNGstatistics}

In our fiducial analysis, we use $\wpp$, $\xilm$, and $\xilq$ to constrain the cosmological parameters. As discussed before, these statistics provide complementary information at different scales. To explore the impact of each statistic and their combinations, we repeat the fits using only $\wpp$, $\xilm$, $\wpp$+$\xilm$, and $\xilm$+$\xilq$, employing the same priors as in the fiducial analysis.

\begin{figure}
                \centering
                 \includegraphics[width=0.48\textwidth]{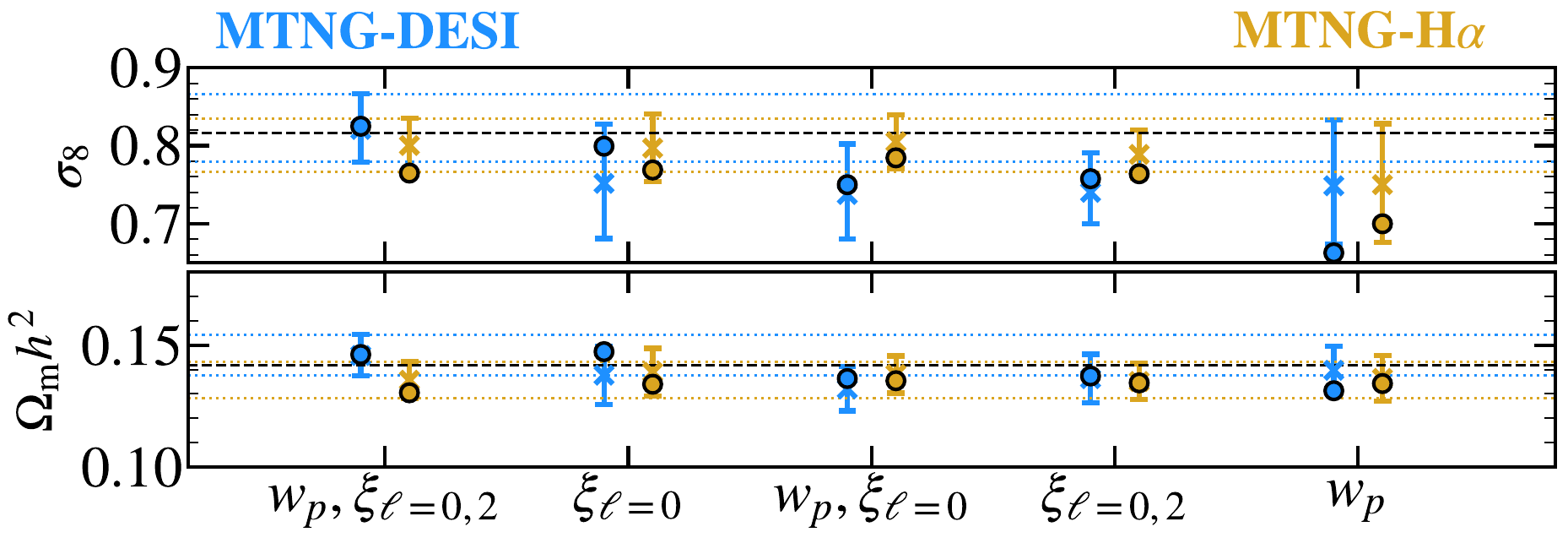}

                 \includegraphics[width=0.48\textwidth]{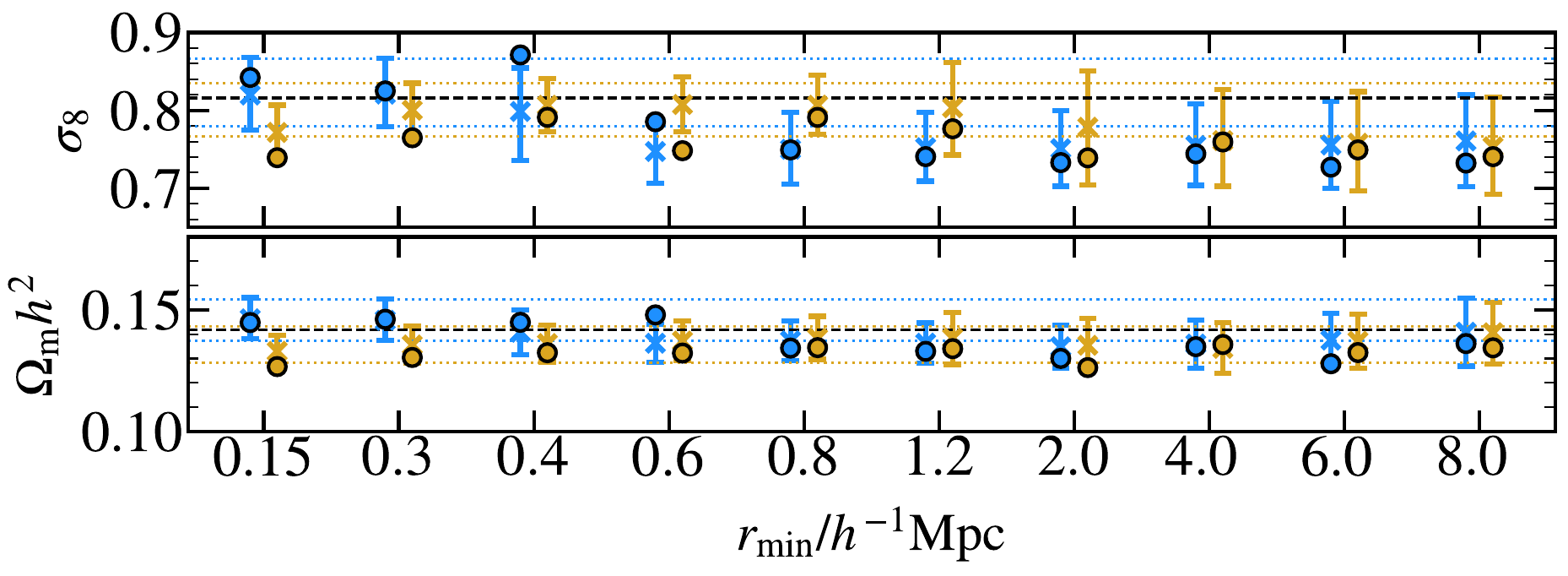}

        \caption{Mean (cross), 68\% (thick lines) and 95\% percent (thin lines) confidence intervals on $\sig$ and $\OmMh$ for different combinations of clustering statistics (upper panel) and minimum scales used for the fit (lower panel) for MTNG-DESI (blue) and MTNG-H$\alpha$ (yellow). The true values of the MTNG simulation are shown with the black dashed line. We add dotted lines for the fiducial analyses to ease the comparison between statistics.}
        \label{fig:constraintsstatsMTNG}
\end{figure}

We show in Figure~\ref{fig:constraintsstatsMTNG} the best-fit, mean, 1$\sigma$, and 2$\sigma$ constraints on $\sig$ and $\OmMh$ for all combinations (computed from the 16th and 84th percentiles and the 5th and 95th percentiles, respectively). We include a dashed line showing the true value of the MTNG simulation. Adding or removing statistics has different effects on the two samples. In the case of MTNG-H$\alpha$, most of the constraining power in $\sig$ comes from the monopole, and we obtain looser constraints when using only $\wpp$. In the case of $\OmMh$, the constraining power is similar for all combinations. For both parameters, we obtain very consistent results when choosing different clustering statistics. This changes when looking at MTNG-DESI. We obtain very wide constraints (biased towards lower values of $\sig$) unless we use the three statistics together. As in the case of MTNG-H$\alpha$, the constraining power on $\OmMh$ does not change much for different combinations, but we obtain slightly lower values when considering only the multipoles of the correlation function.

In summary, inferences are roughly consistent between statistics and their combinations. Even if the projected correlation function alone cannot be used to constrain $\sig$, it is necessary to obtain unbiased constraints when combined with the monopole and the quadrupole, since it helps to break degeneracies between the parameters of the SHAMe-SF model. 

%%%%%%%%%%%%%%%
\begin{figure*}
                \centering
                \includegraphics[width=0.90\textwidth]{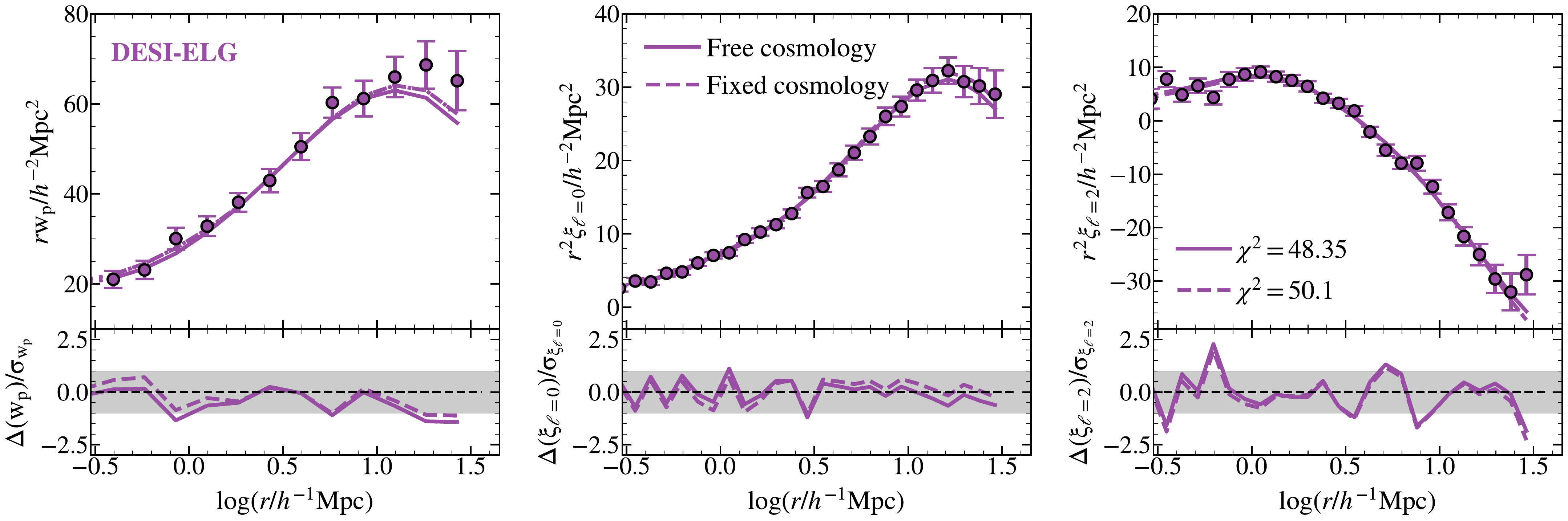}
                        
        \caption{Projected correlation function ($w_p$) and the monopole ($\xi_{\ell=0}$) and quadrupole ($\xi_{\ell=2}$) of the redshift--space correlation function of ELGs in DESI-ELGs, together with the corresponding best-fit SHAMe-SF model (purple lines). Bottom panels: Difference between the data and the fit with SHAMe-SF in units of the diagonal elements of the respective covariance matrix. }
        \label{fig:DESIclustering}
\end{figure*}

\subsection{Scale dependence}
\label{sec:MTNGscales}

We also analyse the effect on the cosmological constraints of excluding the smallest scales. We fit the three clustering statistics using the same priors as in the fiducial analysis, but removing one point at a time. We show the best fits, median, 1$\sigma$, and 2$\sigma$ intervals in the lower panel of Figure~\ref{fig:constraintsstatsMTNG}.  

For MTNG-H$\alpha$, we obtain consistent results across all scales for both $\sig$ and $\OmMh$, but we get looser constraints whenever we remove a point from $r_{\rm min} = 0.8$~$\hMpc$ onwards. For MTNG-DESI, we start to get biased measurements of $\sig$ beyond $r_{\rm min} = 0.6$~$\hMpc$. This seems to be due to both projection effects and degeneracies with the quenching parameters, especially $\alpha_{\rm exp}$. We discuss this further in Appendix~\ref{app:scaleproblems}. As in the case of MTNG-H$\alpha$, we obtain increasingly stringent constraints on both parameters as we include smaller scales. The differences in behaviour between the samples can be attributed to the number of satellites in each sample and the average halo mass and size. The average halo mass and size of halos hosting satellites in the MTNG-H$\alpha$ sample is $\langle M_{\rm host, sat}\rangle= 10^{13.1}~\hMsun$ ($R_{200,c} =  0.54~\ihMpc$), larger than the values for MTNG-DESI: $\langle M_{\rm host, sat}\rangle= 10^{12.6}~\hMsun$ ($R_{200,c} =  0.37~\ihMpc$), which roughly coincides with the scales where the posteriors for $\sig$ start being wider and where projection effects start to appear (note that we expect to have some satellites beyond $R_{200}$, see e.g. \citealt{Hadzhiyska:2022_onehalo, SOM:2025DESI}). This is emphasized by the difference in the number of satellites, with MTNG-DESI having almost 50\% more satellites than MTNG-H$\alpha$ ($f_{\rm sat} = 0.29$ and $f_{\rm sat} = 0.18$, respectively). This highlights the importance of the satellite prescriptions within galaxy-halo population models when obtaining cosmological constraints \citep{ChavesMontero:2023}. We add one point with a minimum scale below our fiducial choice for both samples. We obtain consistent results for MTNG-DESI but lower values in MTNG-H$\alpha$ for both $\sig$ and $\OmMh$. When looking at the full posterior distribution for all SHAMe-SF parameters, we observe a bimodality in the projected posteriors of $\sig$ and some SHAMe-SF parameters ($V_1$, $\sigma$, and $M_{\rm crit}$), causing projection effects (one cluster compatible with measurements for higher scale cuts, the other to lower values of $\sig$).

Thus, we conclude that inferences are consistent for both $\sig$ and $\OmMh$ for both samples, but we obtain biased values of $\sig$ for MTNG-DESI when we stop including scales within the one-halo term. For both samples and parameters, removing those scales yields looser constraints. As in the case of combinations of statistics, this is related to how different scales constrain the parameters of SHAMe-SF.

\begin{figure*}
                \centering
                \includegraphics[width=0.55\textwidth]{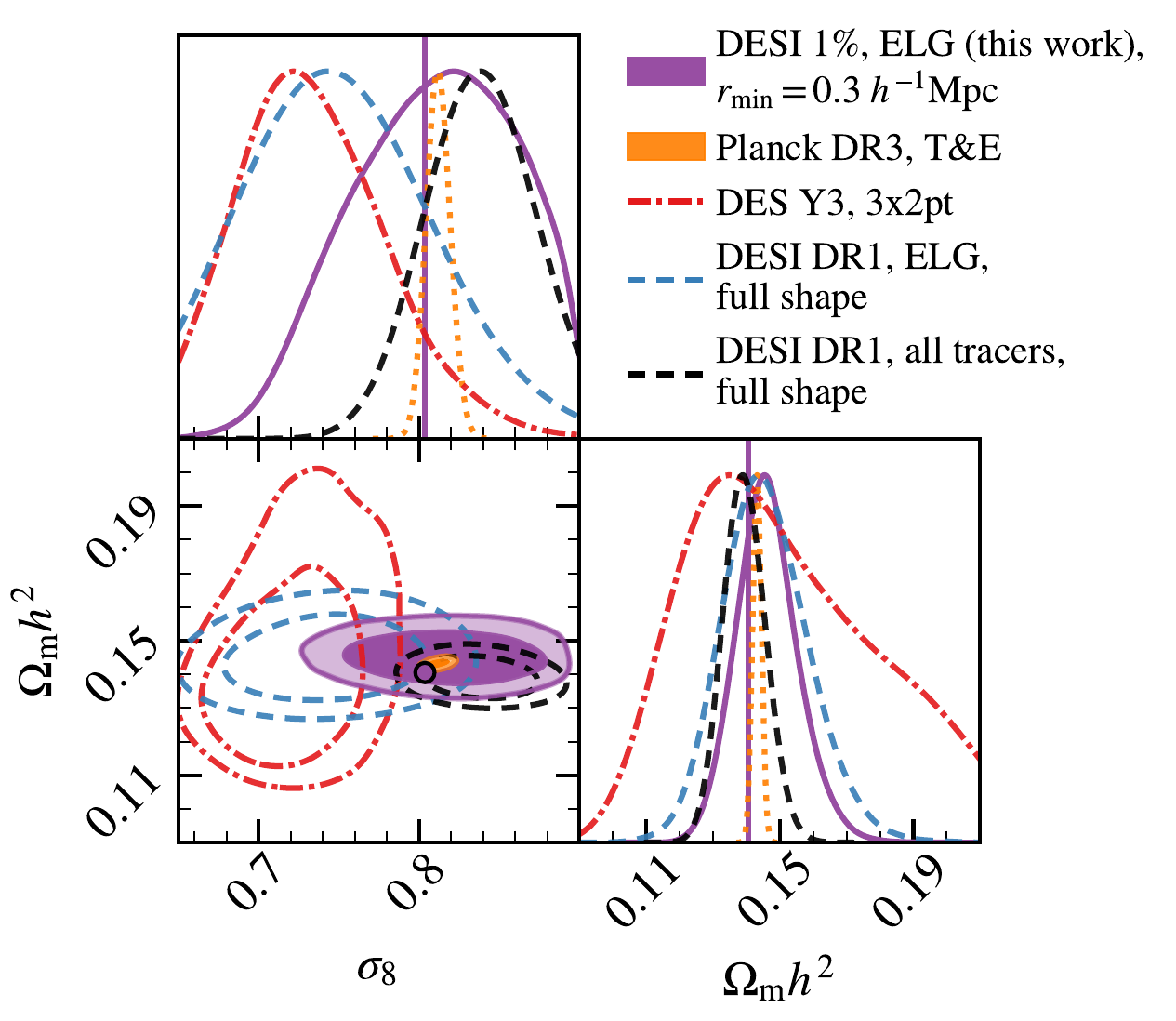}
        \caption{Similar to Figure~\ref{fig:constraintsMTNG}, constraints on $\sig$ and $\OmMh$
        for the DESI-ELG sample using $r_{\rm min} = 0.3~\hMpc$. The best fit is shown with a purple circle (2D) and the solid purple line in the histograms. We add the constraints from DES Y3 (dash-dotted red, \citealt{Abbott:2022}), Planck (dotted orange in the histogram and solid orange in the contours, \citealt{Planck:2018}) and the full-shape analysis of DESI using only ELGs and all tracers combined (blue and black, dashed lines, \citealt{DESI:2025_fullshapeDR1}). }
        \label{fig:DESIcosmo}
\end{figure*}

\begin{figure}
                \centering

                \includegraphics[width=0.45\textwidth]{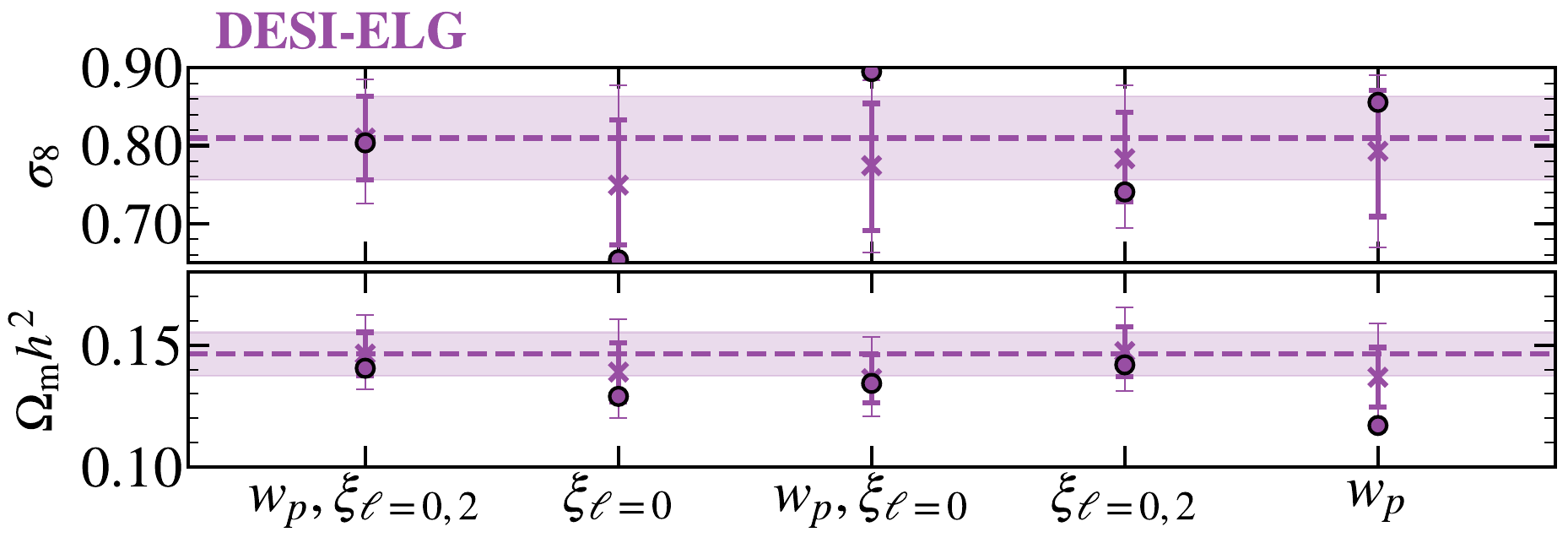}

                \includegraphics[width=0.45\textwidth]{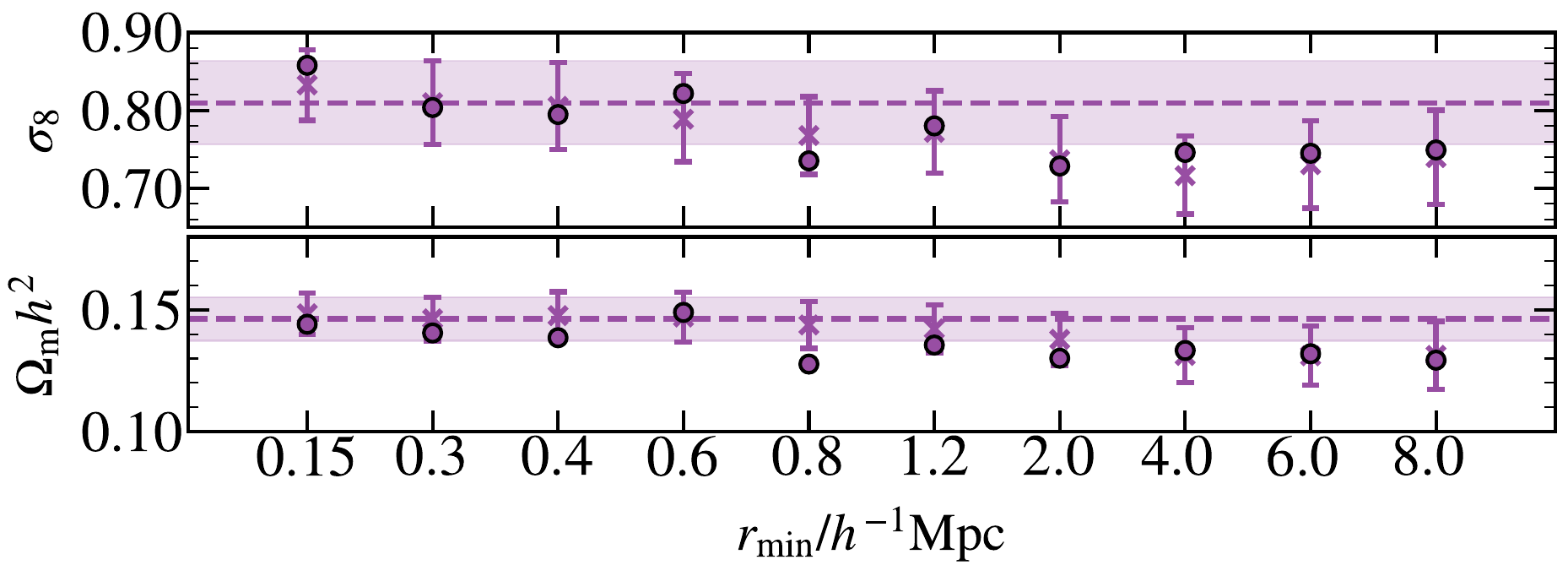}

        \caption{Similar to Figure~\ref{fig:constraintsstatsMTNG}, for DESI-ELG sample when combining different clustering statistics (upper panel) and using different scale cuts (lower panel). The fiducial statistics and scale cut are marked with the highlighted with the shaded region and the purple dashed line. As in Figure~\ref{fig:constraintsstatsMTNG}, we start seeing deviations for $\sig$ when the one-halo term is not included.}
        \label{fig:constraintsscalesstatDESI}
\end{figure}

\section{ELGs from DESI One-Percent data release}
\label{sec:DESIonstraints}
Once the inference pipeline was validated using the MTNG hydrodynamical simulation, we applied it to the ELG clustering data from the DESI One-Percent data release. We take the measurements from \cite{Rocher2023:DESI} for the low-$z$ bin ($0.8<z<1.1$). We add the contributions of the emulator and the scaling errors to their covariance matrix as described in Section~\ref{sec:covariancematrices} for the validation samples, and run the NUTs sampler with the same number of warmup steps (7000), sampling steps (2000), and chains (500).

\label{sec:DESIfidutial}
As in the case of the MTNG analyses, we run two configurations of the sampler: one fixing the cosmological parameters to Planck values, and another freeing cosmological parameters with the priors discussed in Section~\ref{sec:priors} (flat for $\sig$, $\OmM$, and $h$, Gaussian on $\ns$ and $\OmMh$, and fixed $\Mnu$). We set $r_{\rm min} = 0.3$~$\hMpc$ as our minimum scale due to the resolution limit of some simulations after the cosmological scaling. The resulting clustering fits with SHAMe-SF for both runs are shown in Figure~\ref{fig:DESIclustering}, and we show the derived constraints on $\sig$ and $\OmMh$ in Figure~\ref{fig:DESIcosmo}.

As for the validation samples, we find small differences in the clustering fits when we free and fix the cosmological parameters. For the fit varying the cosmological parameters, we measure $\sig=\numerror{\cosmopar{MTNGHalpha_sigma8_mean}}{\cosmopar{MTNGHalpha_sigma8_up}}{\cosmopar{MTNGHalpha_sigma8_down}}$ and $\OmMh=\numerror{\cosmopar{MTNGHalpha_OmMh_mean}}{\cosmopar{MTNGHalpha_OmMh_up}}{\cosmopar{MTNGHalpha_OmMh_down}}$, both compatible with \cite{Planck2015} measurements, shown in Figure~\ref{fig:DESIcosmo} as the black cross and black dashed lines. The aim of fixing the cosmological parameters is two-fold: testing whether Planck cosmology is a good description of the data, and analysing which scales, statistics, or SHAMe-SF parameters are more constrained when the cosmology is fixed. The behaviour of the model is very similar in both cases, with small appreciable differences in Figure~\ref{fig:DESIclustering} for the smaller and larger scales of $\wpp$ and the larger scales of the monopole. The difference in $\chi^2$ between both fits (without normalising by the degrees of freedom) is $\Delta \chi^2=1.75$. The posteriors for the SHAMe-SF parameters, shown in Appendix~\ref{app:fullposteriors}, are similar to those in \cite{SOM:2025DESI} when analysing the same sample, considering the different minimum scales used, and the $1\sigma$ regions are very similar when fixing and freeing cosmology, only finding small differences for $f_{k,\rm{cen}}$ (the parameter controlling the semi-sorting for central subhalos).

We analyse the effect on the constraints of combining different statistics and removing the smallest scales in Figure~\ref{fig:constraintsscalesstatDESI}. For DESI-ELGs, $\sig$ becomes unconstrained when we do not include the quadrupole in the analysis, but constraints are consistent across all combinations of statistics for both $\sig$ and $\OmMh$. Similar to MTNG-DESI, we find looser constraints on $\sig$ when removing any of the statistics, with smaller values of $\sig$ when using only the monopole, tighter constraints when including the quadrupole, and higher values of $\sig$ when combining the three.

Regarding the scale-dependency of the constraints, we observe a similar behaviour to MTNG-DESI in Figure~\ref{fig:constraintsstatsMTNG}: removing scales below $r_{\rm min} = 0.8$~$\hMpc$ yields lower values of $\sig$ and looser constraints. Since we do not know the true cosmology of the Universe, we cannot claim whether this is a bias, as in the case of MTNG-DESI. We note that the change is also related to $\alpha_{\rm exp}$, but due to a displacement in the preferred value rather than projection effects. Constraints in $\OmMh$ are overall consistent, but they shift towards slightly lower values as we stop including smaller scales.

We also show in Figure~\ref{fig:DESIcosmo} the posteriors on $\sig$ and $\OmMh$ from \cite{Planck:2018} \footnote{\url{https://pla.esac.esa.int/}}, DES Y3 MagLim \citep{Abbott:2022}\footnote{\url{https://des.ncsa.illinois.edu/releases/y3a2/Y3key-products}} and the full-shape analysis from DESI for ELGs only and using all tracers \citep{DESI:2025_fullshapeDR1}\footnote{\url{https://data.desi.lbl.gov/doc/releases/dr1/vac/full-shape-cosmo-params/}}. The constraints derived with SHAMe-SF are compatible within 1$\sigma$ with all surveys and have constraining power similar to that of the combination of all tracers in DESI in the full-shape analysis. Our measurements of $\sig$ are slightly higher than those of DESI full shape using only ELGs and DES. Even if our analysis and the full shape analysis of ELGs in DESI target the same tracer, we find it important to remark that they are different samples, and thus comparisons would be unfair. The first difference is in the redshift intervals: in our case, the sample that passed the validation tests was the low-$z$ sample ($0.8<z<1.1$), and for \cite{DESI:2025_fullshapeDR1} was the high-$z$ sample ($1.1<z<1.6$). We show our results for that sample in Appendix~\ref{app:DESIhigh}. Beyond the redshift interval difference, we use only small scales from the One-Percent data release, which have twice the number density but an effective volume 30 times smaller than that of the Y1 data used for the full shape analysis \citep{Yu:2024_DESISHAM, DESI_fullshapecataloguesY1}. Even considering these differences, it is reassuring to obtain comparable constraints between the different approaches using the same tracer.

We additionally compute the constraints on $\sigma_{12}$ and $S_8$. For the former, we measure $\sigma_{\rm 12}=\numerror{\cosmopar{DESI_sigma12_mean}}{\cosmopar{DESI_sigma12_up}}{\cosmopar{DESI_sigma12_down}}$. In the case of $S_8$, we obtain $S_8=\numerror{\cosmopar{DESI_S8_mean}}{\cosmopar{DESI_S8_up}}{\cosmopar{DESI_S8_down}}$, expected given the high values measured for $\OmM$. We show the constraints on these parameters and the comparison with other surveys in Figure~\ref{fig:DESIallcosmo} in Appendix~\ref{app:fullposteriors}.

\label{sec:DESIrobustness}

\section{Conclusions}
\label{sec:conclusions}

In this work, we used the SHAMe-SF model, an extension of Subhalo Abundance Matching specifically designed for star-forming samples, to obtain cosmological constraints from galaxy clustering. 

We applied our inference pipeline to observational data from the DESI ELG survey and to mock catalogues extracted from the MillenniumTNG hydrodynamical simulation. For our fiducial analysis, we freed all the SHAMe-SF parameters (11, discussed in Section~\ref{sec:SHAMeSF}), $\sig$, $h$, and $\OmM$, included {\it Planck}-based priors on $\ns$ and $\Omb h^2$, and fixed the value of $\Mnu$ to 0 MeV (0.06 MeV for DESI-ELGs).

The two mock validation samples were constructed by selecting MillenniumTNG galaxies with stellar masses and star formation rates matching those of DESI ELGs (MTNG-DESI) and of a potential Euclid or Nancy Roman sample (MTNG-H$\alpha$). For both samples, we obtained statistically unbiased cosmological constraints. Additionally, we examined the impact of removing the smallest scales and of varying the combination of summary statistics. For each sample, we find:

\begin{itemize}
\item For MTNG-DESI, removing any of the clustering statistics entering the fiducial analysis, or discarding scales below $r_{\rm min} = 0.6~\ihMpc$, yields broader posteriors and a downward bias in $\sig$ (Figure~\ref{fig:constraintsstatsMTNG}). We attribute this to the loss  of one-halo information: given the typical sizes of satellite-hosting haloes ($R_{200} = 0.37~\ihMpc$), cutting small scales introduces degeneracies between $\sig$ and the quenching parameters governing the satellite population.

\item For MTNG-H$\alpha$, removing small scales broadens the posterior without biasing $\sig$, and different combinations of summary statistics yield consistent results -- with the exception of analyses relying exclusively on $\wpp$, for which $\sig$ remains unconstrained (Figure~\ref{fig:constraintsstatsMTNG}).
\end{itemize}

We also fitted the galaxy clustering with cosmological parameters fixed to their simulation values, finding no significant change in fit quality relative to the free-cosmology case (Figure~\ref{fig:clusteringMTNG} for the validation samples; Figure~\ref{fig:DESIclustering} for DESI ELGs), confirming that cosmological variations are not artificially absorbing residual modelling errors.

Applying our pipeline to $0.8 < z < 1.1$ ELG clustering measurements from the DESI 
One-Percent data release, we obtain

$$ \sig = \numerror{\cosmopar{DESI_sigma8_mean}}{\cosmopar{DESI_sigma8_up}}{\cosmopar{DESI_sigma8_down}}, \quad \OmMh=\numerror{\cosmopar{DESI_OmMh_mean}}{\cosmopar{DESI_OmMh_up}}{\cosmopar{DESI_OmMh_down}}$$

The scale-dependence of the results mirrors the MTNG-DESI behaviour: for $r_{\rm min} > 0.8\ihMpc$ (Figure~\ref{fig:constraintsscalesstatDESI}), $\sig$ becomes 
biased. Scales within the one-halo regime are indispensable for constraining satellite-sensitive parameters and thereby breaking degeneracies with $\sig$ (Section~\ref{sec:MTNGscales}, Appendix~\ref{app:scaleproblems}). Among the summary statistics considered, the quadrupole is essential: without it, $\sig$ cannot be reliably constrained.

We additionally report $\sigma_{12}$, an $h$-independent counterpart to $\sig$, measuring $\sigma_{12} = \numerror{\cosmopar{MTNGDESI_sigma12_mean}}
                        {\cosmopar{MTNGDESI_sigma12_up}}
                        {\cosmopar{MTNGDESI_sigma12_down}}$,
$\sigma_{12} = \numerror{\cosmopar{MTNGHalpha_sigma12_mean}}
                        {\cosmopar{MTNGHalpha_sigma12_up}}
                        {\cosmopar{MTNGHalpha_sigma12_down}}$,
and
$\sigma_{12} = \numerror{\cosmopar{DESI_sigma12_mean}}
                        {\cosmopar{DESI_sigma12_up}}
                        {\cosmopar{DESI_sigma12_down}}$
for MTNG-DESI, MTNG-H$\alpha$, and DESI ELGs, respectively. We further measure $S_8$, obtaining $S_8=\numerror{\cosmopar{MTNGHalpha_S8_mean}}{\cosmopar{MTNGHalpha_S8_up}}{\cosmopar{MTNGHalpha_S8_down}}$, $S_8=\numerror{\cosmopar{MTNGDESI_S8_mean}}{\cosmopar{MTNGDESI_S8_up}}{\cosmopar{MTNGDESI_S8_down}}$, and $S_8=\numerror{\cosmopar{DESI_S8_mean}}{\cosmopar{DESI_S8_up}}{\cosmopar{DESI_S8_down}}$ for MTNG-H$\alpha$, MTNG-DESI, and DESI-ELGs, respectively, all consistent with the cosmology of the simulation.

Our analysis was intentionally restricted to small scales as a proof-of-concept for the ability of SHAMe-SF to separate galaxy--halo connection physics from cosmological information. Natural extensions include combining linear and nonlinear scales within a single self-consistent framework -- for instance, jointly fitting 
the BAO signal alongside deeply nonlinear clustering -- and exploring higher-order summary statistics at small scales.

The constraints presented here were derived from only 1\% of the DESI survey volume, and are comparable to the latest full-shape analysis constraints in DESI. Scaling naively to the complete survey, one would anticipate a factor of $\sqrt{100} \approx 10$ reduction in statistical uncertainties, bringing sub-percent constraints on $\sig$ and $\OmMh$ within reach. Realising this potential will require progress on several fronts.

On the theoretical side, the 
SHAMe-SF model must be stress-tested against a broader suite of hydrodynamical simulations spanning diverse subgrid physics prescriptions, to ensure that the galaxy--halo mapping is not sensitive to the specifics of any single simulation. Furthermore, the simulations required to build SHAMe-SF should be of sufficient volume and resolution to suppress cosmic variance and reliably resolve the relevant subhalo scales, and the emulators underpinning our inference will need to be refined so that their interpolation errors do not become the limiting systematic. 

On the observational side, a rigorous characterisation and mitigation of survey systematics -- fibre assignment, completeness variations, interloper contamination, and target-selection effects -- will be essential, as will the development of accurate covariance matrices on the small scales exploited here. These are substantial but tractable challenges, and the scientific return justifies the investment: the nonlinear regime encodes a wealth of cosmological information that remains largely unexplored, and SHAMe-SF is well positioned to unlock it for the full DESI survey, Euclid, the Nancy Grace Roman Space Telescope, and future spectroscopic surveys targeting star-forming galaxies.

\begin{acknowledgements}
      SOM is grateful to Diego Herrero-Carrion, Irati Lizaso and Irene Flores for their contribution to the emulators, to Tamara Richardson for her inputs in the figures, and to Andrew Hearin and Georgios Zacharegkas for the introduction to the wonders of JAX. SOM is also thankful to the hospitality of people in IATE, specially discussions with Candela Cerdosino and Facundo Rodriguez when constraints were not constraining.  SOM is funded by the Spanish Ministry of Science and Innovation under grant number PRE2020-095788.  REA received support from grant PID2024-161003NB-I00 funded by MICIU/AEI/10.13039/501100011033 and by ERDF/EU. SC acknowledges the support of the “Ramón y Cajal” fellowship (RYC2023-043783-I) and ``Ayudas para Atraccion de Investigadores con Alto Potencial'' (2025/00000640) from Universidad de Sevilla. JCM acknowledges support from the European Union (ERC Consolidator Grant, COSMO-LYA, grant agreement 101044612). IFAE is partially funded by the CERCA program of the Generalitat de Catalunya. MZ acknowledges that the project that gave rise to these results received the support of a fellowship from the "La Caixa" Foundation (ID 100010434). The fellowship code is LCF/BQ/PI25/12100027. S.B. is supported by the UKRI Future Leaders Fellowship [grant numbers MR/V023381/1 and UKRI2044]. The authors also acknowledge the computer resources at MareNostrum and the technical support provided by Barcelona Supercom-puting Center (RES-AECT-AECT-2024-2-0022) Technical and human support provided by DIPC Supercomputing Center is also very gratefully acknowledged. 
\end{acknowledgements}


%-------------------------------------------------------------------

\bibliographystyle{aa} % style aa.bst
\bibliography{aa.bib} % your references 

@ARTICLE{Ivanov:2021,
       author = {{Ivanov}, Mikhail M.},
        title = "{Cosmological constraints from the power spectrum of eBOSS emission line galaxies}",
      journal = {\prd},
         year = 2021,
        month = nov,
       volume = {104},
       number = {10},
          eid = {103514},
        pages = {103514},
          doi = {10.1103/PhysRevD.104.103514},
archivePrefix = {arXiv},
       eprint = {2106.12580},
 primaryClass = {astro-ph.CO},
       adsurl = {https://ui.adsabs.harvard.edu/abs/2021PhRvD.104j3514I}
}

@ARTICLE{Dawson:2016eboss,
       author = {{Dawson}, Kyle S. and {Kneib}, Jean-Paul and {Percival}, Will J. and {Alam}, Shadab and {Albareti}, Franco D. and {Anderson}, Scott F. and {Armengaud}, Eric and {Aubourg}, {\'E}ric and {Bailey}, Stephen and {Bautista}, Julian E. and {Berlind}, Andreas A. and {Bershady}, Matthew A. and {Beutler}, Florian and {Bizyaev}, Dmitry and {Blanton}, Michael R. and {Blomqvist}, Michael and {Bolton}, Adam S. and {Bovy}, Jo and {Brandt}, W.~N. and {Brinkmann}, Jon and {Brownstein}, Joel R. and {Burtin}, Etienne and {Busca}, N.~G. and {Cai}, Zheng and {Chuang}, Chia-Hsun and {Clerc}, Nicolas and {Comparat}, Johan and {Cope}, Frances and {Croft}, Rupert A.~C. and {Cruz-Gonzalez}, Irene and {da Costa}, Luiz N. and {Cousinou}, Marie-Claude and {Darling}, Jeremy and {de la Macorra}, Axel and {de la Torre}, Sylvain and {Delubac}, Timoth{\'e}e and {du Mas des Bourboux}, H{\'e}lion and {Dwelly}, Tom and {Ealet}, Anne and {Eisenstein}, Daniel J. and {Eracleous}, Michael and {Escoffier}, S. and {Fan}, Xiaohui and {Finoguenov}, Alexis and {Font-Ribera}, Andreu and {Frinchaboy}, Peter and {Gaulme}, Patrick and {Georgakakis}, Antonis and {Green}, Paul and {Guo}, Hong and {Guy}, Julien and {Ho}, Shirley and {Holder}, Diana and {Huehnerhoff}, Joe and {Hutchinson}, Timothy and {Jing}, Yipeng and {Jullo}, Eric and {Kamble}, Vikrant and {Kinemuchi}, Karen and {Kirkby}, David and {Kitaura}, Francisco-Shu and {Klaene}, Mark A. and {Laher}, Russ R. and {Lang}, Dustin and {Laurent}, Pierre and {Le Goff}, Jean-Marc and {Li}, Cheng and {Liang}, Yu and {Lima}, Marcos and {Lin}, Qiufan and {Lin}, Weipeng and {Lin}, Yen-Ting and {Long}, Daniel C. and {Lundgren}, Britt and {MacDonald}, Nicholas and {Geimba Maia}, Marcio Antonio and {Malanushenko}, Elena and {Malanushenko}, Viktor and {Mariappan}, Vivek and {McBride}, Cameron K. and {McGreer}, Ian D. and {M{\'e}nard}, Brice and {Merloni}, Andrea and {Meza}, Andres and {Montero-Dorta}, Antonio D. and {Muna}, Demitri and {Myers}, Adam D. and {Nandra}, Kirpal and {Naugle}, Tracy and {Newman}, Jeffrey A. and {Noterdaeme}, Pasquier and {Nugent}, Peter and {Ogando}, Ricardo and {Olmstead}, Matthew D. and {Oravetz}, Audrey and {Oravetz}, Daniel J. and {Padmanabhan}, Nikhil and {Palanque-Delabrouille}, Nathalie and {Pan}, Kaike and {Parejko}, John K. and {P{\^a}ris}, Isabelle and {Peacock}, John A. and {Petitjean}, Patrick and {Pieri}, Matthew M. and {Pisani}, Alice and {Prada}, Francisco and {Prakash}, Abhishek and {Raichoor}, Anand and {Reid}, Beth and {Rich}, James and {Ridl}, Jethro and {Rodriguez-Torres}, Sergio and {Carnero Rosell}, Aurelio and {Ross}, Ashley J. and {Rossi}, Graziano and {Ruan}, John and {Salvato}, Mara and {Sayres}, Conor and {Schneider}, Donald P. and {Schlegel}, David J. and {Seljak}, Uros and {Seo}, Hee-Jong and {Sesar}, Branimir and {Shandera}, Sarah and {Shu}, Yiping and {Slosar}, An{\v{z}}e and {Sobreira}, Flavia and {Streblyanska}, Alina and {Suzuki}, Nao and {Taylor}, Donna and {Tao}, Charling and {Tinker}, Jeremy L. and {Tojeiro}, Rita and {Vargas-Maga{\~n}a}, Mariana and {Wang}, Yuting and {Weaver}, Benjamin A. and {Weinberg}, David H. and {White}, Martin and {Wood-Vasey}, W.~M. and {Yeche}, Christophe and {Zhai}, Zhongxu and {Zhao}, Cheng and {Zhao}, Gong-bo and {Zheng}, Zheng and {Ben Zhu}, Guangtun and {Zou}, Hu},
        title = "{The SDSS-IV Extended Baryon Oscillation Spectroscopic Survey: Overview and Early Data}",
      journal = {\aj},
         year = 2016,
        month = feb,
       volume = {151},
       number = {2},
          eid = {44},
        pages = {44},
          doi = {10.3847/0004-6256/151/2/44},
archivePrefix = {arXiv},
       eprint = {1508.04473},
 primaryClass = {astro-ph.CO},
       adsurl = {https://ui.adsabs.harvard.edu/abs/2016AJ....151...44D}
}

@ARTICLE{Knebe:2022,
       author = {{Knebe}, Alexander and {Lopez-Cano}, Daniel and {Avila}, Santiago and {Favole}, Ginevra and {Stevens}, Adam R.~H. and {Gonzalez-Perez}, Violeta and {Reyes-Peraza}, Guillermo and {Yepes}, Gustavo and {Chuang}, Chia-Hsun and {Kitaura}, Francisco-Shu},
        title = "{UNITSIM-Galaxies: data release and clustering of emission-line galaxies}",
      journal = {\mnras},
         year = 2022,
        month = mar,
       volume = {510},
       number = {4},
        pages = {5392-5407},
          doi = {10.1093/mnras/stac006},
archivePrefix = {arXiv},
       eprint = {2103.13088},
 primaryClass = {astro-ph.GA},
       adsurl = {https://ui.adsabs.harvard.edu/abs/2022MNRAS.510.5392K}
}

@ARTICLE{Lange:2023,
       author = {{Lange}, Johannes U. and {Hearin}, Andrew P. and {Leauthaud}, Alexie and {van den Bosch}, Frank C. and {Xhakaj}, Enia and {Guo}, Hong and {Wechsler}, Risa H. and {DeRose}, Joseph},
        title = "{Constraints on S$_{8}$ from a full-scale and full-shape analysis of redshift-space clustering and galaxy-galaxy lensing in BOSS}",
      journal = {\mnras},
         year = 2023,
        month = apr,
       volume = {520},
       number = {4},
        pages = {5373-5393},
          doi = {10.1093/mnras/stad473},
archivePrefix = {arXiv},
       eprint = {2301.08692},
 primaryClass = {astro-ph.CO},
       adsurl = {https://ui.adsabs.harvard.edu/abs/2023MNRAS.520.5373L}
}

@ARTICLE{DESI_fullshapecataloguesY1,
       author = {{Adame}, A.~G. and {Aguilar}, J. and {Ahlen}, S. and {Alam}, S. and {Alexander}, D.~M. and {Alvarez}, M. and {Alves}, O. and {Anand}, A. and {Andrade}, U. and {Armengaud}, E. and {Avila}, S. and {Aviles}, A. and {Awan}, H. and {Bailey}, S. and {Baltay}, C. and {Bault}, A. and {Behera}, J. and {BenZvi}, S. and {Beutler}, F. and {Bianchi}, D. and {Blake}, C. and {Blum}, R. and {Brieden}, S. and {Brodzeller}, A. and {Brooks}, D. and {Buckley-Geer}, E. and {Burtin}, E. and {Calderon}, R. and {Canning}, R. and {Carnero Rosell}, A. and {Cereskaite}, R. and {Cervantes-Cota}, J.~L. and {Chabanier}, S. and {Chaussidon}, E. and {Chaves-Montero}, J. and {Chen}, S. and {Chen}, X. and {Claybaugh}, T. and {Cole}, S. and {Cuceu}, A. and {Davis}, T.~M. and {Dawson}, K. and {de la Macorra}, A. and {de Mattia}, A. and {Deiosso}, N. and {Dey}, A. and {Dey}, B. and {Ding}, Z. and {Doel}, P. and {Edelstein}, J. and {Eftekharzadeh}, S. and {Eisenstein}, D.~J. and {Elliott}, A. and {Fagrelius}, P. and {Fanning}, K. and {Ferraro}, S. and {Ereza}, J. and {Findlay}, N. and {Flaugher}, B. and {Font-Ribera}, A. and {Forero-S{\'a}nchez}, D. and {Forero-Romero}, J.~E. and {Garcia-Quintero}, C. and {Garrison}, L.~H. and {Gazta{\~n}aga}, E. and {Gil-Mar{\'\i}n}, H. and {Gontcho}, S. Gontcho A. and {Gonzalez-Morales}, A.~X. and {Gonzalez-Perez}, V. and {Gordon}, C. and {Green}, D. and {Gruen}, D. and {Gsponer}, R. and {Gutierrez}, G. and {Guy}, J. and {Hadzhiyska}, B. and {Hahn}, C. and {Hanif}, M.~M.~S. and {Herrera-Alcantar}, H.~K. and {Honscheid}, K. and {Howlett}, C. and {Huterer}, D. and {Ir{\v{s}}i{\v{c}}}, V. and {Ishak}, M. and {Juneau}, S. and {Kara{\c{c}}ayl{\i}}, N.~G. and {Kehoe}, R. and {Kent}, S. and {Kirkby}, D. and {Kong}, H. and {Koposov}, S.~E. and {Kremin}, A. and {Krolewski}, A. and {Lai}, Y. and {Lan}, T.-W. and {Landriau}, M. and {Lang}, D. and {Lasker}, J. and {Le Goff}, J.~M. and {Le Guillou}, L. and {Leauthaud}, A. and {Levi}, M.~E. and {Li}, T.~S. and {Lodha}, K. and {Magneville}, C. and {Manera}, M. and {Margala}, D. and {Martini}, P. and {Maus}, M. and {McDonald}, P. and {Medina-Varela}, L. and {Meisner}, A. and {Mena-Fern{\'a}ndez}, J. and {Miquel}, R. and {Moon}, J. and {Moore}, S. and {Moustakas}, J. and {Mueller}, E. and {Mu{\~n}oz-Guti{\'e}rrez}, A. and {Myers}, A.~D. and {Nadathur}, S. and {Napolitano}, L. and {Neveux}, R. and {Newman}, J.~A. and {Nguyen}, N.~M. and {Nie}, J. and {Niz}, G. and {Noriega}, H.~E. and {Padmanabhan}, N. and {Paillas}, E. and {Palanque-Delabrouille}, N. and {Pan}, J. and {Penmetsa}, S. and {Percival}, W.~J. and {Pieri}, M.~M. and {Pinon}, M. and {Poppett}, C. and {Porredon}, A. and {Prada}, F. and {P{\'e}rez-Fern{\'a}ndez}, A. and {P{\'e}rez-R{\`a}fols}, I. and {Rabinowitz}, D. and {Raichoor}, A. and {Ram{\'\i}rez-P{\'e}rez}, C. and {Ramirez-Solano}, S. and {Rashkovetskyi}, M. and {Ravoux}, C. and {Rezaie}, M. and {Rich}, J. and {Rocher}, A. and {Rockosi}, C. and {Rodr{\'\i}guez-Mart{\'\i}nez}, F. and {Roe}, N.~A. and {Rosado-Marin}, A. and {Ross}, A.~J. and {Rossi}, G. and {Ruggeri}, R. and {Ruhlmann-Kleider}, V. and {Samushia}, L. and {Sanchez}, E. and {Saulder}, C. and {Schlafly}, E.~F. and {Schlegel}, D. and {Schubnell}, M. and {Seo}, H. and {Sharples}, R. and {Silber}, J. and {Slosar}, A. and {Smith}, A. and {Sprayberry}, D. and {Tan}, T. and {Tarl{\'e}}, G. and {Trusov}, S. and {Vaisakh}, R. and {Valcin}, D. and {Valdes}, F. and {Vargas-Maga{\~n}a}, M. and {Verde}, L. and {Walther}, M. and {Wang}, B. and {Wang}, M.~S. and {Weaver}, B.~A. and {Weaverdyck}, N. and {Wechsler}, R.~H. and {Weinberg}, D.~H. and {White}, M. and {Wilson}, M.~J. and {Yu}, J. and {Yu}, Y. and {Yuan}, S. and {Y{\`e}che}, C. and {Zaborowski}, E.~A. and {Zarrouk}, P. and {Zhang}, H. and {Zhao}, C. and {Zhao}, R. and {Zhou}, R. and {Zou}, H. and {The DESI collaboration}},
        title = "{DESI 2024 V: Full-Shape galaxy clustering from galaxies and quasars}",
      journal = {\jcap},
         year = 2025,
        month = sep,
       volume = {2025},
       number = {9},
          eid = {008},
        pages = {008},
          doi = {10.1088/1475-7516/2025/09/008},
archivePrefix = {arXiv},
       eprint = {2411.12021},
 primaryClass = {astro-ph.CO},
       adsurl = {https://ui.adsabs.harvard.edu/abs/2025JCAP...09..008A}
}

@ARTICLE{Gao:2026,
       author = {{Gao}, Wenhao and {Liu}, Zhenjie and {Zhai}, Zhongxu and {Tinker}, Jeremy L. and {Zhang}, Jun and {Banerjee}, Arka and {DeRose}, Joseph and {Guo}, Hong and {Mao}, Yao-Yuan and {Storey-Fisher}, Kate and {Wechsler}, Risa H.},
        title = "{Joint Analysis of Small-scale Galaxy Clustering and Galaxy─Galaxy Lensing from BOSS Galaxies}",
      journal = {\apj},
         year = 2026,
        month = feb,
       volume = {998},
       number = {2},
          eid = {268},
        pages = {268},
          doi = {10.3847/1538-4357/ae3bd7},
archivePrefix = {arXiv},
       eprint = {2601.12976},
 primaryClass = {astro-ph.CO},
       adsurl = {https://ui.adsabs.harvard.edu/abs/2026ApJ...998..268G}
}

@ARTICLE{Madar:2024,
       author = {{Madar}, Makun S. and {Baugh}, Carlton M. and {Shi}, Difu},
        title = "{Predictions for the abundance and clustering of H {\ensuremath{\alpha}} emitting galaxies}",
      journal = {\mnras},
         year = 2024,
        month = dec,
       volume = {535},
       number = {4},
        pages = {3324-3341},
          doi = {10.1093/mnras/stae2560},
archivePrefix = {arXiv},
       eprint = {2405.04601},
 primaryClass = {astro-ph.CO},
       adsurl = {https://ui.adsabs.harvard.edu/abs/2024MNRAS.535.3324M}
}

@ARTICLE{Osato:2026,
       author = {{Osato}, Ken and {Okumura}, Teppei},
        title = "{Clustering of emission line galaxies with IllustrisTNG -- II. cosmology challenge with anisotropic correlation functions and ELG-halo connections}",
      journal = {arXiv e-prints},
         year = 2026,
        month = feb,
          eid = {arXiv:2602.01925},
        pages = {arXiv:2602.01925},
          doi = {10.48550/arXiv.2602.01925},
archivePrefix = {arXiv},
       eprint = {2602.01925},
 primaryClass = {astro-ph.CO},
       adsurl = {https://ui.adsabs.harvard.edu/abs/2026arXiv260201925O}
}

@ARTICLE{Rapoport:2025,
       author = {{Rapoport}, Ivan and {Desjacques}, Vincent and {Parimbelli}, Gabriele and {Behar}, Ehud and {Crocce}, Martin},
        title = "{Spatially Resolved Modeling of Galactic H{\ensuremath{\alpha}} Emission for Galaxy Clustering}",
      journal = {\apj},
         year = 2025,
        month = jul,
       volume = {988},
       number = {1},
          eid = {44},
        pages = {44},
          doi = {10.3847/1538-4357/adde4c},
archivePrefix = {arXiv},
       eprint = {2502.08778},
 primaryClass = {astro-ph.GA},
       adsurl = {https://ui.adsabs.harvard.edu/abs/2025ApJ...988...44R}
}

@ARTICLE{Sevilla-Noarbe:2021,
       author = {{Sevilla-Noarbe}, I. and {Bechtol}, K. and {Carrasco Kind}, M. and {Carnero Rosell}, A. and {Becker}, M.~R. and {Drlica-Wagner}, A. and {Gruendl}, R.~A. and {Rykoff}, E.~S. and {Sheldon}, E. and {Yanny}, B. and {Alarcon}, A. and {Allam}, S. and {Amon}, A. and {Benoit-L{\'e}vy}, A. and {Bernstein}, G.~M. and {Bertin}, E. and {Burke}, D.~L. and {Carretero}, J. and {Choi}, A. and {Diehl}, H.~T. and {Everett}, S. and {Flaugher}, B. and {Gaztanaga}, E. and {Gschwend}, J. and {Harrison}, I. and {Hartley}, W.~G. and {Hoyle}, B. and {Jarvis}, M. and {Johnson}, M.~D. and {Kessler}, R. and {Kron}, R. and {Kuropatkin}, N. and {Leistedt}, B. and {Li}, T.~S. and {Menanteau}, F. and {Morganson}, E. and {Ogando}, R.~L.~C. and {Palmese}, A. and {Paz-Chinch{\'o}n}, F. and {Pieres}, A. and {Pond}, C. and {Rodriguez-Monroy}, M. and {Smith}, J. Allyn and {Stringer}, K.~M. and {Troxel}, M.~A. and {Tucker}, D.~L. and {de Vicente}, J. and {Wester}, W. and {Zhang}, Y. and {Abbott}, T.~M.~C. and {Aguena}, M. and {Annis}, J. and {Avila}, S. and {Bhargava}, S. and {Bridle}, S.~L. and {Brooks}, D. and {Brout}, D. and {Castander}, F.~J. and {Cawthon}, R. and {Chang}, C. and {Conselice}, C. and {Costanzi}, M. and {Crocce}, M. and {da Costa}, L.~N. and {Pereira}, M.~E.~S. and {Davis}, T.~M. and {Desai}, S. and {Dietrich}, J.~P. and {Doel}, P. and {Eckert}, K. and {Evrard}, A.~E. and {Ferrero}, I. and {Fosalba}, P. and {Garc{\'\i}a-Bellido}, J. and {Gerdes}, D.~W. and {Giannantonio}, T. and {Gruen}, D. and {Gutierrez}, G. and {Hinton}, S.~R. and {Hollowood}, D.~L. and {Honscheid}, K. and {Huff}, E.~M. and {Huterer}, D. and {James}, D.~J. and {Jeltema}, T. and {Kuehn}, K. and {Lahav}, O. and {Lidman}, C. and {Lima}, M. and {Lin}, H. and {Maia}, M.~A.~G. and {Marshall}, J.~L. and {Martini}, P. and {Melchior}, P. and {Miquel}, R. and {Mohr}, J.~J. and {Morgan}, R. and {Neilsen}, E. and {Plazas}, A.~A. and {Romer}, A.~K. and {Roodman}, A. and {Sanchez}, E. and {Scarpine}, V. and {Schubnell}, M. and {Serrano}, S. and {Smith}, M. and {Suchyta}, E. and {Tarle}, G. and {Thomas}, D. and {To}, C. and {Varga}, T.~N. and {Wechsler}, R.~H. and {Weller}, J. and {Wilkinson}, R.~D. and {DES Collaboration}},
        title = "{Dark Energy Survey Year 3 Results: Photometric Data Set for Cosmology}",
      journal = {\apjs},
         year = 2021,
        month = jun,
       volume = {254},
       number = {2},
          eid = {24},
        pages = {24},
          doi = {10.3847/1538-4365/abeb66},
archivePrefix = {arXiv},
       eprint = {2011.03407},
 primaryClass = {astro-ph.CO},
       adsurl = {https://ui.adsabs.harvard.edu/abs/2021ApJS..254...24S}
}

@ARTICLE{Gatti:2021,
       author = {{Gatti}, M. and {Sheldon}, E. and {Amon}, A. and {Becker}, M. and {Troxel}, M. and {Choi}, A. and {Doux}, C. and {MacCrann}, N. and {Navarro-Alsina}, A. and {Harrison}, I. and {Gruen}, D. and {Bernstein}, G. and {Jarvis}, M. and {Secco}, L.~F. and {Fert{\'e}}, A. and {Shin}, T. and {McCullough}, J. and {Rollins}, R.~P. and {Chen}, R. and {Chang}, C. and {Pandey}, S. and {Tutusaus}, I. and {Prat}, J. and {Elvin-Poole}, J. and {Sanchez}, C. and {Plazas}, A.~A. and {Roodman}, A. and {Zuntz}, J. and {Abbott}, T.~M.~C. and {Aguena}, M. and {Allam}, S. and {Annis}, J. and {Avila}, S. and {Bacon}, D. and {Bertin}, E. and {Bhargava}, S. and {Brooks}, D. and {Burke}, D.~L. and {Carnero Rosell}, A. and {Carrasco Kind}, M. and {Carretero}, J. and {Castander}, F.~J. and {Conselice}, C. and {Costanzi}, M. and {Crocce}, M. and {da Costa}, L.~N. and {Davis}, T.~M. and {De Vicente}, J. and {Desai}, S. and {Diehl}, H.~T. and {Dietrich}, J.~P. and {Doel}, P. and {Drlica-Wagner}, A. and {Eckert}, K. and {Everett}, S. and {Ferrero}, I. and {Frieman}, J. and {Garc{\'\i}a-Bellido}, J. and {Gerdes}, D.~W. and {Giannantonio}, T. and {Gruendl}, R.~A. and {Gschwend}, J. and {Gutierrez}, G. and {Hartley}, W.~G. and {Hinton}, S.~R. and {Hollowood}, D.~L. and {Honscheid}, K. and {Hoyle}, B. and {Huff}, E.~M. and {Huterer}, D. and {Jain}, B. and {James}, D.~J. and {Jeltema}, T. and {Krause}, E. and {Kron}, R. and {Kuropatkin}, N. and {Lima}, M. and {Maia}, M.~A.~G. and {Marshall}, J.~L. and {Miquel}, R. and {Morgan}, R. and {Myles}, J. and {Palmese}, A. and {Paz-Chinch{\'o}n}, F. and {Rykoff}, E.~S. and {Samuroff}, S. and {Sanchez}, E. and {Scarpine}, V. and {Schubnell}, M. and {Serrano}, S. and {Sevilla-Noarbe}, I. and {Smith}, M. and {Soares-Santos}, M. and {Suchyta}, E. and {Swanson}, M.~E.~C. and {Tarle}, G. and {Thomas}, D. and {To}, C. and {Tucker}, D.~L. and {Varga}, T.~N. and {Wechsler}, R.~H. and {Weller}, J. and {Wester}, W. and {Wilkinson}, R.~D.},
        title = "{Dark energy survey year 3 results: weak lensing shape catalogue}",
      journal = {\mnras},
         year = 2021,
        month = jul,
       volume = {504},
       number = {3},
        pages = {4312-4336},
          doi = {10.1093/mnras/stab918},
archivePrefix = {arXiv},
       eprint = {2011.03408},
 primaryClass = {astro-ph.CO},
       adsurl = {https://ui.adsabs.harvard.edu/abs/2021MNRAS.504.4312G}
}

@ARTICLE{Kuijken:2019,
       author = {{Kuijken}, K. and {Heymans}, C. and {Dvornik}, A. and {Hildebrandt}, H. and {de Jong}, J.~T.~A. and {Wright}, A.~H. and {Erben}, T. and {Bilicki}, M. and {Giblin}, B. and {Shan}, H.-Y. and {Getman}, F. and {Grado}, A. and {Hoekstra}, H. and {Miller}, L. and {Napolitano}, N. and {Paolilo}, M. and {Radovich}, M. and {Schneider}, P. and {Sutherland}, W. and {Tewes}, M. and {Tortora}, C. and {Valentijn}, E.~A. and {Verdoes Kleijn}, G.~A.},
        title = "{The fourth data release of the Kilo-Degree Survey: ugri imaging and nine-band optical-IR photometry over 1000 square degrees}",
      journal = {\aap},
         year = 2019,
        month = may,
       volume = {625},
          eid = {A2},
        pages = {A2},
          doi = {10.1051/0004-6361/201834918},
archivePrefix = {arXiv},
       eprint = {1902.11265},
 primaryClass = {astro-ph.GA},
       adsurl = {https://ui.adsabs.harvard.edu/abs/2019A&A...625A...2K}
}

@ARTICLE{Driver:2022,
       author = {{Driver}, Simon P. and {Bellstedt}, Sabine and {Robotham}, Aaron S.~G. and {Baldry}, Ivan K. and {Davies}, Luke J. and {Liske}, Jochen and {Obreschkow}, Danail and {Taylor}, Edward N. and {Wright}, Angus H. and {Alpaslan}, Mehmet and {Bamford}, Steven P. and {Bauer}, Amanda E. and {Bland-Hawthorn}, Joss and {Bilicki}, Maciej and {Bravo}, Mat{\'\i}as and {Brough}, Sarah and {Casura}, Sarah and {Cluver}, Michelle E. and {Colless}, Matthew and {Conselice}, Christopher J. and {Croom}, Scott M. and {de Jong}, Jelte and {D'Eugenio}, Franceso and {De Propris}, Roberto and {Dogruel}, Burak and {Drinkwater}, Michael J. and {Dvornik}, Andrej and {Farrow}, Daniel J. and {Frenk}, Carlos S. and {Giblin}, Benjamin and {Graham}, Alister W. and {Grootes}, Meiert W. and {Gunawardhana}, Madusha L.~P. and {Hashemizadeh}, Abdolhosein and {H{\"a}u{\ss}ler}, Boris and {Heymans}, Catherine and {Hildebrandt}, Hendrik and {Holwerda}, Benne W. and {Hopkins}, Andrew M. and {Jarrett}, Tom H. and {Heath Jones}, D. and {Kelvin}, Lee S. and {Koushan}, Soheil and {Kuijken}, Konrad and {Lara-L{\'o}pez}, Maritza A. and {Lange}, Rebecca and {L{\'o}pez-S{\'a}nchez}, {\'A}ngel R. and {Loveday}, Jon and {Mahajan}, Smriti and {Meyer}, Martin and {Moffett}, Amanda J. and {Napolitano}, Nicola R. and {Norberg}, Peder and {Owers}, Matt S. and {Radovich}, Mario and {Raouf}, Mojtaba and {Peacock}, John A. and {Phillipps}, Steven and {Pimbblet}, Kevin A. and {Popescu}, Cristina and {Said}, Khaled and {Sansom}, Anne E. and {Seibert}, Mark and {Sutherland}, Will J. and {Thorne}, Jessica E. and {Tuffs}, Richard J. and {Turner}, Ryan and {van der Wel}, Arjen and {van Kampen}, Eelco and {Wilkins}, Steve M.},
        title = "{Galaxy And Mass Assembly (GAMA): Data Release 4 and the z < 0.1 total and z < 0.08 morphological galaxy stellar mass functions}",
      journal = {\mnras},
         year = 2022,
        month = jun,
       volume = {513},
       number = {1},
        pages = {439-467},
          doi = {10.1093/mnras/stac472},
archivePrefix = {arXiv},
       eprint = {2203.08539},
 primaryClass = {astro-ph.GA},
       adsurl = {https://ui.adsabs.harvard.edu/abs/2022MNRAS.513..439D}
}

@ARTICLE{Abbott:2022,
       author = {{Abbott}, T.~M.~C. and {Aguena}, M. and {Alarcon}, A. and {Allam}, S. and {Alves}, O. and {Amon}, A. and {Andrade-Oliveira}, F. and {Annis}, J. and {Avila}, S. and {Bacon}, D. and {Baxter}, E. and {Bechtol}, K. and {Becker}, M.~R. and {Bernstein}, G.~M. and {Bhargava}, S. and {Birrer}, S. and {Blazek}, J. and {Brandao-Souza}, A. and {Bridle}, S.~L. and {Brooks}, D. and {Buckley-Geer}, E. and {Burke}, D.~L. and {Camacho}, H. and {Campos}, A. and {Carnero Rosell}, A. and {Carrasco Kind}, M. and {Carretero}, J. and {Castander}, F.~J. and {Cawthon}, R. and {Chang}, C. and {Chen}, A. and {Chen}, R. and {Choi}, A. and {Conselice}, C. and {Cordero}, J. and {Costanzi}, M. and {Crocce}, M. and {da Costa}, L.~N. and {da Silva Pereira}, M.~E. and {Davis}, C. and {Davis}, T.~M. and {De Vicente}, J. and {DeRose}, J. and {Desai}, S. and {Di Valentino}, E. and {Diehl}, H.~T. and {Dietrich}, J.~P. and {Dodelson}, S. and {Doel}, P. and {Doux}, C. and {Drlica-Wagner}, A. and {Eckert}, K. and {Eifler}, T.~F. and {Elsner}, F. and {Elvin-Poole}, J. and {Everett}, S. and {Evrard}, A.~E. and {Fang}, X. and {Farahi}, A. and {Fernandez}, E. and {Ferrero}, I. and {Fert{\'e}}, A. and {Fosalba}, P. and {Friedrich}, O. and {Frieman}, J. and {Garc{\'\i}a-Bellido}, J. and {Gatti}, M. and {Gaztanaga}, E. and {Gerdes}, D.~W. and {Giannantonio}, T. and {Giannini}, G. and {Gruen}, D. and {Gruendl}, R.~A. and {Gschwend}, J. and {Gutierrez}, G. and {Harrison}, I. and {Hartley}, W.~G. and {Herner}, K. and {Hinton}, S.~R. and {Hollowood}, D.~L. and {Honscheid}, K. and {Hoyle}, B. and {Huff}, E.~M. and {Huterer}, D. and {Jain}, B. and {James}, D.~J. and {Jarvis}, M. and {Jeffrey}, N. and {Jeltema}, T. and {Kovacs}, A. and {Krause}, E. and {Kron}, R. and {Kuehn}, K. and {Kuropatkin}, N. and {Lahav}, O. and {Leget}, P.-F. and {Lemos}, P. and {Liddle}, A.~R. and {Lidman}, C. and {Lima}, M. and {Lin}, H. and {MacCrann}, N. and {Maia}, M.~A.~G. and {Marshall}, J.~L. and {Martini}, P. and {McCullough}, J. and {Melchior}, P. and {Mena-Fern{\'a}ndez}, J. and {Menanteau}, F. and {Miquel}, R. and {Mohr}, J.~J. and {Morgan}, R. and {Muir}, J. and {Myles}, J. and {Nadathur}, S. and {Navarro-Alsina}, A. and {Nichol}, R.~C. and {Ogando}, R.~L.~C. and {Omori}, Y. and {Palmese}, A. and {Pandey}, S. and {Park}, Y. and {Paz-Chinch{\'o}n}, F. and {Petravick}, D. and {Pieres}, A. and {Plazas Malag{\'o}n}, A.~A. and {Porredon}, A. and {Prat}, J. and {Raveri}, M. and {Rodriguez-Monroy}, M. and {Rollins}, R.~P. and {Romer}, A.~K. and {Roodman}, A. and {Rosenfeld}, R. and {Ross}, A.~J. and {Rykoff}, E.~S. and {Samuroff}, S. and {S{\'a}nchez}, C. and {Sanchez}, E. and {Sanchez}, J. and {Sanchez Cid}, D. and {Scarpine}, V. and {Schubnell}, M. and {Scolnic}, D. and {Secco}, L.~F. and {Serrano}, S. and {Sevilla-Noarbe}, I. and {Sheldon}, E. and {Shin}, T. and {Smith}, M. and {Soares-Santos}, M. and {Suchyta}, E. and {Swanson}, M.~E.~C. and {Tabbutt}, M. and {Tarle}, G. and {Thomas}, D. and {To}, C. and {Troja}, A. and {Troxel}, M.~A. and {Tucker}, D.~L. and {Tutusaus}, I. and {Varga}, T.~N. and {Walker}, A.~R. and {Weaverdyck}, N. and {Wechsler}, R. and {Weller}, J. and {Yanny}, B. and {Yin}, B. and {Zhang}, Y. and {Zuntz}, J. and {DES Collaboration}},
        title = "{Dark Energy Survey Year 3 results: Cosmological constraints from galaxy clustering and weak lensing}",
      journal = {\prd},
         year = 2022,
        month = jan,
       volume = {105},
       number = {2},
          eid = {023520},
        pages = {023520},
          doi = {10.1103/PhysRevD.105.023520},
archivePrefix = {arXiv},
       eprint = {2105.13549},
 primaryClass = {astro-ph.CO},
       adsurl = {https://ui.adsabs.harvard.edu/abs/2022PhRvD.105b3520A}
}

@ARTICLE{Scodeggio:2018,
       author = {{Scodeggio}, M. and {Guzzo}, L. and {Garilli}, B. and {Granett}, B.~R. and {Bolzonella}, M. and {de la Torre}, S. and {Abbas}, U. and {Adami}, C. and {Arnouts}, S. and {Bottini}, D. and {Cappi}, A. and {Coupon}, J. and {Cucciati}, O. and {Davidzon}, I. and {Franzetti}, P. and {Fritz}, A. and {Iovino}, A. and {Krywult}, J. and {Le Brun}, V. and {Le F{\`e}vre}, O. and {Maccagni}, D. and {Ma{\l}ek}, K. and {Marchetti}, A. and {Marulli}, F. and {Polletta}, M. and {Pollo}, A. and {Tasca}, L.~A.~M. and {Tojeiro}, R. and {Vergani}, D. and {Zanichelli}, A. and {Bel}, J. and {Branchini}, E. and {De Lucia}, G. and {Ilbert}, O. and {McCracken}, H.~J. and {Moutard}, T. and {Peacock}, J.~A. and {Zamorani}, G. and {Burden}, A. and {Fumana}, M. and {Jullo}, E. and {Marinoni}, C. and {Mellier}, Y. and {Moscardini}, L. and {Percival}, W.~J.},
        title = "{The VIMOS Public Extragalactic Redshift Survey (VIPERS). Full spectroscopic data and auxiliary information release (PDR-2)}",
      journal = {\aap},
         year = 2018,
        month = jan,
       volume = {609},
          eid = {A84},
        pages = {A84},
          doi = {10.1051/0004-6361/201630114},
archivePrefix = {arXiv},
       eprint = {1611.07048},
 primaryClass = {astro-ph.GA},
       adsurl = {https://ui.adsabs.harvard.edu/abs/2018A&A...609A..84S}
}

@ARTICLE{Guzzo:2014,
       author = {{Guzzo}, L. and {Scodeggio}, M. and {Garilli}, B. and {Granett}, B.~R. and {Fritz}, A. and {Abbas}, U. and {Adami}, C. and {Arnouts}, S. and {Bel}, J. and {Bolzonella}, M. and {Bottini}, D. and {Branchini}, E. and {Cappi}, A. and {Coupon}, J. and {Cucciati}, O. and {Davidzon}, I. and {De Lucia}, G. and {de la Torre}, S. and {Franzetti}, P. and {Fumana}, M. and {Hudelot}, P. and {Ilbert}, O. and {Iovino}, A. and {Krywult}, J. and {Le Brun}, V. and {Le F{\`e}vre}, O. and {Maccagni}, D. and {Ma{\l}ek}, K. and {Marulli}, F. and {McCracken}, H.~J. and {Paioro}, L. and {Peacock}, J.~A. and {Polletta}, M. and {Pollo}, A. and {Schlagenhaufer}, H. and {Tasca}, L.~A.~M. and {Tojeiro}, R. and {Vergani}, D. and {Zamorani}, G. and {Zanichelli}, A. and {Burden}, A. and {Di Porto}, C. and {Marchetti}, A. and {Marinoni}, C. and {Mellier}, Y. and {Moscardini}, L. and {Nichol}, R.~C. and {Percival}, W.~J. and {Phleps}, S. and {Wolk}, M.},
        title = "{The VIMOS Public Extragalactic Redshift Survey (VIPERS). An unprecedented view of galaxies and large-scale structure at 0.5 < z < 1.2}",
      journal = {\aap},
         year = 2014,
        month = jun,
       volume = {566},
          eid = {A108},
        pages = {A108},
          doi = {10.1051/0004-6361/201321489},
archivePrefix = {arXiv},
       eprint = {1303.2623},
 primaryClass = {astro-ph.CO},
       adsurl = {https://ui.adsabs.harvard.edu/abs/2014A&A...566A.108G}
}

@inproceedings{PSO,
author={Kennedy,James and Eberhart,Russell},
year={1995},
title={Particle swarm optimization},
booktitle={IEEE International Conference on Neural Networks - Conference Proceedings},
volume={4},
pages={1942-1948},
note={Cited By :50803},
language={English},
url={www.scopus.com},
}

@ARTICLE{Lin:2023,
       author = {{Lin}, Sicheng and {Tinker}, Jeremy L. and {Blanton}, Michael R. and {Guo}, Hong and {Raichoor}, Anand and {Comparat}, Johan and {Brownstein}, Joel R.},
        title = "{Abundance matching analysis of the emission-line galaxy sample in the extended Baryon Oscillation Spectroscopic Survey}",
      journal = {\mnras},
         year = 2023,
        month = mar,
       volume = {519},
       number = {3},
        pages = {4253-4262},
          doi = {10.1093/mnras/stac2793},
       adsurl = {https://ui.adsabs.harvard.edu/abs/2023MNRAS.519.4253L}
}

@ARTICLE{ChavesMontero:2023,
       author = {{Chaves-Montero}, Jon{\'a}s and {Angulo}, Raul E. and {Contreras}, Sergio},
        title = "{The galaxy formation origin of the lensing is low problem}",
      journal = {\mnras},
         year = 2023,
        month = may,
       volume = {521},
       number = {1},
        pages = {937-951},
          doi = {10.1093/mnras/stad243},
archivePrefix = {arXiv},
       eprint = {2211.01744},
 primaryClass = {astro-ph.CO},
       adsurl = {https://ui.adsabs.harvard.edu/abs/2023MNRAS.521..937C}
}

@article{Simha:2012,
    author = {Simha, Vimal and Weinberg, David H. and Davé, Romeel and Fardal, Mark and Katz, Neal and Oppenheimer, Benjamin D.},
    title = "{Testing subhalo abundance matching in cosmological smoothed particle hydrodynamics simulations}",
    journal = {Monthly Notices of the Royal Astronomical Society},
    volume = {423},
    number = {4},
    pages = {3458-3473},
    year = {2012},
    month = {07},
    issn = {0035-8711},
    doi = {10.1111/j.1365-2966.2012.21142.x},
    url = {https://doi.org/10.1111/j.1365-2966.2012.21142.x},
    eprint = {https://academic.oup.com/mnras/article-pdf/423/4/3458/4905216/mnras0423-3458.pdf},
}

@ARTICLE{Nelson:2017TNGcolors,
       author = {{Nelson}, Dylan and {Pillepich}, Annalisa and {Springel}, Volker and {Weinberger}, Rainer and {Hernquist}, Lars and {Pakmor}, R{\"u}diger and {Genel}, Shy and {Torrey}, Paul and {Vogelsberger}, Mark and {Kauffmann}, Guinevere and {Marinacci}, Federico and {Naiman}, Jill},
        title = "{First results from the IllustrisTNG simulations: the galaxy colour bimodality}",
      journal = {\mnras},
         year = 2018,
        month = mar,
       volume = {475},
       number = {1},
        pages = {624-647},
          doi = {10.1093/mnras/stx3040},
archivePrefix = {arXiv},
       eprint = {1707.03395},
 primaryClass = {astro-ph.GA},
       adsurl = {https://ui.adsabs.harvard.edu/abs/2018MNRAS.475..624N}
}

@ARTICLE{Alam:2020,
       author = {{Alam}, Shadab and {Peacock}, John A. and {Kraljic}, Katarina and {Ross}, Ashley J. and {Comparat}, Johan},
        title = "{Multitracer extension of the halo model: probing quenching and conformity in eBOSS}",
      journal = {\mnras},
         year = 2020,
        month = sep,
       volume = {497},
       number = {1},
        pages = {581-595},
          doi = {10.1093/mnras/staa1956},
archivePrefix = {arXiv},
       eprint = {1910.05095},
 primaryClass = {astro-ph.CO},
       adsurl = {https://ui.adsabs.harvard.edu/abs/2020MNRAS.497..581A}
}

@ARTICLE{VosGines:2024,
       author = {{Vos-Gin{\'e}s}, Bernhard and {Avila}, Santiago and {Gonzalez-Perez}, Violeta and {Yepes}, Gustavo},
        title = "{Improving and extending non-Poissonian distributions for satellite galaxies sampling in HOD: applications to eBOSS ELGs}",
      journal = {\mnras},
         year = 2024,
        month = may,
       volume = {530},
       number = {3},
        pages = {3458-3476},
          doi = {10.1093/mnras/stae1096},
archivePrefix = {arXiv},
       eprint = {2310.18189},
 primaryClass = {astro-ph.CO},
       adsurl = {https://ui.adsabs.harvard.edu/abs/2024MNRAS.530.3458V}
}

@ARTICLE{Gao1:2022,
       author = {{Gao}, Hongyu and {Jing}, Y.~P. and {Zheng}, Yun and {Xu}, Kun},
        title = "{Constructing the Emission-line Galaxy-Host Halo Connection through Auto and Cross Correlations}",
      journal = {\apj},
         year = 2022,
        month = mar,
       volume = {928},
       number = {1},
          eid = {10},
        pages = {10},
          doi = {10.3847/1538-4357/ac501b},
archivePrefix = {arXiv},
       eprint = {2111.11657},
 primaryClass = {astro-ph.GA},
       adsurl = {https://ui.adsabs.harvard.edu/abs/2022ApJ...928...10G}
}

@ARTICLE{Gao2:2023,
       author = {{Gao}, Hongyu and {Jing}, Y.~P. and {Gui}, Shanquan and {Xu}, Kun and {Zheng}, Yun and {Zhao}, Donghai and {Aguilar}, Jessica Nicole and {Ahlen}, Steven and {Brooks}, David and {Claybaugh}, Todd and {Dawson}, Kyle and {xde la Macorra}, Axel and {Doel}, Peter and {Fanning}, Kevin and {Forero-Romero}, Jaime E. and {A Gontcho}, Satya Gontcho and {Guy}, Julien and {Honscheid}, Klaus and {Kehoe}, Robert and {Landriau}, Martin and {Manera}, Marc and {Meisner}, Aaron and {Miquel}, Ramon and {Moustakas}, John and {Newman}, Jeffrey A. and {Nie}, Jundan and {Percival}, Will and {Rossi}, Graziano and {Schubnell}, Michael and {Seo}, Hee-Jong and {Tarl{\'e}}, Gregory and {Weaver}, Benjamin Alan and {Yu}, Jiaxi and {Zhou}, Zhimin},
        title = "{The DESI One-Percent Survey: Constructing Galaxy-Halo Connections for ELGs and LRGs Using Auto and Cross Correlations}",
      journal = {\apj},
         year = 2023,
        month = sep,
       volume = {954},
       number = {2},
          eid = {207},
        pages = {207},
          doi = {10.3847/1538-4357/ace90a},
archivePrefix = {arXiv},
       eprint = {2306.06317},
 primaryClass = {astro-ph.GA},
       adsurl = {https://ui.adsabs.harvard.edu/abs/2023ApJ...954..207G}
}

@ARTICLE{Gao3:2024,
       author = {{Gao}, Hongyu and {Jing}, Y.~P. and {Xu}, Kun and {Zhao}, Donghai and {Gui}, Shanquan and {Zheng}, Yun and {Luo}, Xiaolin and {Aguilar}, Jessica Nicole and {Ahlen}, Steven and {Brooks}, David and {Claybaugh}, Todd and {Cole}, Shaun and {de la Macorra}, Axel and {Forero-Romero}, Jaime E. and {Gontcho A Gontcho}, Satya and {Ishak}, Mustapha and {Lambert}, Andrew and {Landriau}, Martin and {Manera}, Marc and {Meisner}, Aaron and {Miquel}, Ramon and {Nie}, Jundan and {Rezaie}, Mehdi and {Rossi}, Graziano and {Sanchez}, Eusebio and {Schubnell}, Michael and {Seo}, Hee-Jong and {Tarl{\'e}}, Gregory and {Weaver}, Benjamin Alan and {Zhou}, Zhimin},
        title = "{The DESI One-Percent Survey: A Concise Model for the Galactic Conformity of Emission-line Galaxies}",
      journal = {\apj},
         year = 2024,
        month = jan,
       volume = {961},
       number = {1},
          eid = {74},
        pages = {74},
          doi = {10.3847/1538-4357/ad09d6},
archivePrefix = {arXiv},
       eprint = {2309.03802},
 primaryClass = {astro-ph.GA},
       adsurl = {https://ui.adsabs.harvard.edu/abs/2024ApJ...961...74G}
}

@ARTICLE{Hadzhiyska:2022_onehalo,
       author = {{Hadzhiyska}, Boryana and {Hernquist}, Lars and {Eisenstein}, Daniel and {Delgado}, Ana Maria and {Bose}, Sownak and {Kannan}, Rahul and {Pakmor}, R{\"u}diger and {Springel}, Volker and {Contreras}, Sergio and {Barrera}, Monica and {Ferlito}, Fulvio and {Hern{\'a}ndez-Aguayo}, C{\'e}sar and {White}, Simon D.~M. and {Frenk}, Carlos},
        title = "{The MillenniumTNG Project: refining the one-halo model of red and blue galaxies at different redshifts}",
      journal = {\mnras},
         year = 2023,
        month = sep,
       volume = {524},
       number = {2},
        pages = {2524-2538},
          doi = {10.1093/mnras/stad279},
archivePrefix = {arXiv},
       eprint = {2210.10068},
 primaryClass = {astro-ph.CO},
       adsurl = {https://ui.adsabs.harvard.edu/abs/2023MNRAS.524.2524H}
}

@ARTICLE{Hadzhiyska:2022,
       author = {{Hadzhiyska}, Boryana and {Eisenstein}, Daniel and {Hernquist}, Lars and {Pakmor}, R{\"u}diger and {Bose}, Sownak and {Delgado}, Ana Maria and {Contreras}, Sergio and {Kannan}, Rahul and {White}, Simon D.~M. and {Springel}, Volker and {Frenk}, Carlos and {Hern{\'a}ndez-Aguayo}, C{\'e}sar and {Barrera}, Fulvio Ferlito and {Monica}},
        title = "{The MillenniumTNG Project: an improved two-halo model for the galaxy-halo connection of red and blue galaxies}",
      journal = {\mnras},
         year = 2023,
        month = sep,
       volume = {524},
       number = {2},
        pages = {2507-2523},
          doi = {10.1093/mnras/stad731},
archivePrefix = {arXiv},
       eprint = {2210.10072},
 primaryClass = {astro-ph.CO},
       adsurl = {https://ui.adsabs.harvard.edu/abs/2023MNRAS.524.2507H}
}

@ARTICLE{Favole:2022,
       author = {{Favole}, Ginevra and {Montero-Dorta}, Antonio D. and {Artale}, M. Celeste and {Contreras}, Sergio and {Zehavi}, Idit and {Xu}, Xiaoju},
        title = "{Subhalo abundance matching through the lens of a hydrodynamical simulation}",
      journal = {\mnras},
         year = 2022,
        month = jan,
       volume = {509},
       number = {2},
        pages = {1614-1625},
          doi = {10.1093/mnras/stab3006},
archivePrefix = {arXiv},
       eprint = {2101.10733},
 primaryClass = {astro-ph.GA},
       adsurl = {https://ui.adsabs.harvard.edu/abs/2022MNRAS.509.1614F}
}

@ARTICLE{Hadzhiyska:2021,
       author = {{Hadzhiyska}, Boryana and {Tacchella}, Sandro and {Bose}, Sownak and {Eisenstein}, Daniel J.},
        title = "{The galaxy-halo connection of emission-line galaxies in IllustrisTNG}",
      journal = {\mnras},
         year = 2021,
        month = apr,
       volume = {502},
       number = {3},
        pages = {3599-3617},
          doi = {10.1093/mnras/stab243},
archivePrefix = {arXiv},
       eprint = {2011.05331},
 primaryClass = {astro-ph.GA},
       adsurl = {https://ui.adsabs.harvard.edu/abs/2021MNRAS.502.3599H}
}

@ARTICLE{DESI:2016,
       author = {{DESI Collaboration} and {Aghamousa}, Amir and {Aguilar}, Jessica and {Ahlen}, Steve and {Alam}, Shadab and {Allen}, Lori E. and {Allende Prieto}, Carlos and {Annis}, James and {Bailey}, Stephen and {Balland}, Christophe and {Ballester}, Otger and {Baltay}, Charles and {Beaufore}, Lucas and {Bebek}, Chris and {Beers}, Timothy C. and {Bell}, Eric F. and {Bernal}, Jos{\'e} Luis and {Besuner}, Robert and {Beutler}, Florian and {Blake}, Chris and {Bleuler}, Hannes and {Blomqvist}, Michael and {Blum}, Robert and {Bolton}, Adam S. and {Briceno}, Cesar and {Brooks}, David and {Brownstein}, Joel R. and {Buckley-Geer}, Elizabeth and {Burden}, Angela and {Burtin}, Etienne and {Busca}, Nicolas G. and {Cahn}, Robert N. and {Cai}, Yan-Chuan and {Cardiel-Sas}, Laia and {Carlberg}, Raymond G. and {Carton}, Pierre-Henri and {Casas}, Ricard and {Castander}, Francisco J. and {Cervantes-Cota}, Jorge L. and {Claybaugh}, Todd M. and {Close}, Madeline and {Coker}, Carl T. and {Cole}, Shaun and {Comparat}, Johan and {Cooper}, Andrew P. and {Cousinou}, M. -C. and {Crocce}, Martin and {Cuby}, Jean-Gabriel and {Cunningham}, Daniel P. and {Davis}, Tamara M. and {Dawson}, Kyle S. and {de la Macorra}, Axel and {De Vicente}, Juan and {Delubac}, Timoth{\'e}e and {Derwent}, Mark and {Dey}, Arjun and {Dhungana}, Govinda and {Ding}, Zhejie and {Doel}, Peter and {Duan}, Yutong T. and {Ealet}, Anne and {Edelstein}, Jerry and {Eftekharzadeh}, Sarah and {Eisenstein}, Daniel J. and {Elliott}, Ann and {Escoffier}, St{\'e}phanie and {Evatt}, Matthew and {Fagrelius}, Parker and {Fan}, Xiaohui and {Fanning}, Kevin and {Farahi}, Arya and {Farihi}, Jay and {Favole}, Ginevra and {Feng}, Yu and {Fernandez}, Enrique and {Findlay}, JosephDark Energy Survey year 1 R. and {Finkbeiner}, Douglas P. and {Fitzpatrick}, Michael J. and {Flaugher}, Brenna and {Flender}, Samuel and {Font-Ribera}, Andreu and {Forero-Romero}, Jaime E. and {Fosalba}, Pablo and {Frenk}, Carlos S. and {Fumagalli}, Michele and {Gaensicke}, Boris T. and {Gallo}, Giuseppe and {Garcia-Bellido}, Juan and {Gaztanaga}, Enrique and {Pietro Gentile Fusillo}, Nicola and {Gerard}, Terry and {Gershkovich}, Irena and {Giannantonio}, Tommaso and {Gillet}, Denis and {Gonzalez-de-Rivera}, Guillermo and {Gonzalez-Perez}, Violeta and {Gott}, Shelby and {Graur}, Or and {Gutierrez}, Gaston and {Guy}, Julien and {Habib}, Salman and {Heetderks}, Henry and {Heetderks}, Ian and {Heitmann}, Katrin and {Hellwing}, Wojciech A. and {Herrera}, David A. and {Ho}, Shirley and {Holland}, Stephen and {Honscheid}, Klaus and {Huff}, Eric and {Hutchinson}, Timothy A. and {Huterer}, Dragan and {Hwang}, Ho Seong and {Illa Laguna}, Joseph Maria and {Ishikawa}, Yuzo and {Jacobs}, Dianna and {Jeffrey}, Niall and {Jelinsky}, Patrick and {Jennings}, Elise and {Jiang}, Linhua and {Jimenez}, Jorge and {Johnson}, Jennifer and {Joyce}, Richard and {Jullo}, Eric and {Juneau}, St{\'e}phanie and {Kama}, Sami and {Karcher}, Armin and {Karkar}, Sonia and {Kehoe}, Robert and {Kennamer}, Noble and {Kent}, Stephen and {Kilbinger}, Martin and {Kim}, Alex G. and {Kirkby}, David and {Kisner}, Theodore and {Kitanidis}, Ellie and {Kneib}, Jean-Paul and {Koposov}, Sergey and {Kovacs}, Eve and {Koyama}, Kazuya and {Kremin}, Anthony and {Kron}, Richard and {Kronig}, Luzius and {Kueter-Young}, Andrea and {Lacey}, Cedric G. and {Lafever}, Robin and {Lahav}, Ofer and {Lambert}, Andrew and {Lampton}, Michael and {Landriau}, Martin and {Lang}, Dustin and {Lauer}, Tod R. and {Le Goff}, Jean-Marc and {Le Guillou}, Laurent and {Le Van Suu}, Auguste and {Lee}, Jae Hyeon and {Lee}, Su-Jeong and {Leitner}, Daniela and {Lesser}, Michael and {Levi}, Michael E. and {L'Huillier}, Benjamin and {Li}, Baojiu and {Liang}, Ming and {Lin}, Huan and {Linder}, Eric and {Loebman}, Sarah R. and {Luki{\'c}}, Zarija and {Ma}, Jun and {MacCrann}, Niall and {Magneville}, Christophe and {Makarem}, Laleh and {Manera}, Marc and {Manser}, Christopher J. and {Marshall}, Robert and {Martini}, Paul and {Massey}, Richard and {Matheson}, Thomas and {McCauley}, Jeremy and {McDonald}, Patrick and {McGreer}, Ian D. and {Meisner}, Aaron and {Metcalfe}, Nigel and {Miller}, Timothy N. and {Miquel}, Ramon and {Moustakas}, John and {Myers}, Adam and {Naik}, Milind and {Newman}, Jeffrey A. and {Nichol}, Robert C. and {Nicola}, Andrina and {Nicolati da Costa}, Luiz and {Nie}, Jundan and {Niz}, Gustavo and {Norberg}, Peder and {Nord}, Brian and {Norman}, Dara and {Nugent}, Peter and {O'Brien}, Thomas and {Oh}, Minji and {Olsen}, Knut A.~G.},
        title = "{The DESI Experiment Part I: Science,Targeting, and Survey Design}",
      journal = {arXiv e-prints},
         year = 2016,
        month = oct,
          eid = {arXiv:1611.00036},
        pages = {arXiv:1611.00036},
          doi = {10.48550/arXiv.1611.00036},
archivePrefix = {arXiv},
       eprint = {1611.00036},
 primaryClass = {astro-ph.IM},
       adsurl = {https://ui.adsabs.harvard.edu/abs/2016arXiv161100036D}
}

@ARTICLE{C2023:lensing,
       author = {{Contreras}, Sergio and {Angulo}, Raul E. and {Chaves-Montero}, Jon{\'a}s and {White}, Simon D.~M. and {Aric{\`o}}, Giovanni},
        title = "{Consistent and simultaneous modelling of galaxy clustering and galaxy-galaxy lensing with subhalo abundance matching}",
      journal = {\mnras},
         year = 2023,
        month = mar,
       volume = {520},
       number = {1},
        pages = {489-502},
          doi = {10.1093/mnras/stad122},
archivePrefix = {arXiv},
       eprint = {2211.11745},
 primaryClass = {astro-ph.CO},
       adsurl = {https://ui.adsabs.harvard.edu/abs/2023MNRAS.520..489C}
}

@ARTICLE{C22:MTNGCosmology,
       author = {{Contreras}, Sergio and {Angulo}, Raul E. and {Springel}, Volker and {White}, Simon D.~M. and {Hadzhiyska}, Boryana and {Hernquist}, Lars and {Pakmor}, R{\"u}diger and {Kannan}, Rahul and {Hern{\'a}ndez-Aguayo}, C{\'e}sar and {Barrera}, Monica and {Ferlito}, Fulvio and {Delgado}, Ana Maria and {Bose}, Sownak and {Frenk}, Carlos},
        title = "{The MillenniumTNG Project: inferring cosmology from galaxy clustering with accelerated N-body scaling and subhalo abundance matching}",
      journal = {\mnras},
         year = 2023,
        month = sep,
       volume = {524},
       number = {2},
        pages = {2489-2506},
          doi = {10.1093/mnras/stac3699},
archivePrefix = {arXiv},
       eprint = {2210.10075},
 primaryClass = {astro-ph.GA},
       adsurl = {https://ui.adsabs.harvard.edu/abs/2023MNRAS.524.2489C}
}

@ARTICLE{C22a,
       author = {{Contreras}, Sergio and {Angulo}, Raul E. and {Springel}, Volker and {White}, Simon D.~M. and {Hadzhiyska}, Boryana and {Hernquist}, Lars and {Pakmor}, R{\"u}diger and {Kannan}, Rahul and {Hern{\'a}ndez-Aguayo}, C{\'e}sar and {Barrera}, Monica and {Ferlito}, Fulvio and {Delgado}, Ana Maria and {Bose}, Sownak and {Frenk}, Carlos},
        title = "{The MillenniumTNG Project: inferring cosmology from galaxy clustering with accelerated N-body scaling and subhalo abundance matching}",
      journal = {\mnras},
         year = 2023,
        month = sep,
       volume = {524},
       number = {2},
        pages = {2489-2506},
          doi = {10.1093/mnras/stac3699},
archivePrefix = {arXiv},
       eprint = {2210.10075},
 primaryClass = {astro-ph.GA},
       adsurl = {https://ui.adsabs.harvard.edu/abs/2023MNRAS.524.2489C}
}

@article{HernandezAguayo2022,
       author = {{Hern{\'a}ndez-Aguayo}, C{\'e}sar and {Springel}, Volker and {Pakmor}, R{\"u}diger and {Barrera}, Monica and {Ferlito}, Fulvio and {White}, Simon D.~M. and {Hernquist}, Lars and {Hadzhiyska}, Boryana and {Delgado}, Ana Maria and {Kannan}, Rahul and {Bose}, Sownak and {Frenk}, Carlos},
        title = "{The MillenniumTNG Project: high-precision predictions for matter clustering and halo statistics}",
      journal = {\mnras},
         year = 2023,
        month = sep,
       volume = {524},
       number = {2},
        pages = {2556-2578},
          doi = {10.1093/mnras/stad1657},
archivePrefix = {arXiv},
       eprint = {2210.10059},
 primaryClass = {astro-ph.CO},
       adsurl = {https://ui.adsabs.harvard.edu/abs/2023MNRAS.524.2556H}
}

@article{Pakmor2022,
       author = {{Pakmor}, R{\"u}diger and {Springel}, Volker and {Coles}, Jonathan P. and {Guillet}, Thomas and {Pfrommer}, Christoph and {Bose}, Sownak and {Barrera}, Monica and {Delgado}, Ana Maria and {Ferlito}, Fulvio and {Frenk}, Carlos and {Hadzhiyska}, Boryana and {Hern{\'a}ndez-Aguayo}, C{\'e}sar and {Hernquist}, Lars and {Kannan}, Rahul and {White}, Simon D.~M.},
        title = "{The MillenniumTNG Project: the hydrodynamical full physics simulation and a first look at its galaxy clusters}",
      journal = {\mnras},
         year = 2023,
        month = sep,
       volume = {524},
       number = {2},
        pages = {2539-2555},
          doi = {10.1093/mnras/stac3620},
archivePrefix = {arXiv},
       eprint = {2210.10060},
 primaryClass = {astro-ph.CO},
       adsurl = {https://ui.adsabs.harvard.edu/abs/2023MNRAS.524.2539P}
}

@article{Barrera2022,
       author = {{Barrera}, Monica and {Springel}, Volker and {White}, Simon D.~M. and {Hern{\'a}ndez-Aguayo}, C{\'e}sar and {Hernquist}, Lars and {Frenk}, Carlos and {Pakmor}, R{\"u}diger and {Ferlito}, Fulvio and {Hadzhiyska}, Boryana and {Delgado}, Ana Maria and {Kannan}, Rahul and {Bose}, Sownak},
        title = "{The MillenniumTNG Project: semi-analytic galaxy formation models on the past lightcone}",
      journal = {\mnras},
         year = 2023,
        month = nov,
       volume = {525},
       number = {4},
        pages = {6312-6335},
          doi = {10.1093/mnras/stad2688},
archivePrefix = {arXiv},
       eprint = {2210.10419},
 primaryClass = {astro-ph.CO},
       adsurl = {https://ui.adsabs.harvard.edu/abs/2023MNRAS.525.6312B}
}

@ARTICLE{Ferlito2022,
       author = {{Ferlito}, Fulvio and {Springel}, Volker and {Davies}, Christopher T. and {Hern{\'a}ndez-Aguayo}, C{\'e}sar and {Pakmor}, R{\"u}diger and {Barrera}, Monica and {White}, Simon D.~M. and {Delgado}, Ana Maria and {Hadzhiyska}, Boryana and {Hernquist}, Lars and {Kannan}, Rahul and {Bose}, Sownak and {Frenk}, Carlos},
        title = "{The MillenniumTNG Project: the impact of baryons and massive neutrinos on high-resolution weak gravitational lensing convergence maps}",
      journal = {\mnras},
         year = 2023,
        month = oct,
       volume = {524},
       number = {4},
        pages = {5591-5606},
          doi = {10.1093/mnras/stad2205},
archivePrefix = {arXiv},
       eprint = {2304.12338},
 primaryClass = {astro-ph.CO},
       adsurl = {https://ui.adsabs.harvard.edu/abs/2023MNRAS.524.5591F}
}

@article{Hadzhiyska2022b,
       author = {{Hadzhiyska}, Boryana and {Eisenstein}, Daniel and {Hernquist}, Lars and {Pakmor}, R{\"u}diger and {Bose}, Sownak and {Delgado}, Ana Maria and {Contreras}, Sergio and {Kannan}, Rahul and {White}, Simon D.~M. and {Springel}, Volker and {Frenk}, Carlos and {Hern{\'a}ndez-Aguayo}, C{\'e}sar and {Ferlito}, Fulvio and {Barrera}, Monica},
        title = "{The MillenniumTNG Project: An improved two-halo model for the galaxy-halo connection of red and blue galaxies}",
      journal = {arXiv e-prints},
         year = 2022,
        month = oct,
          eid = {arXiv:2210.10072},
        pages = {arXiv:2210.10072},
archivePrefix = {arXiv},
       eprint = {2210.10072},
 primaryClass = {astro-ph.CO},
       adsurl = {https://ui.adsabs.harvard.edu/abs/2022arXiv221010072H}
}

@ARTICLE{Delgado2022,
       author = {{Delgado}, Ana Maria and {Hadzhiyska}, Boryana and {Bose}, Sownak and {Springel}, Volker and {Hernquist}, Lars and {Barrera}, Monica and {Pakmor}, R{\"u}diger and {Ferlito}, Fulvio and {Kannan}, Rahul and {Hern{\'a}ndez-Aguayo}, C{\'e}sar and {White}, Simon D.~M. and {Frenk}, Carlos},
        title = "{The MillenniumTNG project: intrinsic alignments of galaxies and haloes}",
      journal = {\mnras},
         year = 2023,
        month = aug,
       volume = {523},
       number = {4},
        pages = {5899-5914},
          doi = {10.1093/mnras/stad1781},
archivePrefix = {arXiv},
       eprint = {2304.12346},
 primaryClass = {astro-ph.CO},
       adsurl = {https://ui.adsabs.harvard.edu/abs/2023MNRAS.523.5899D}
}

@article{Kannan2022,
       author = {{Kannan}, Rahul and {Springel}, Volker and {Hernquist}, Lars and {Pakmor}, R{\"u}diger and {Delgado}, Ana Maria and {Hadzhiyska}, Boryana and {Hern{\'a}ndez-Aguayo}, C{\'e}sar and {Barrera}, Monica and {Ferlito}, Fulvio and {Bose}, Sownak and {White}, Simon D.~M. and {Frenk}, Carlos and {Smith}, Aaron and {Garaldi}, Enrico},
        title = "{The MillenniumTNG project: the galaxy population at z {\ensuremath{\geq}} 8}",
      journal = {\mnras},
         year = 2023,
        month = sep,
       volume = {524},
       number = {2},
        pages = {2594-2605},
          doi = {10.1093/mnras/stac3743},
archivePrefix = {arXiv},
       eprint = {2210.10066},
 primaryClass = {astro-ph.GA},
       adsurl = {https://ui.adsabs.harvard.edu/abs/2023MNRAS.524.2594K}
}

@article{Bose2022,
         author = {{Bose}, Sownak and {Hadzhiyska}, Boryana and {Barrera}, Monica and {Delgado}, Ana Maria and {Ferlito}, Fulvio and {Frenk}, Carlos and {Hern{\'a}ndez-Aguayo}, C{\'e}sar and {Hernquist}, Lars and {Kannan}, Rahul and {Pakmor}, R{\"u}diger and {Springel}, Volker and {White}, Simon D.~M.},
        title = "{The MillenniumTNG Project: the large-scale clustering of galaxies}",
      journal = {\mnras},
         year = 2023,
        month = sep,
       volume = {524},
       number = {2},
        pages = {2579-2593},
          doi = {10.1093/mnras/stad1097},
archivePrefix = {arXiv},
       eprint = {2210.10065},
 primaryClass = {astro-ph.CO},
       adsurl = {https://ui.adsabs.harvard.edu/abs/2023MNRAS.524.2579B}
}

@ARTICLE{Raveri:2019,
       author = {{Raveri}, Marco and {Hu}, Wayne},
        title = "{Concordance and discordance in cosmology}",
      journal = {\prd},
         year = 2019,
        month = feb,
       volume = {99},
       number = {4},
          eid = {043506},
        pages = {043506},
          doi = {10.1103/PhysRevD.99.043506},
archivePrefix = {arXiv},
       eprint = {1806.04649},
 primaryClass = {astro-ph.CO},
       adsurl = {https://ui.adsabs.harvard.edu/abs/2019PhRvD..99d3506R}
}

@ARTICLE{Simha:2013,
       author = {{Simha}, Vimal and {Cole}, Shaun},
        title = "{Cosmological constraints from applying SHAM to rescaled cosmological simulations}",
      journal = {\mnras},
         year = 2013,
        month = dec,
       volume = {436},
       number = {2},
        pages = {1142-1151},
          doi = {10.1093/mnras/stt1643},
archivePrefix = {arXiv},
       eprint = {1302.0852},
 primaryClass = {astro-ph.CO},
       adsurl = {https://ui.adsabs.harvard.edu/abs/2013MNRAS.436.1142S}
}

@misc{tensorflow,
title={ {TensorFlow}: Large-Scale Machine Learning on Heterogeneous Systems},
url={https://www.tensorflow.org/},
note={Software available from tensorflow.org},
author={
    Mart\'{\i}n~Abadi and
    Ashish~Agarwal and
    Paul~Barham and
    Eugene~Brevdo and
    Zhifeng~Chen and
    Craig~Citro and
    Greg~S.~Corrado and
    Andy~Davis and
    Jeffrey~Dean and
    Matthieu~Devin and
    Sanjay~Ghemawat and
    Ian~Goodfellow and
    Andrew~Harp and
    Geoffrey~Irving and
    Michael~Isard and
    Yangqing Jia and
    Rafal~Jozefowicz and
    Lukasz~Kaiser and
    Manjunath~Kudlur and
    Josh~Levenberg and
    Dandelion~Man\'{e} and
    Rajat~Monga and
    Sherry~Moore and
    Derek~Murray and
    Chris~Olah and
    Mike~Schuster and
    Jonathon~Shlens and
    Benoit~Steiner and
    Ilya~Sutskever and
    Kunal~Talwar and
    Paul~Tucker and
    Vincent~Vanhoucke and
    Vijay~Vasudevan and
    Fernanda~Vi\'{e}gas and
    Oriol~Vinyals and
    Pete~Warden and
    Martin~Wattenberg and
    Martin~Wicke and
    Yuan~Yu and
    Xiaoqiang~Zheng},
  year={2015},
}

@ARTICLE{PSOBACCO,
       author = {{Aric{\`o}}, Giovanni and {Angulo}, Raul E. and {Hern{\'a}ndez-Monteagudo}, Carlos and {Contreras}, Sergio and {Zennaro}, Matteo},
        title = "{Simultaneous modelling of matter power spectrum and bispectrum in the presence of baryons}",
      journal = {\mnras},
         year = {2021},
        month = may,
       volume = {503},
       number = {3},
        pages = {3596-3609},
          doi = {10.1093/mnras/stab699},
archivePrefix = {arXiv},
       eprint = {2009.14225},
 primaryClass = {astro-ph.CO},
       adsurl = {https://ui.adsabs.harvard.edu/abs/2021MNRAS.503.3596A}
}

@ARTICLE{Arico:2021,
       author = {{Aric{\`o}}, Giovanni and {Angulo}, Raul E. and {Contreras}, Sergio and {Ondaro-Mallea}, Lurdes and {Pellejero-Iba{\~n}ez}, Marcos and {Zennaro}, Matteo},
        title = "{The BACCO simulation project: a baryonification emulator with neural networks}",
      journal = {\mnras},
         year = 2021,
        month = sep,
       volume = {506},
       number = {3},
        pages = {4070-4082},
          doi = {10.1093/mnras/stab1911},
archivePrefix = {arXiv},
       eprint = {2011.15018},
 primaryClass = {astro-ph.CO},
       adsurl = {https://ui.adsabs.harvard.edu/abs/2021MNRAS.506.4070A}
}

@ARTICLE{C21c,
       author = {{Contreras}, S. and {Angulo}, R.~E. and {Zennaro}, M.},
        title = "{A flexible subhalo abundance matching model for galaxy clustering in redshift space}",
      journal = {\mnras},
         year = 2021,
        month = nov,
       volume = {508},
       number = {1},
        pages = {175-189},
          doi = {10.1093/mnras/stab2560},
archivePrefix = {arXiv},
       eprint = {2012.06596},
 primaryClass = {astro-ph.CO},
       adsurl = {https://ui.adsabs.harvard.edu/abs/2021MNRAS.508..175C}
}

@ARTICLE{AlcockPaczynski,
       author = {{Alcock}, C. and {Paczynski}, B.},
        title = "{An evolution free test for non-zero cosmological constant}",
      journal = {\nat},
         year = 1979,
        month = oct,
       volume = {281},
        pages = {358},
          doi = {10.1038/281358a0},
       adsurl = {https://ui.adsabs.harvard.edu/abs/1979Natur.281..358A}
}

@ARTICLE{Angulo:2021,
       author = {{Angulo}, Raul E. and {Zennaro}, Matteo and {Contreras}, Sergio and {Aric{\`o}}, Giovanni and {Pellejero-Iba{\~n}ez}, Marcos and {St{\"u}cker}, Jens},
        title = "{The BACCO simulation project: exploiting the full power of large-scale structure for cosmology}",
      journal = {\mnras},
         year = 2021,
        month = nov,
       volume = {507},
       number = {4},
        pages = {5869-5881},
          doi = {10.1093/mnras/stab2018},
archivePrefix = {arXiv},
       eprint = {2004.06245},
 primaryClass = {astro-ph.CO},
       adsurl = {https://ui.adsabs.harvard.edu/abs/2021MNRAS.507.5869A}
}

@ARTICLE{Planck:2018,
       author = {{Planck Collaboration} and {Aghanim}, N. and {Akrami}, Y. and
         {Ashdown}, M. and {Aumont}, J. and {Baccigalupi}, C. and
         {Ballardini}, M. and {Banday}, A.~J. and {Barreiro}, R.~B. and
         {Bartolo}, N. and {Basak}, S. and {Battye}, R. and {Benabed}, K. and
         {Bernard}, J. -P. and {Bersanelli}, M. and {Bielewicz}, P. and
         {Bock}, J.~J. and {Bond}, J.~R. and {Borrill}, J. and {Bouchet}, F.~R. and
         {Boulanger}, F. and {Bucher}, M. and {Burigana}, C. and
         {Butler}, R.~C. and {Calabrese}, E. and {Cardoso}, J. -F. and
         {Carron}, J. and {Challinor}, A. and {Chiang}, H.~C. and {Chluba}, J. and
         {Colombo}, L.~P.~L. and {Combet}, C. and {Contreras}, D. and
         {Crill}, B.~P. and {Cuttaia}, F. and {de Bernardis}, P. and
         {de Zotti}, G. and {Delabrouille}, J. and {Delouis}, J. -M. and
         {Di Valentino}, E. and {Diego}, J.~M. and {Dor{\'e}}, O. and
         {Douspis}, M. and {Ducout}, A. and {Dupac}, X. and {Dusini}, S. and
         {Efstathiou}, G. and {Elsner}, F. and {En{\ss}lin}, T.~A. and
         {Eriksen}, H.~K. and {Fantaye}, Y. and {Farhang}, M. and
         {Fergusson}, J. and {Fernandez-Cobos}, R. and {Finelli}, F. and
         {Forastieri}, F. and {Frailis}, M. and {Fraisse}, A.~A. and
         {Franceschi}, E. and {Frolov}, A. and {Galeotta}, S. and {Galli}, S. and
         {Ganga}, K. and {G{\'e}nova-Santos}, R.~T. and {Gerbino}, M. and
         {Ghosh}, T. and {Gonz{\'a}lez-Nuevo}, J. and {G{\'o}rski}, K.~M. and
         {Gratton}, S. and {Gruppuso}, A. and {Gudmundsson}, J.~E. and
         {Hamann}, J. and {Handley}, W. and {Hansen}, F.~K. and {Herranz}, D. and
         {Hildebrandt}, S.~R. and {Hivon}, E. and {Huang}, Z. and
         {Jaffe}, A.~H. and {Jones}, W.~C. and {Karakci}, A. and
         {Keih{\"a}nen}, E. and {Keskitalo}, R. and {Kiiveri}, K. and {Kim}, J. and
         {Kisner}, T.~S. and {Knox}, L. and {Krachmalnicoff}, N. and {Kunz}, M. and
         {Kurki-Suonio}, H. and {Lagache}, G. and {Lamarre}, J. -M. and
         {Lasenby}, A. and {Lattanzi}, M. and {Lawrence}, C.~R. and
         {Le Jeune}, M. and {Lemos}, P. and {Lesgourgues}, J. and {Levrier}, F. and
         {Lewis}, A. and {Liguori}, M. and {Lilje}, P.~B. and {Lilley}, M. and
         {Lindholm}, V. and {L{\'o}pez-Caniego}, M. and {Lubin}, P.~M. and
         {Ma}, Y. -Z. and {Mac{\'\i}as-P{\'e}rez}, J.~F. and {Maggio}, G. and
         {Maino}, D. and {Mandolesi}, N. and {Mangilli}, A. and
         {Marcos-Caballero}, A. and {Maris}, M. and {Martin}, P.~G. and
         {Martinelli}, M. and {Mart{\'\i}nez-Gonz{\'a}lez}, E. and
         {Matarrese}, S. and {Mauri}, N. and {McEwen}, J.~D. and
         {Meinhold}, P.~R. and {Melchiorri}, A. and {Mennella}, A. and
         {Migliaccio}, M. and {Millea}, M. and {Mitra}, S. and
         {Miville-Desch{\^e}nes}, M. -A. and {Molinari}, D. and {Montier}, L. and
         {Morgante}, G. and {Moss}, A. and {Natoli}, P. and
         {N{\o}rgaard-Nielsen}, H.~U. and {Pagano}, L. and {Paoletti}, D. and
         {Partridge}, B. and {Patanchon}, G. and {Peiris}, H.~V. and
         {Perrotta}, F. and {Pettorino}, V. and {Piacentini}, F. and
         {Polastri}, L. and {Polenta}, G. and {Puget}, J. -L. and
         {Rachen}, J.~P. and {Reinecke}, M. and {Remazeilles}, M. and
         {Renzi}, A. and {Rocha}, G. and {Rosset}, C. and {Roudier}, G. and
         {Rubi{\~n}o-Mart{\'\i}n}, J.~A. and {Ruiz-Granados}, B. and
         {Salvati}, L. and {Sandri}, M. and {Savelainen}, M. and {Scott}, D. and
         {Shellard}, E.~P.~S. and {Sirignano}, C. and {Sirri}, G. and
         {Spencer}, L.~D. and {Sunyaev}, R. and {Suur-Uski}, A. -S. and
         {Tauber}, J.~A. and {Tavagnacco}, D. and {Tenti}, M. and
         {Toffolatti}, L. and {Tomasi}, M. and {Trombetti}, T. and
         {Valenziano}, L. and {Valiviita}, J. and {Van Tent}, B. and
         {Vibert}, L. and {Vielva}, P. and {Villa}, F. and {Vittorio}, N. and {Wand
        elt}, B.~D. and {Wehus}, I.~K. and {White}, M. and {White}, S.~D.~M. and
         {Zacchei}, A. and {Zonca}, A.},
        title = "{Planck 2018 results. VI. Cosmological parameters}",
      journal = {\aap},
         year = 2020,
        month = sep,
       volume = {641},
          eid = {A6},
        pages = {A6},
          doi = {10.1051/0004-6361/201833910},
archivePrefix = {arXiv},
       eprint = {1807.06209},
 primaryClass = {astro-ph.CO},
       adsurl = {https://ui.adsabs.harvard.edu/abs/2020A&A...641A...6P}
}

@ARTICLE{Zehavi:2002,
       author = {{Zehavi}, Idit and {Blanton}, Michael R. and {Frieman}, Joshua A. and
         {Weinberg}, David H. and {Mo}, Houjun J. and {Strauss}, Michael A. and
         {Anderson}, Scott F. and {Annis}, James and {Bahcall}, Neta A. and
         {Bernardi}, Mariangela and {Briggs}, John W. and {Brinkmann}, Jon and
         {Burles}, Scott and {Carey}, Larry and {Castander}, Francisco J. and
         {Connolly}, Andrew J. and {Csabai}, Istvan and
         {Dalcanton}, Julianne J. and {Dodelson}, Scott and {Doi}, Mamoru and
         {Eisenstein}, Daniel and {Evans}, Michael L. and
         {Finkbeiner}, Douglas P. and {Friedman}, Scott and
         {Fukugita}, Masataka and {Gunn}, James E. and {Hennessy}, Greg S. and
         {Hindsley}, Robert B. and {Ivezi{\'c}}, {\v{Z}}eljko and
         {Kent}, Stephen and {Knapp}, Gillian R. and {Kron}, Richard and
         {Kunszt}, Peter and {Lamb}, Donald Q. and {Leger}, R. French and
         {Long}, Daniel C. and {Loveday}, Jon and {Lupton}, Robert H. and
         {McKay}, Timothy and {Meiksin}, Avery and {Merrelli}, Aronne and
         {Munn}, Jeffrey A. and {Narayanan}, Vijay and {Newcomb}, Matt and
         {Nichol}, Robert C. and {Owen}, Russell and {Peoples}, John and
         {Pope}, Adrian and {Rockosi}, Constance M. and {Schlegel}, David and
         {Schneider}, Donald P. and {Scoccimarro}, Roman and {Sheth}, Ravi K. and
         {Siegmund}, Walter and {Smee}, Stephen and {Snir}, Yehuda and
         {Stebbins}, Albert and {Stoughton}, Christopher and {SubbaRao}, Mark and
         {Szalay}, Alexander S. and {Szapudi}, Istvan and {Tegmark}, Max and
         {Tucker}, Douglas L. and {Uomoto}, Alan and {Vanden Berk}, Dan and
         {Vogeley}, Michael S. and {Waddell}, Patrick and {Yanny}, Brian and
         {York}, Donald G.},
        title = "{Galaxy Clustering in Early Sloan Digital Sky Survey Redshift Data}",
      journal = {\apj},
         year = 2002,
        month = may,
       volume = {571},
       number = {1},
        pages = {172-190},
          doi = {10.1086/339893},
archivePrefix = {arXiv},
       eprint = {astro-ph/0106476},
 primaryClass = {astro-ph},
       adsurl = {https://ui.adsabs.harvard.edu/abs/2002ApJ...571..172Z}
}

@ARTICLE{Avila:2020,
       author = {{Avila}, S. and {Gonzalez-Perez}, V. and {Mohammad}, F.~G. and
         {de Mattia}, A. and {Zhao}, C. and {Raichoor}, A. and {Tamone}, A. and
         {Alam}, S. and {Bautista}, J. and {Bianchi}, D. and {Burtin}, E. and
         {Chapman}, M.~J. and {Chuang}, C. -H. and {Comparat}, J. and
         {Dawson}, K. and {Divers}, T. and {du Mas des Bourboux}, H. and
         {Gil-Marin}, H. and {Mueller}, E.~M. and {Habib}, S. and
         {Heitmann}, K. and {Ruhlmann-Kleider}, V. and {Padilla}, N. and
         {Percival}, W.~J. and {Ross}, A.~J. and {Seo}, H.~J. and
         {Schneider}, D.~P. and {Zhao}, G.},
        title = "{The Completed SDSS-IV extended Baryon Oscillation Spectroscopic Survey: exploring the halo occupation distribution model for emission line galaxies}",
      journal = {\mnras},
         year = 2020,
        month = sep,
       volume = {499},
       number = {4},
        pages = {5486-5507},
          doi = {10.1093/mnras/staa2951},
archivePrefix = {arXiv},
       eprint = {2007.09012},
 primaryClass = {astro-ph.CO},
       adsurl = {https://ui.adsabs.harvard.edu/abs/2020MNRAS.499.5486A}
}

@ARTICLE{C20,
       author = {{Contreras}, S. and {Angulo}, R.~E. and {Zennaro}, M. and
         {Aric{\`o}}, G. and {Pellejero-Iba{\~n}ez}, M.},
        title = "{3 per cent-accurate predictions for the clustering of dark matter, haloes, and subhaloes, over a wide range of cosmologies and scales}",
      journal = {\mnras},
         year = 2020,
        month = oct,
       volume = {499},
       number = {4},
        pages = {4905-4917},
          doi = {10.1093/mnras/staa3117},
archivePrefix = {arXiv},
       eprint = {2001.03176},
 primaryClass = {astro-ph.CO},
       adsurl = {https://ui.adsabs.harvard.edu/abs/2020MNRAS.499.4905C}
}

@ARTICLE{Planck2015,
       author = {{Planck Collaboration} and {Ade}, P.~A.~R. and {Aghanim}, N. and
         {Arnaud}, M. and {Ashdown}, M. and {Aumont}, J. and {Baccigalupi}, C. and
         {Banday}, A.~J. and {Barreiro}, R.~B. and {Bartlett}, J.~G. and
         {Bartolo}, N. and {Battaner}, E. and {Battye}, R. and {Benabed}, K. and
         {Beno{\^\i}t}, A. and {Benoit-L{\'e}vy}, A. and {Bernard}, J. -P. and
         {Bersanelli}, M. and {Bielewicz}, P. and {Bock}, J.~J. and
         {Bonaldi}, A. and {Bonavera}, L. and {Bond}, J.~R. and {Borrill}, J. and
         {Bouchet}, F.~R. and {Boulanger}, F. and {Bucher}, M. and
         {Burigana}, C. and {Butler}, R.~C. and {Calabrese}, E. and
         {Cardoso}, J. -F. and {Catalano}, A. and {Challinor}, A. and
         {Chamballu}, A. and {Chary}, R. -R. and {Chiang}, H.~C. and
         {Chluba}, J. and {Christensen}, P.~R. and {Church}, S. and
         {Clements}, D.~L. and {Colombi}, S. and {Colombo}, L.~P.~L. and
         {Combet}, C. and {Coulais}, A. and {Crill}, B.~P. and {Curto}, A. and
         {Cuttaia}, F. and {Danese}, L. and {Davies}, R.~D. and {Davis}, R.~J. and
         {de Bernardis}, P. and {de Rosa}, A. and {de Zotti}, G. and
         {Delabrouille}, J. and {D{\'e}sert}, F. -X. and {Di Valentino}, E. and
         {Dickinson}, C. and {Diego}, J.~M. and {Dolag}, K. and {Dole}, H. and
         {Donzelli}, S. and {Dor{\'e}}, O. and {Douspis}, M. and {Ducout}, A. and
         {Dunkley}, J. and {Dupac}, X. and {Efstathiou}, G. and {Elsner}, F. and
         {En{\ss}lin}, T.~A. and {Eriksen}, H.~K. and {Farhang}, M. and
         {Fergusson}, J. and {Finelli}, F. and {Forni}, O. and {Frailis}, M. and
         {Fraisse}, A.~A. and {Franceschi}, E. and {Frejsel}, A. and
         {Galeotta}, S. and {Galli}, S. and {Ganga}, K. and {Gauthier}, C. and
         {Gerbino}, M. and {Ghosh}, T. and {Giard}, M. and
         {Giraud-H{\'e}raud}, Y. and {Giusarma}, E. and {Gjerl{\o}w}, E. and
         {Gonz{\'a}lez-Nuevo}, J. and {G{\'o}rski}, K.~M. and {Gratton}, S. and
         {Gregorio}, A. and {Gruppuso}, A. and {Gudmundsson}, J.~E. and
         {Hamann}, J. and {Hansen}, F.~K. and {Hanson}, D. and
         {Harrison}, D.~L. and {Helou}, G. and {Henrot-Versill{\'e}}, S. and
         {Hern{\'a}ndez-Monteagudo}, C. and {Herranz}, D. and {Hildebrand
        t}, S.~R. and {Hivon}, E. and {Hobson}, M. and {Holmes}, W.~A. and
         {Hornstrup}, A. and {Hovest}, W. and {Huang}, Z. and
         {Huffenberger}, K.~M. and {Hurier}, G. and {Jaffe}, A.~H. and
         {Jaffe}, T.~R. and {Jones}, W.~C. and {Juvela}, M. and
         {Keih{\"a}nen}, E. and {Keskitalo}, R. and {Kisner}, T.~S. and
         {Kneissl}, R. and {Knoche}, J. and {Knox}, L. and {Kunz}, M. and
         {Kurki-Suonio}, H. and {Lagache}, G. and {L{\"a}hteenm{\"a}ki}, A. and
         {Lamarre}, J. -M. and {Lasenby}, A. and {Lattanzi}, M. and
         {Lawrence}, C.~R. and {Leahy}, J.~P. and {Leonardi}, R. and
         {Lesgourgues}, J. and {Levrier}, F. and {Lewis}, A. and {Liguori}, M. and
         {Lilje}, P.~B. and {Linden-V{\o}rnle}, M. and {L{\'o}pez-Caniego}, M. and
         {Lubin}, P.~M. and {Mac{\'\i}as-P{\'e}rez}, J.~F. and {Maggio}, G. and
         {Maino}, D. and {Mandolesi}, N. and {Mangilli}, A. and {Marchini}, A. and
         {Maris}, M. and {Martin}, P.~G. and {Martinelli}, M. and
         {Mart{\'\i}nez-Gonz{\'a}lez}, E. and {Masi}, S. and {Matarrese}, S. and
         {McGehee}, P. and {Meinhold}, P.~R. and {Melchiorri}, A. and
         {Melin}, J. -B. and {Mendes}, L. and {Mennella}, A. and
         {Migliaccio}, M. and {Millea}, M. and {Mitra}, S. and
         {Miville-Desch{\^e}nes}, M. -A. and {Moneti}, A. and {Montier}, L. and
         {Morgante}, G. and {Mortlock}, D. and {Moss}, A. and {Munshi}, D. and
         {Murphy}, J.~A. and {Naselsky}, P. and {Nati}, F. and {Natoli}, P. and
         {Netterfield}, C.~B. and {N{\o}rgaard-Nielsen}, H.~U. and
         {Noviello}, F. and {Novikov}, D. and {Novikov}, I. and
         {Oxborrow}, C.~A. and {Paci}, F. and {Pagano}, L. and {Pajot}, F. and
         {Paladini}, R. and {Paoletti}, D. and {Partridge}, B. and {Pasian}, F. and
         {Patanchon}, G. and {Pearson}, T.~J. and {Perdereau}, O. and
         {Perotto}, L. and {Perrotta}, F. and {Pettorino}, V. and
         {Piacentini}, F. and {Piat}, M. and {Pierpaoli}, E. and
         {Pietrobon}, D. and {Plaszczynski}, S. and {Pointecouteau}, E. and
         {Polenta}, G. and {Popa}, L. and {Pratt}, G.~W. and {Pr{\'e}zeau}, G. and
         {Prunet}, S. and {Puget}, J. -L. and {Rachen}, J.~P. and
         {Reach}, W.~T. and {Rebolo}, R. and {Reinecke}, M. and
         {Remazeilles}, M. and {Renault}, C. and {Renzi}, A. and
         {Ristorcelli}, I. and {Rocha}, G. and {Rosset}, C. and {Rossetti}, M. and
         {Roudier}, G. and {Rouill{\'e} d'Orfeuil}, B. and {Rowan-Robinson}, M. and
         {Rubi{\~n}o-Mart{\'\i}n}, J.~A. and {Rusholme}, B. and {Said}, N. and
         {Salvatelli}, V. and {Salvati}, L. and {Sandri}, M. and {Santos}, D. and
         {Savelainen}, M. and {Savini}, G. and {Scott}, D. and
         {Seiffert}, M.~D. and {Serra}, P. and {Shellard}, E.~P.~S. and
         {Spencer}, L.~D. and {Spinelli}, M. and {Stolyarov}, V. and
         {Stompor}, R. and {Sudiwala}, R. and {Sunyaev}, R. and {Sutton}, D. and
         {Suur-Uski}, A. -S. and {Sygnet}, J. -F. and {Tauber}, J.~A. and
         {Terenzi}, L. and {Toffolatti}, L. and {Tomasi}, M. and {Tristram}, M. and
         {Trombetti}, T. and {Tucci}, M. and {Tuovinen}, J. and
         {T{\"u}rler}, M. and {Umana}, G. and {Valenziano}, L. and
         {Valiviita}, J. and {Van Tent}, F. and {Vielva}, P. and {Villa}, F. and
         {Wade}, L.~A. and {Wandelt}, B.~D. and {Wehus}, I.~K. and {White}, M. and
         {White}, S.~D.~M. and {Wilkinson}, A. and {Yvon}, D. and {Zacchei}, A. and
         {Zonca}, A.},
        title = "{Planck 2015 results. XIII. Cosmological parameters}",
      journal = {\aap},
         year = 2016,
        month = sep,
       volume = {594},
          eid = {A13},
        pages = {A13},
          doi = {10.1051/0004-6361/201525830},
archivePrefix = {arXiv},
       eprint = {1502.01589},
 primaryClass = {astro-ph.CO},
       adsurl = {https://ui.adsabs.harvard.edu/abs/2016A&A...594A..13P}
}

@ARTICLE{AREPO,
       author = {{Springel}, Volker},
        title = "{E pur si muove: Galilean-invariant cosmological hydrodynamical simulations on a moving mesh}",
      journal = {\mnras},
         year = 2010,
        month = jan,
       volume = {401},
       number = {2},
        pages = {791-851},
          doi = {10.1111/j.1365-2966.2009.15715.x},
archivePrefix = {arXiv},
       eprint = {0901.4107},
 primaryClass = {astro-ph.CO},
       adsurl = {https://ui.adsabs.harvard.edu/abs/2010MNRAS.401..791S}
}

@ARTICLE{TNGb,
       author = {{Springel}, Volker and {Pakmor}, R{\"u}diger and {Pillepich}, Annalisa and
         {Weinberger}, Rainer and {Nelson}, Dylan and {Hernquist}, Lars and
         {Vogelsberger}, Mark and {Genel}, Shy and {Torrey}, Paul and
         {Marinacci}, Federico and {Naiman}, Jill},
        title = "{First results from the IllustrisTNG simulations: matter and galaxy clustering}",
      journal = {\mnras},
         year = 2018,
        month = mar,
       volume = {475},
       number = {1},
        pages = {676-698},
          doi = {10.1093/mnras/stx3304},
archivePrefix = {arXiv},
       eprint = {1707.03397},
 primaryClass = {astro-ph.GA},
       adsurl = {https://ui.adsabs.harvard.edu/abs/2018MNRAS.475..676S}
}

@ARTICLE{TNGc,
       author = {{Marinacci}, Federico and {Vogelsberger}, Mark and
         {Pakmor}, R{\"u}diger and {Torrey}, Paul and {Springel}, Volker and
         {Hernquist}, Lars and {Nelson}, Dylan and {Weinberger}, Rainer and
         {Pillepich}, Annalisa and {Naiman}, Jill and {Genel}, Shy},
        title = "{First results from the IllustrisTNG simulations: radio haloes and magnetic fields}",
      journal = {\mnras},
         year = 2018,
        month = nov,
       volume = {480},
       number = {4},
        pages = {5113-5139},
          doi = {10.1093/mnras/sty2206},
archivePrefix = {arXiv},
       eprint = {1707.03396},
 primaryClass = {astro-ph.CO},
       adsurl = {https://ui.adsabs.harvard.edu/abs/2018MNRAS.480.5113M}
}

@ARTICLE{TNGd,
       author = {{Pillepich}, Annalisa and {Nelson}, Dylan and {Hernquist}, Lars and
         {Springel}, Volker and {Pakmor}, R{\"u}diger and {Torrey}, Paul and
         {Weinberger}, Rainer and {Genel}, Shy and {Naiman}, Jill P. and
         {Marinacci}, Federico and {Vogelsberger}, Mark},
        title = "{First results from the IllustrisTNG simulations: the stellar mass content of groups and clusters of galaxies}",
      journal = {\mnras},
         year = 2018,
        month = mar,
       volume = {475},
       number = {1},
        pages = {648-675},
          doi = {10.1093/mnras/stx3112},
archivePrefix = {arXiv},
       eprint = {1707.03406},
 primaryClass = {astro-ph.GA},
       adsurl = {https://ui.adsabs.harvard.edu/abs/2018MNRAS.475..648P}
}

@ARTICLE{TNGe,
       author = {{Naiman}, Jill P. and {Pillepich}, Annalisa and {Springel}, Volker and
         {Ramirez-Ruiz}, Enrico and {Torrey}, Paul and {Vogelsberger}, Mark and
         {Pakmor}, R{\"u}diger and {Nelson}, Dylan and {Marinacci}, Federico and
         {Hernquist}, Lars and {Weinberger}, Rainer and {Genel}, Shy},
        title = "{First results from the IllustrisTNG simulations: a tale of two elements - chemical evolution of magnesium and europium}",
      journal = {\mnras},
         year = 2018,
        month = jun,
       volume = {477},
       number = {1},
        pages = {1206-1224},
          doi = {10.1093/mnras/sty618},
archivePrefix = {arXiv},
       eprint = {1707.03401},
 primaryClass = {astro-ph.GA},
       adsurl = {https://ui.adsabs.harvard.edu/abs/2018MNRAS.477.1206N}
}

@ARTICLE{Arico:2020,
       author = {{Aric{\`o}}, Giovanni and {Angulo}, Raul E. and {Hern{\'a}ndez-Monteagudo}, Carlos and {Contreras}, Sergio and {Zennaro}, Matteo and {Pellejero-Iba{\~n}ez}, Marcos and {Rosas-Guevara}, Yetli},
        title = "{Modelling the large-scale mass density field of the universe as a function of cosmology and baryonic physics}",
      journal = {\mnras},
         year = 2020,
        month = jul,
       volume = {495},
       number = {4},
        pages = {4800-4819},
          doi = {10.1093/mnras/staa1478},
archivePrefix = {arXiv},
       eprint = {1911.08471},
 primaryClass = {astro-ph.CO},
       adsurl = {https://ui.adsabs.harvard.edu/abs/2020MNRAS.495.4800A}
}

@ARTICLE{Pellejero:2023,
       author = {{Pellejero Iba{\~n}ez}, Marcos and {Angulo}, Raul E. and {Zennaro}, Matteo and {St{\"u}cker}, Jens and {Contreras}, Sergio and {Aric{\`o}}, Giovanni and {Maion}, Francisco},
        title = "{The bacco simulation project: bacco hybrid Lagrangian bias expansion model in redshift space}",
      journal = {\mnras},
         year = 2023,
        month = apr,
       volume = {520},
       number = {3},
        pages = {3725-3741},
          doi = {10.1093/mnras/stad368},
archivePrefix = {arXiv},
       eprint = {2207.06437},
 primaryClass = {astro-ph.CO},
       adsurl = {https://ui.adsabs.harvard.edu/abs/2023MNRAS.520.3725P}
}

@ARTICLE{Ruiz:2011,
       author = {{Ruiz}, Andr{\'e}s. N. and {Padilla}, Nelson D. and
         {Dom{\'\i}nguez}, Mariano J. and {Cora}, Sof{\'\i}a. A.},
        title = "{How accurate is it to update the cosmology of your halo catalogues?}",
      journal = {\mnras},
         year = "2011",
        month = "Dec",
       volume = {418},
       number = {4},
        pages = {2422-2434},
          doi = {10.1111/j.1365-2966.2011.19635.x},
archivePrefix = {arXiv},
       eprint = {1103.5074},
 primaryClass = {astro-ph.CO},
       adsurl = {https://ui.adsabs.harvard.edu/abs/2011MNRAS.418.2422R}
}

@ARTICLE{Angulo:2016,
       author = {{Angulo}, Raul E. and {Pontzen}, Andrew},
        title = "{Cosmological N-body simulations with suppressed variance}",
      journal = {\mnras},
         year = "2016",
        month = "Oct",
       volume = {462},
       number = {1},
        pages = {L1-L5},
          doi = {10.1093/mnrasl/slw098},
archivePrefix = {arXiv},
       eprint = {1603.05253},
 primaryClass = {astro-ph.CO},
       adsurl = {https://ui.adsabs.harvard.edu/abs/2016MNRAS.462L...1A}
}

@ARTICLE{Zennaro:2019,
       author = {{Zennaro}, Matteo and {Angulo}, Ra{\'u}l E. and {Aric{\`o}}, Giovanni and {Contreras}, Sergio and {Pellejero-Ib{\'a}{\~n}ez}, Marcos},
        title = "{How to add massive neutrinos to your {\ensuremath{\Lambda}}CDM simulation - extending cosmology rescaling algorithms}",
      journal = {\mnras},
         year = 2019,
        month = nov,
       volume = {489},
       number = {4},
        pages = {5938-5951},
          doi = {10.1093/mnras/stz2612},
archivePrefix = {arXiv},
       eprint = {1905.08696},
 primaryClass = {astro-ph.CO},
       adsurl = {https://ui.adsabs.harvard.edu/abs/2019MNRAS.489.5938Z}
}

@ARTICLE{emcee,
       author = {{Foreman-Mackey}, Daniel and {Hogg}, David W. and {Lang}, Dustin and
         {Goodman}, Jonathan},
        title = "{emcee: The MCMC Hammer}",
      journal = {Publications of the Astronomical Society of the Pacific},
         year = "2013",
        month = "Mar",
       volume = {125},
       number = {925},
        pages = {306},
          doi = {10.1086/670067},
archivePrefix = {arXiv},
       eprint = {1202.3665},
 primaryClass = {astro-ph.IM},
       adsurl = {https://ui.adsabs.harvard.edu/abs/2013PASP..125..306F}
}

@ARTICLE{Angulo:2010,
   author = {{Angulo}, R.~E. and {White}, S.~D.~M.},
    title = "{One simulation to fit them all - changing the background parameters of a cosmological N-body simulation}",
  journal = {\mnras},
archivePrefix = "arXiv",
   eprint = {0912.4277},
     year = 2010,
    month = jun,
   volume = 405,
    pages = {143-154},
      doi = {10.1111/j.1365-2966.2010.16459.x},
   adsurl = {http://adsabs.harvard.edu/abs/2010MNRAS.405..143A}
}

@ARTICLE{Wechsler:2018,
       author = {{Wechsler}, Risa H. and {Tinker}, Jeremy L.},
        title = "{The Connection Between Galaxies and Their Dark Matter Halos}",
      journal = {\araa},
         year = 2018,
        month = sep,
       volume = {56},
        pages = {435-487},
          doi = {10.1146/annurev-astro-081817-051756},
archivePrefix = {arXiv},
       eprint = {1804.03097},
 primaryClass = {astro-ph.GA},
       adsurl = {https://ui.adsabs.harvard.edu/abs/2018ARA&A..56..435W}
}

@ARTICLE{Zennaro:2023,
       author = {{Zennaro}, Matteo and {Angulo}, Raul E. and {Pellejero-Ib{\'a}{\~n}ez}, Marcos and {St{\"u}cker}, Jens and {Contreras}, Sergio and {Aric{\`o}}, Giovanni},
        title = "{The BACCO simulation project: biased tracers in real space}",
      journal = {\mnras},
         year = 2023,
        month = sep,
       volume = {524},
       number = {2},
        pages = {2407-2419},
          doi = {10.1093/mnras/stad2008},
archivePrefix = {arXiv},
       eprint = {2101.12187},
 primaryClass = {astro-ph.CO},
       adsurl = {https://ui.adsabs.harvard.edu/abs/2023MNRAS.524.2407Z}
}

@ARTICLE{C23_HOD,
       author = {{Contreras}, Sergio and {Zehavi}, Idit},
        title = "{On the origin of the evolution of the halo occupation distribution}",
      journal = {\mnras},
         year = 2023,
        month = nov,
       volume = {525},
       number = {3},
        pages = {4257-4269},
          doi = {10.1093/mnras/stad2452},
archivePrefix = {arXiv},
       eprint = {2305.19628},
 primaryClass = {astro-ph.CO},
       adsurl = {https://ui.adsabs.harvard.edu/abs/2023MNRAS.525.4257C}
}

@ARTICLE{Tinker:2012,
   author = {{Tinker}, J.~L. and {Sheldon}, E.~S. and {Wechsler}, R.~H. and 
	{Becker}, M.~R. and {Rozo}, E. and {Zu}, Y. and {Weinberg}, D.~H. and 
	{Zehavi}, I. and {Blanton}, M.~R. and {Busha}, M.~T. and {Koester}, B.~P.
	},
    title = "{Cosmological Constraints from Galaxy Clustering and the Mass-to-number Ratio of Galaxy Clusters}",
  journal = {\apj},
archivePrefix = "arXiv",
   eprint = {1104.1635},
     year = 2012,
    month = jan,
   volume = 745,
      eid = {16},
    pages = {16},
      doi = {10.1088/0004-637X/745/1/16},
   adsurl = {http://adsabs.harvard.edu/abs/2012ApJ...745...16T}
}

@ARTICLE{Jing:1998a,
   author = {{Jing}, Y.~P. and {Mo}, H.~J. and {B{\"o}rner}, G.},
    title = "{Spatial Correlation Function and Pairwise Velocity Dispersion of Galaxies: Cold Dark Matter Models versus the Las Campanas Survey}",
  journal = {\apj},
   eprint = {astro-ph/9707106},
     year = 1998,
    month = feb,
   volume = 494,
    pages = {1-12},
      doi = {10.1086/305209},
   adsurl = {http://adsabs.harvard.edu/abs/1998ApJ...494....1J}
}

@ARTICLE{ChavesMontero:2016,
       author = {{Chaves-Montero}, Jon{\'a}s and {Angulo}, Raul E. and {Schaye}, Joop and {Schaller}, Matthieu and {Crain}, Robert A. and {Furlong}, Michelle and {Theuns}, Tom},
        title = "{Subhalo abundance matching and assembly bias in the EAGLE simulation}",
      journal = {\mnras},
         year = 2016,
        month = aug,
       volume = {460},
       number = {3},
        pages = {3100-3118},
          doi = {10.1093/mnras/stw1225},
archivePrefix = {arXiv},
       eprint = {1507.01948},
 primaryClass = {astro-ph.GA},
       adsurl = {https://ui.adsabs.harvard.edu/abs/2016MNRAS.460.3100C}
}

@ARTICLE{Paviot:2024,
       author = {{Paviot}, R. and {Rocher}, A. and {Codis}, S. and {de Mattia}, A. and {Jullo}, E. and {de la Torre}, S.},
        title = "{Impact of assembly bias on clustering plus weak lensing cosmological analysis}",
      journal = {\aap},
         year = 2024,
        month = oct,
       volume = {690},
          eid = {A221},
        pages = {A221},
          doi = {10.1051/0004-6361/202449574},
archivePrefix = {arXiv},
       eprint = {2402.07715},
 primaryClass = {astro-ph.CO},
       adsurl = {https://ui.adsabs.harvard.edu/abs/2024A&A...690A.221P}
}

@ARTICLE{SOM:2025DESI,
       author = {{Ortega-Martinez}, Sara and {Contreras}, Sergio and {Angulo}, Raul E. and {Chaves-Montero}, Jon{\'a}s},
        title = "{Investigating the galaxy{\textendash}halo connection of DESI emission-line galaxies with SHAMe-SF}",
      journal = {\aap},
         year = 2025,
        month = may,
       volume = {697},
          eid = {A226},
        pages = {A226},
          doi = {10.1051/0004-6361/202453086},
archivePrefix = {arXiv},
       eprint = {2411.11830},
 primaryClass = {astro-ph.CO},
       adsurl = {https://ui.adsabs.harvard.edu/abs/2025A&A...697A.226O}
}

@ARTICLE{Granett:2019,
       author = {{Granett}, Benjamin R. and {Favole}, Ginevra and {Montero-Dorta}, Antonio D. and {Branchini}, Enzo and {Guzzo}, Luigi and {de la Torre}, Sylvain},
        title = "{Measuring the growth of structure by matching dark matter haloes to galaxies with VIPERS and SDSS}",
      journal = {\mnras},
         year = 2019,
        month = oct,
       volume = {489},
       number = {1},
        pages = {653-662},
          doi = {10.1093/mnras/stz2152},
archivePrefix = {arXiv},
       eprint = {1905.10375},
 primaryClass = {astro-ph.CO},
       adsurl = {https://ui.adsabs.harvard.edu/abs/2019MNRAS.489..653G}
}

@ARTICLE{Peacock:2000,
   author = {{Peacock}, J.~A. and {Smith}, R.~E.},
    title = "{Halo occupation numbers and galaxy bias}",
  journal = {\mnras},
   eprint = {astro-ph/0005010},
     year = 2000,
    month = nov,
   volume = 318,
    pages = {1144-1156},
      doi = {10.1046/j.1365-8711.2000.03779.x},
   adsurl = {http://adsabs.harvard.edu/abs/2000MNRAS.318.1144P}
}

@ARTICLE{Favole:2025,
       author = {{Favole}, Ginevra and {Kitaura}, Francisco-Shu and {Hadzhiyska}, Boryana and {Eisenstein}, Daniel J. and {Garrison}, Lehman H. and {Bose}, Sownak},
        title = "{ELG$\times$LRG distribution through dark matter halo dynamics}",
      journal = {arXiv e-prints},
         year = 2025,
        month = dec,
          eid = {arXiv:2512.04362},
        pages = {arXiv:2512.04362},
          doi = {10.48550/arXiv.2512.04362},
archivePrefix = {arXiv},
       eprint = {2512.04362},
 primaryClass = {astro-ph.GA},
       adsurl = {https://ui.adsabs.harvard.edu/abs/2025arXiv251204362F}
}

@ARTICLE{Guo:2015a,
   author = {{Guo}, H. and {Zheng}, Z. and {Zehavi}, I. and {Dawson}, K. and 
	{Skibba}, R.~A. and {Tinker}, J.~L. and {Weinberg}, D.~H. and 
	{White}, M. and {Schneider}, D.~P.},
    title = "{Velocity bias from the small-scale clustering of SDSS-III BOSS galaxies}",
  journal = {\mnras},
archivePrefix = "arXiv",
   eprint = {1407.4811},
     year = 2015,
    month = jan,
   volume = 446,
    pages = {578-594},
      doi = {10.1093/mnras/stu2120},
   adsurl = {http://adsabs.harvard.edu/abs/2015MNRAS.446..578G}
}

@ARTICLE{Zehavi:2011,
   author = {{Zehavi}, I. and {Zheng}, Z. and {Weinberg}, D.~H. and {Blanton}, M.~R. and 
	{Bahcall}, N.~A. and {Berlind}, A.~A. and {Brinkmann}, J. and 
	{Frieman}, J.~A. and {Gunn}, J.~E. and {Lupton}, R.~H. and {Nichol}, R.~C. and 
	{Percival}, W.~J. and {Schneider}, D.~P. and {Skibba}, R.~A. and 
	{Strauss}, M.~A. and {Tegmark}, M. and {York}, D.~G.},
    title = "{Galaxy Clustering in the Completed SDSS Redshift Survey: The Dependence on Color and Luminosity}",
  journal = {\apj},
archivePrefix = "arXiv",
   eprint = {1005.2413},
     year = 2011,
    month = jul,
   volume = 736,
      eid = {59},
    pages = {59},
      doi = {10.1088/0004-637X/736/1/59},
   adsurl = {http://adsabs.harvard.edu/abs/2011ApJ...736...59Z}
}

@ARTICLE{Zheng:2007,
   author = {{Zheng}, Z. and {Coil}, A.~L. and {Zehavi}, I.},
    title = "{Galaxy Evolution from Halo Occupation Distribution Modeling of DEEP2 and SDSS Galaxy Clustering}",
  journal = {\apj},
   eprint = {astro-ph/0703457},
     year = 2007,
    month = oct,
   volume = 667,
    pages = {760-779},
      doi = {10.1086/521074},
   adsurl = {http://adsabs.harvard.edu/abs/2007ApJ...667..760Z}
}

@ARTICLE{Benson:2000,
   author = {{Benson}, A.~J. and {Cole}, S. and {Frenk}, C.~S. and {Baugh}, C.~M. and 
	{Lacey}, C.~G.},
    title = "{The nature of galaxy bias and clustering}",
  journal = {\mnras},
   eprint = {astro-ph/9903343},
     year = 2000,
    month = feb,
   volume = 311,
    pages = {793-808},
      doi = {10.1046/j.1365-8711.2000.03101.x},
   adsurl = {http://adsabs.harvard.edu/abs/2000MNRAS.311..793B}
}

@string{ aj = "AJ" }

@ARTICLE{Guo:2013a,
   author = {{Guo}, Q. and {White}, S. and {Angulo}, R.~E. and {Henriques}, B. and 
	{Lemson}, G. and {Boylan-Kolchin}, M. and {Thomas}, P. and {Short}, C.
	},
    title = "{Galaxy formation in WMAP1 and WMAP7 cosmologies}",
  journal = {\mnras},
archivePrefix = "arXiv",
   eprint = {1206.0052},
     year = 2013,
    month = jan,
   volume = 428,
    pages = {1351-1365},
      doi = {10.1093/mnras/sts115},
   adsurl = {http://adsabs.harvard.edu/abs/2013MNRAS.428.1351G}
}

@ARTICLE{Trujillo-Gomez:2011,
   author = {{Trujillo-Gomez}, Sebastian and {Klypin}, Anatoly and {Primack}, Joel and {Romanowsky}, Aaron J.},
    title = "{Galaxies in ??CDM with Halo Abundance Matching: Luminosity-Velocity Relation, Baryonic Mass-Velocity Relation, Velocity Function, and Clustering}",
  journal = {\apj},
     year = 2011,
    month = nov,
   volume = 742,
    pages = {16, 24 pp},
   adsurl = {http://adsabs.harvard.edu/abs/2011ApJ...742...16T}
}

@ARTICLE{Davis:1985,
   author = {{Davis}, M. and {Efstathiou}, G. and {Frenk}, C. S. and {White}, S. D. M.},
    title = "{The evolution of large-scale structure in a universe dominated by cold dark matter}",
  journal = {\apj},
     year = 1985,
    month = may,
   volume = 292,
    pages = { 371-394},
   adsurl = {http://adsabs.harvard.edu/abs/1985ApJ...292..371D}
}

@ARTICLE{Vale:2006,
   author = {{Vale}, A. and {Ostriker}, J. P.},
    title = "{The non-parametric model for linking galaxy luminosity with halo/subhalo mass}",
  journal = {\mnras},
     year = 2006,
    month = sep,
   volume = 371,
    pages = {1173-1187},
   adsurl = {http://adsabs.harvard.edu/abs/2006MNRAS.371.1173V}
}

@ARTICLE{Shankar:2006,
   author = {{Shankar}, F. and {Lapi}, A. and {Salucci}, P. and {De Zotti}, G. and {Danese}, L.},
    title = "{Modeling Luminosity-dependent Galaxy Clustering through Cosmic Time}",
  journal = {\apj},
     year = 2006,
    month = may,
   volume = 643,
    pages = {14-25},
      doi = {10.1086/502794},
   adsurl = {http://adsabs.harvard.edu/abs/2006ApJ...643...14S}
}

@ARTICLE{Conroy:2006,
   author = {{Conroy}, Charlie and {Wechsler}, Risa H. and {Kravtsov}, Andrey V.},
    title = "{Modeling Luminosity-dependent Galaxy Clustering through Cosmic Time}",
  journal = {\apj},
     year = 2006,
    month = aug,
   volume = 647,
    pages = {201-214},
      doi = {10.1086/503602},
   adsurl = {http://adsabs.harvard.edu/abs/2006ApJ...647..201C}
}

@ARTICLE{Zheng:2005,
   author = {{Zheng}, Z. and {Berlind}, A.~A. and {Weinberg}, D.~H. and {Benson}, A.~J. and 
	{Baugh}, C.~M. and {Cole}, S. and {Dav{\'e}}, R. and {Frenk}, C.~S. and 
	{Katz}, N. and {Lacey}, C.~G.},
    title = "{Theoretical Models of the Halo Occupation Distribution: Separating Central and Satellite Galaxies}",
  journal = {\apj},
   eprint = {arXiv:astro-ph/0408564},
     year = 2005,
    month = nov,
   volume = 633,
    pages = {791-809},
      doi = {10.1086/466510},
   adsurl = {http://adsabs.harvard.edu/abs/2005ApJ...633..791Z}
}

@ARTICLE{Springel:2001,
   author = {{Springel}, Volker and {White}, Simon D. M. and {Tormen}, Giuseppe and {Kauffmann}, Guinevere},
    title = "{Dynamical friction and galaxy merging time-scales}",
  journal = {\mnras},
   eprint = {arXiv:astro-ph/0012055},
     year = 2001,
    month = dec,
   volume = 328,
    pages = {726-750},
      doi = {10.1046/j.1365-8711.2001.04912.x},
   adsurl = {http://adsabs.harvard.edu/abs/2001MNRAS.328..726S}
}

@ARTICLE{Hearin:2013b,
       author = {{Hearin}, Andrew P. and {Watson}, Douglas F.},
        title = "{The dark side of galaxy colour}",
      journal = {\mnras},
         year = 2013,
        month = oct,
       volume = {435},
       number = {2},
        pages = {1313-1324},
          doi = {10.1093/mnras/stt1374},
archivePrefix = {arXiv},
       eprint = {1304.5557},
 primaryClass = {astro-ph.CO},
       adsurl = {https://ui.adsabs.harvard.edu/abs/2013MNRAS.435.1313H}
}

@ARTICLE{Norberg:2009,
   author = {{Norberg}, P. and {Baugh}, C.~M. and {Gazta{n}aga}, E. and 
	{Croton}, D.~J.},
    title = "{Statistical analysis of galaxy surveys - I. Robust error estimation for two-point clustering statistics}",
  journal = {\mnras},
archivePrefix = "arXiv",
   eprint = {0810.1885},
     year = 2009,
    month = jun,
   volume = 396,
    pages = {19-38},
      doi = {10.1111/j.1365-2966.2009.14389.x},
   adsurl = {http://adsabs.harvard.edu/abs/2009MNRAS.396...19N}
}

@ARTICLE{Abazajian:2009,
   author = {{Abazajian}, K.~N. and {Adelman-McCarthy}, J.~K. and {Ag{\"u}eros}, M.~A. and 
	{Allam}, S.~S. and {Allende Prieto}, C. and {An}, D. and {Anderson}, K.~S.~J. and 
	{Anderson}, S.~F. and {Annis}, J. and {Bahcall}, N.~A. and et al.},
    title = "{The Seventh Data Release of the Sloan Digital Sky Survey}",
  journal = apjs,
     year = 2009,
    month = jun,
   volume = 182,
    pages = {543-558},
      doi = {10.1088/0067-0049/182/2/543},
   adsurl = {http://adsabs.harvard.edu/abs/2009ApJS..182..543A}
}

@ARTICLE{Berlind:2003,
   author = {{Berlind}, A.~A. and {Weinberg}, D.~H. and {Benson}, A.~J. and 
	{Baugh}, C.~M. and {Cole}, S. and {Dav{\'e}}, R. and {Frenk}, C.~S. and 
	{Jenkins}, A. and {Katz}, N. and {Lacey}, C.~G.},
    title = "{The Halo Occupation Distribution and the Physics of Galaxy Formation}",
  journal = {\apj},
   eprint = {arXiv:astro-ph/0212357},
     year = 2003,
    month = aug,
   volume = 593,
    pages = {1-25},
      doi = {10.1086/376517},
   adsurl = {http://esoads.eso.org/abs/2003ApJ...593....1B}
}

@ARTICLE{Prada:2023,
       author = {{Prada}, F. and {Ereza}, J. and {Smith}, A. and {Lasker}, J. and {Vaisakh}, R. and {Kehoe}, R. and {Dong-P{\'a}ez}, C.~A. and {Siudek}, M. and {Wang}, M.~S. and {Alam}, S. and {Beutler}, F. and {Bianchi}, D. and {Cole}, S. and {Dey}, B. and {Kirkby}, D. and {Norberg}, P. and {Aguilar}, J. and {Ahlen}, S. and {Brooks}, D. and {Claybaugh}, T. and {Dawson}, K. and {de la Macorra}, A. and {Fanning}, K. and {Forero-Romero}, J.~E. and {Gontcho}, S. Gontcho A and {Hahn}, C. and {Honscheid}, K. and {Ishak}, M. and {Kisner}, T. and {Landriau}, M. and {Manera}, M. and {Meisner}, A. and {Miquel}, R. and {Moustakas}, J. and {Mueller}, E. and {Nie}, J. and {Percival}, W.~J. and {Poppett}, C. and {Rezaie}, M. and {Rossi}, G. and {Sanchez}, E. and {Schubnell}, M. and {Tarl{\'e}}, G. and {Vargas-Maga{\~n}a}, M. and {Weaver}, B.~A. and {Yuan}, S. and {Zhou}, Z.},
        title = "{The DESI One-Percent Survey: Modelling the clustering and halo occupation of all four DESI tracers with Uchuu}",
      journal = {arXiv e-prints},
         year = 2023,
        month = jun,
          eid = {arXiv:2306.06315},
        pages = {arXiv:2306.06315},
          doi = {10.48550/arXiv.2306.06315},
archivePrefix = {arXiv},
       eprint = {2306.06315},
 primaryClass = {astro-ph.CO},
       adsurl = {https://ui.adsabs.harvard.edu/abs/2023arXiv230606315P}
}

@ARTICLE{Hagen:2025,
       author = {{Hagen}, T. and {Dawson}, K.~S. and {Zheng}, Z. and {Aguilar}, J. and {Ahlen}, S. and {BenZvi}, S. and {Bianchi}, D. and {Brooks}, D. and {Castander}, F.~J. and {Claybaugh}, T. and {Cuceu}, A. and {de la Macorra}, A. and {Doel}, P. and {Ferraro}, S. and {Font-Ribera}, A. and {Forero-Romero}, J.~E. and {Gaztanaga}, E. and {Gontcho}, S. Gontcho A and {Gonzalez-Perez}, V. and {Gutierrez}, G. and {Hahn}, C. and {Honscheid}, K. and {Ishak}, M. and {Juneau}, S. and {Kehoe}, R. and {Kisner}, T. and {Kremin}, A. and {Lamman}, C. and {Landriau}, M. and {Le Guillou}, L. and {Leauthaud}, A. and {Levi}, M.~E. and {Manera}, M. and {Meisner}, A. and {Miquel}, R. and {Moustakas}, J. and {Nadathur}, S. and {Palanque-Delabrouille}, N. and {Prada}, F. and {Perez-Rafols}, I. and {Ross}, A.~J. and {Rossi}, G. and {Saito}, S. and {Sanchez}, E. and {Schlegel}, D. and {Schubnell}, M. and {Silber}, J. and {Sprayberry}, D. and {Tarle}, G. and {Weaver}, B.~A. and {Zhou}, R. and {Zou}, H.},
        title = "{DESI Emission Line Galaxies: Clustering Dependence on Stellar Mass and [OII] Luminosity}",
      journal = {arXiv e-prints},
         year = 2025,
        month = may,
          eid = {arXiv:2505.20430},
        pages = {arXiv:2505.20430},
          doi = {10.48550/arXiv.2505.20430},
archivePrefix = {arXiv},
       eprint = {2505.20430},
 primaryClass = {astro-ph.GA},
       adsurl = {https://ui.adsabs.harvard.edu/abs/2025arXiv250520430H}
}

@ARTICLE{Rocher2023:DESI,
       author = {{Rocher}, Antoine and {Ruhlmann-Kleider}, Vanina and {Burtin}, Etienne and {Yuan}, Sihan and {de Mattia}, Arnaud and {Ross}, Ashley J. and {Aguilar}, Jessica and {Ahlen}, Steven and {Alam}, Shadab and {Bianchi}, Davide and {Brooks}, David and {Cole}, Shaun and {Dawson}, Kyle and {de la Macorra}, Axel and {Doel}, Peter and {Eisenstein}, Daniel J. and {Fanning}, Kevin and {Forero-Romero}, Jaime E. and {Garrison}, Lehman H. and {Gontcho A Gontcho}, Satya and {Gonzalez-Perez}, Violeta and {Guy}, Julien and {Hadzhiyska}, Boryana and {Hahn}, ChangHoon and {Honscheid}, Klaus and {Kisner}, Theodore and {Landriau}, Martin and {Lasker}, James and {E. Levi}, Michael and {Manera}, Marc and {Meisner}, Aaron and {Miquel}, Ramon and {Moustakas}, John and {Mueller}, Eva-Maria and {Newman}, Jeffrey A. and {Nie}, Jundan and {Percival}, Will J. and {Poppett}, Claire and {Qin}, Fei and {Rossi}, Graziano and {Samushia}, Lado and {Sanchez}, Eusebio and {Schlegel}, David and {Schubnell}, Michael and {Seo}, Hee-Jong and {Tarl{\'e}}, Gregory and {Vargas-Maga{\~n}a}, Mariana and {Weaver}, Benjamin A. and {Yu}, Jiaxi and {Zhang}, Hanyu and {Zheng}, Zheng and {Zhou}, Zhimin and {Zou}, Hu},
        title = "{The DESI One-Percent survey: exploring the Halo Occupation Distribution of Emission Line Galaxies with ABACUSSUMMIT simulations}",
      journal = {\jcap},
         year = 2023,
        month = oct,
       volume = {2023},
       number = {10},
          eid = {016},
        pages = {016},
          doi = {10.1088/1475-7516/2023/10/016},
archivePrefix = {arXiv},
       eprint = {2306.06319},
 primaryClass = {astro-ph.CO},
       adsurl = {https://ui.adsabs.harvard.edu/abs/2023JCAP...10..016R}
}

@ARTICLE{Yuan:2024_knn,
       author = {{Yuan}, Sihan and {Abel}, Tom and {Wechsler}, Risa H.},
        title = "{Robust cosmological inference from non-linear scales with k-th nearest neighbour statistics}",
      journal = {\mnras},
         year = 2024,
        month = jan,
       volume = {527},
       number = {2},
        pages = {1993-2009},
          doi = {10.1093/mnras/stad3359},
archivePrefix = {arXiv},
       eprint = {2310.04501},
 primaryClass = {astro-ph.CO},
       adsurl = {https://ui.adsabs.harvard.edu/abs/2024MNRAS.527.1993Y}
}

@ARTICLE{SOM2024,
       author = {{Ortega-Martinez}, S. and {Contreras}, S. and {Angulo}, R.},
        title = "{SHAMe-SF: Predicting the clustering of star-forming galaxies with an enhanced abundance matching model}",
      journal = {\aap},
         year = 2024,
        month = sep,
       volume = {689},
          eid = {A66},
        pages = {A66},
          doi = {10.1051/0004-6361/202449597},
archivePrefix = {arXiv},
       eprint = {2401.17374},
 primaryClass = {astro-ph.CO},
       adsurl = {https://ui.adsabs.harvard.edu/abs/2024A&A...689A..66O}
}

\begin{appendix} %First appendix
\onecolumn 

\section{Inference pipeline testing}
\label{app:emulator}
In this appendix, we begin by evaluating the emulator's performance, and analyse the effects of the emulator and the scaling algorithm on the sampling of the posteriors (Sect.~\ref{sec:NUTS}) and the PSO algorithm (Sect.~\ref{sec:PSO}). As mentioned in Section~\ref{sec:emulator}, we use two test samples (not seen by the neural networks during training): one with fixed Planck cosmology and another that also varies the cosmological parameters. 

\begin{figure*}[h!]
                \centering
                            \includegraphics[width=0.95\textwidth]{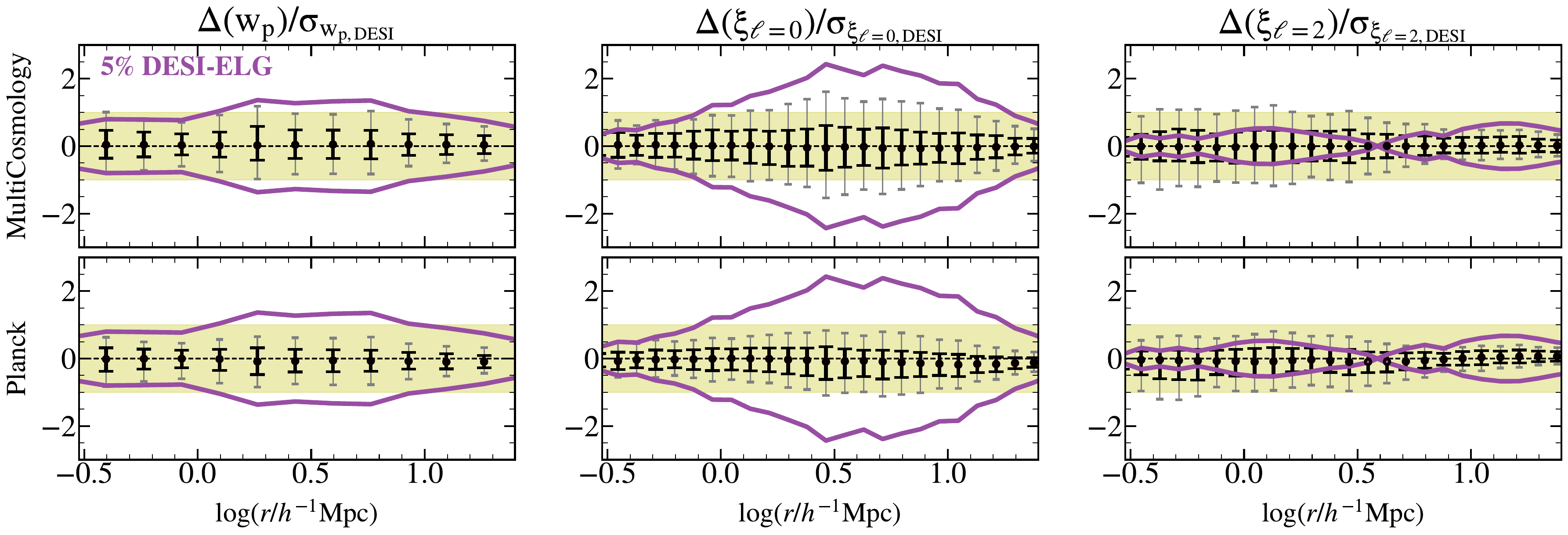}

        \caption{Difference between the test sample and the emulator evaluation for the same parameters normalised by the error of the DESI-ELG sample for all the range of cosmologies (upper panel) and fixed Planck cosmology (lower panel). We show the values of the projected correlation function (left), monopole (centre) and quadrupole (right) of the correlation function. The black and gray errorbars show the $1\sigma$ and $2\sigma$ differences, respectively. For comparison, we add the 5\% of the signal on each statistic normalised by the error of the DESI-ELG sample (purple). We add the yellow region and the dashed black line to guide the eye for deviations of 1$\sigma_{\rm DESI}$ and no deviation, respectively.}
        \label{fig:emulatorerror}
\end{figure*}

In Figure~\ref{fig:emulatorerror}, we compare the emulator predictions for a set of SHAMe-SF parameters and cosmological parameters (upper panel) and fixed Planck cosmology (lower panel) to the real mocks for those parameters, normalised by the error of the DESI-ELG sample. We scale the errors by the ratio between the SHAMe-SF evaluation and DESI-ELG clustering to account for differences in amplitude. Emulator errors are subdominant within 1$\sigma$, and either subdominant or comparable within 2$\sigma$, and we do not observe significant changes in behaviour between fixed Planck cosmology and considering all cosmologies.

\subsection{Posterior estimation}
\label{app:posterior}
After running the fiducial analysis on the MTNG-DESI, MTNG-H$\alpha$ and DESI-ELGs samples, we test the best fits for free cosmological parameters and Planck cosmology. The aim is two-fold: analyse potential changes on the posteriors introduced by the emulator or the scaling approximations, and to validate the inferences obtained for each sample.

\begin{figure*}[h!]
                \centering
                \includegraphics[width=0.95\textwidth]{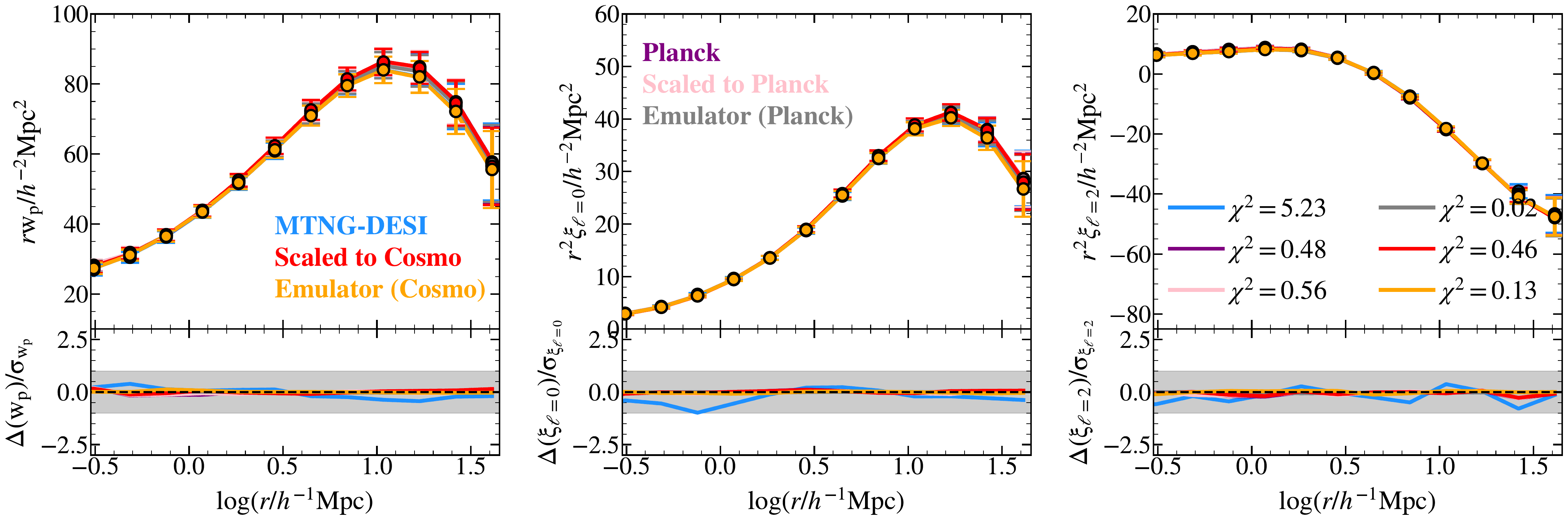}

        \caption{Projected correlation function (left), monopole (centre) and quadrupole (right) of the projected correlation function  measured for the four test samples built from the best fit to MTNG-DESI (errorbars). The fits with the SHAMe-SF model to each test sample are shown with solid lines in the lower panels. }
        \label{fig:TestMTNGclustering}
\end{figure*}

\twocolumn

\begin{figure}
                \centering
                \includegraphics[width=0.45\textwidth]{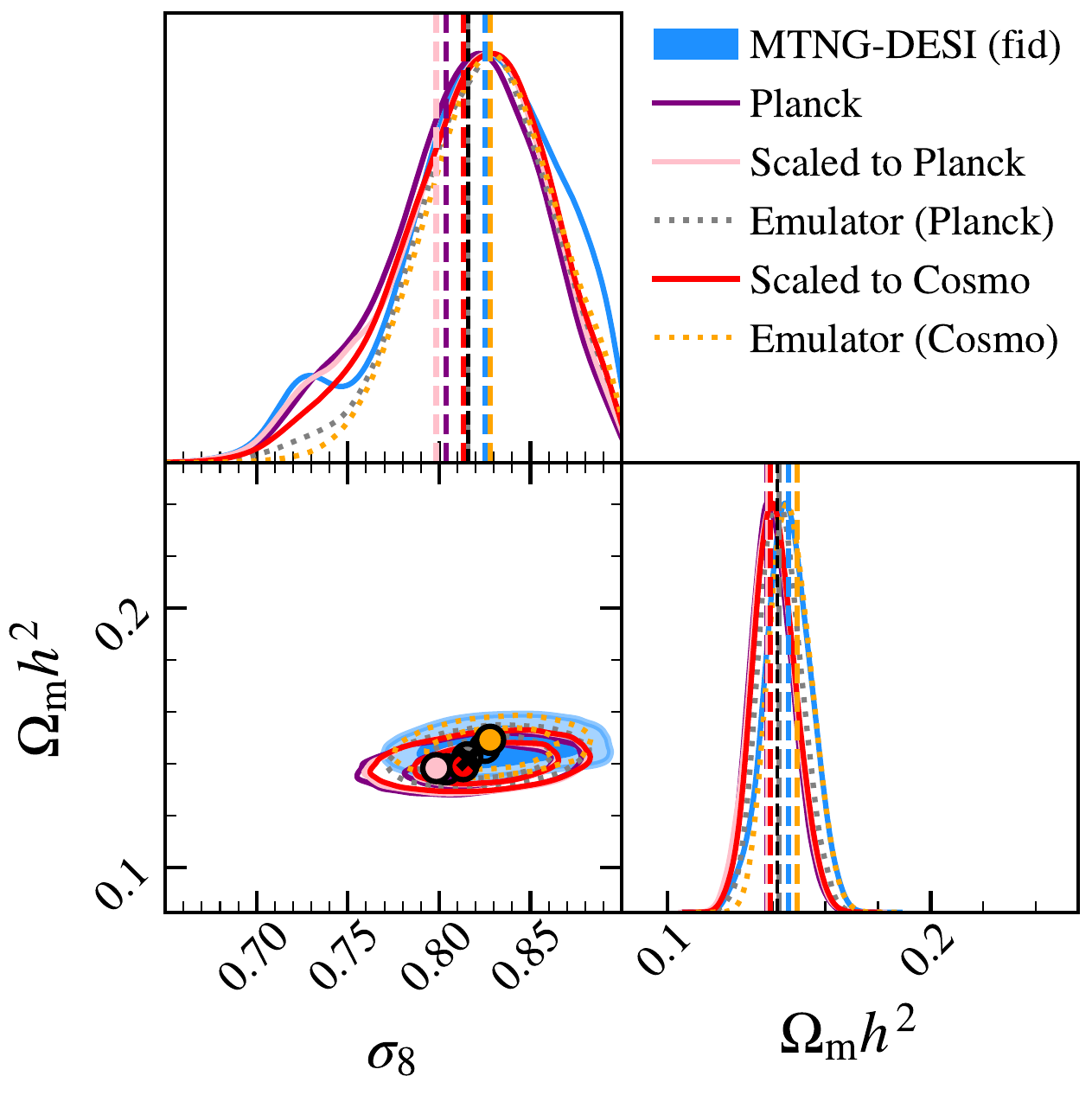}

        \caption{Projected constraints on $\sig$ and $\OmMh$,  for the test samples built from MTNG-DESI (lines) and MTNG-DESI (filled contour, blue). The black cross and black dashed lines on the histograms mark the true cosmology of the MTNG simulation. The best-fit of each sample is marked using coloured circles for the 2D distributions and dashed lines on the histograms. }
        \label{fig:posteriorerror}
\end{figure}

For each of the validation samples, we build five auxiliary test samples:

\begin{itemize}
\item Planck emulator: for the best fit with Planck cosmology, we evaluate the emulator using the best-fit SHAMe-SF parameters. This is the best-case scenario, where we have only uncertainties from the covariance matrix, allowing us to evaluate the sampling algorithm's accuracy by fitting an instance of the model.

\item Planck Scaled: for the best fit with Planck cosmology, we populate TheOne simulation scaled to Planck cosmology. Along with the sampling accuracy, this test also includes the effect of the emulator error.

\item Planck: for the best fit with Planck cosmology, we populate a simulation run with the same cosmology and use it to quantify deviations introduced by the emulator error and the scaling algorithm.

\item Cosmology emulator: Equivalent to Planck emulator, but using the best fit for the run with free cosmological parameters. 

\item Cosmology Scaled: Equivalent to Planck Scaled, but scaling the closer simulation to the best fit with free cosmological parameters.

\end{itemize}

We fit all the samples using the inference pipeline with the fiducial setup (priors on $\Omb h^2$, $\ns$ and $\Mnu$, $r_{\rm min}=0.3~\ihMpc$), and show the clustering measurements for MTNG-DESI and the five test samples in Figure~\ref{fig:TestMTNGclustering}. Note that the points are almost indistinguishable, and so the sample passes the emulator+scaling test, and we are not introducing biases in the clustering measurements. We show the posteriors for each sample for $\sig$ and $\OmMh$ in Figure~\ref{fig:posteriorerror}. The shape and location of the distributions are almost identical across all samples. We note that the small ``bump” for lower values of $\sig$ only appears in the samples generated by populating simulations with the SHAMe-SF model, but not in the samples that are instances of the emulator. MTNG-H$\alpha$ and DESI-ELG (low-$z$) also pass the test, but it is not the case for DESI-high$z$. We discuss this sample and the tests further in Appendix~\ref{app:DESIhigh}.

\subsection{Best fit estimation}
\label{app:bestfit}
\begin{figure}
                \centering
                \includegraphics[width=0.45\textwidth]{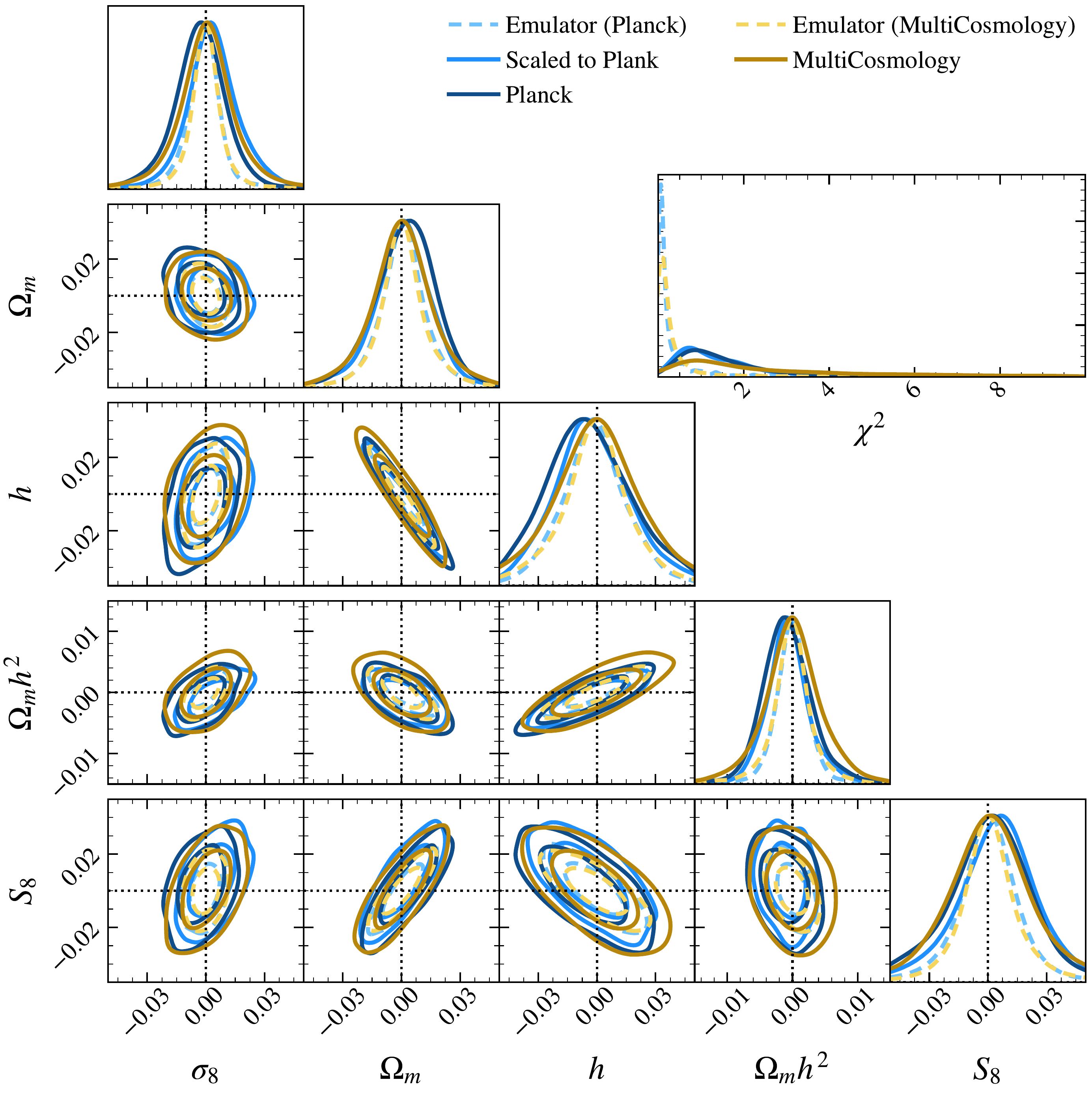}

        \caption{Distribution of the differences between the true value of the cosmological parameters and the best-fit using PSO for the five test samples defined in Section~\ref{app:bestfit}. We use dotted lines for the samples built using emulator instances, shades of blue for Planck cosmology, and yellow-ish colours for the two samples that also vary cosmology. The inset panel shows the histogram of $\chi^2$ values for each sample.}
        \label{fig:PSOerror}
\end{figure}

\begin{figure}
                \centering
                \includegraphics[width=0.45\textwidth]{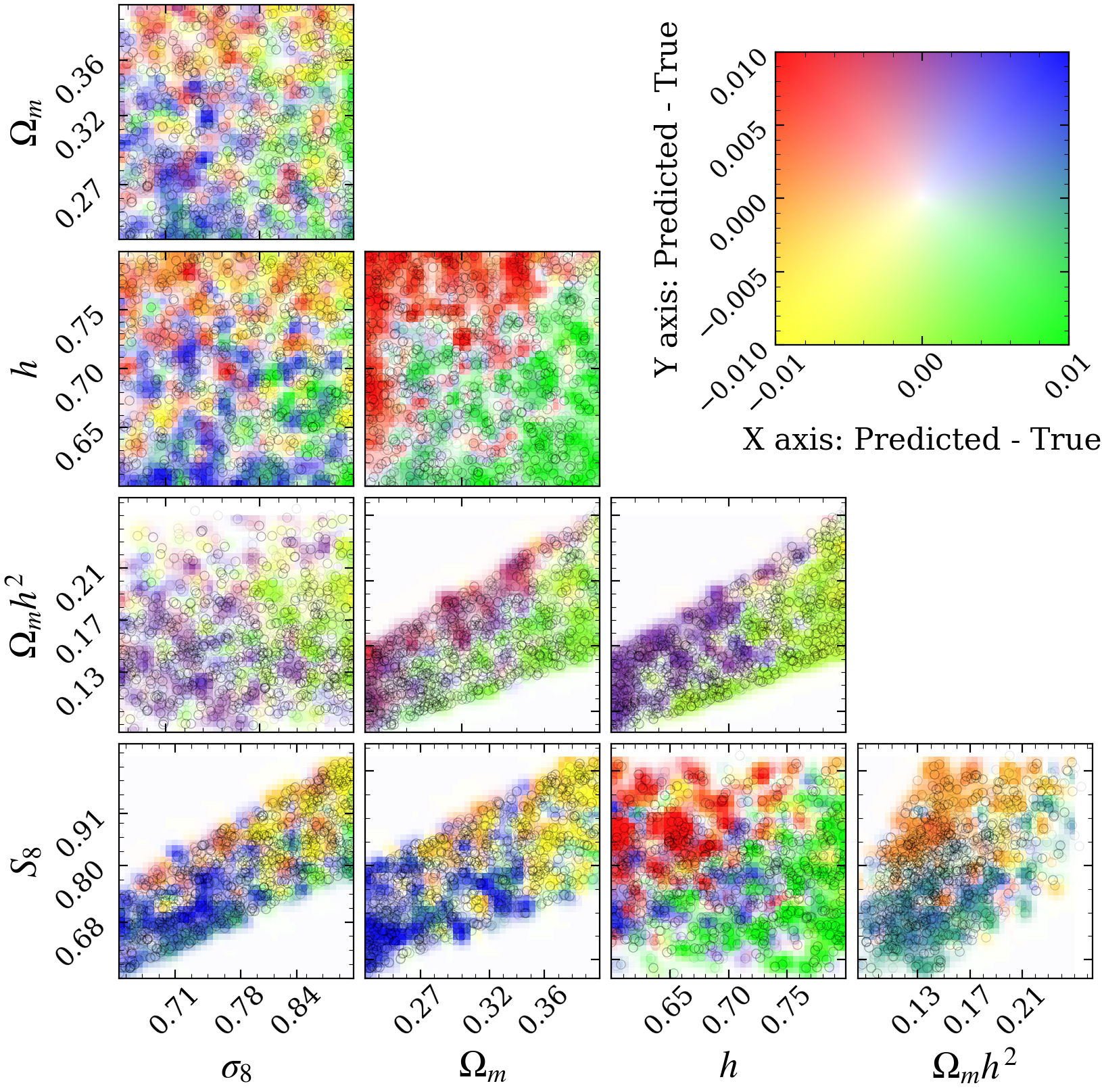}

        \caption{Distribution of the cosmological parameters of the test sample, coloured by the variation for each point between the true value and the PSO estimate for the best-fit. We show the locations of the true parameter values in the test using grey circles. }
        \label{fig:CosmoparamPSOerror}
\end{figure}

It is also necessary to analyse the accuracy of the PSO-based best-fit estimation, also considering the effects of the emulator and the scaling.  This provides further information on parameter degeneracies. Again, we use five test samples:

\begin{itemize}
\item Planck Scaled: SHAMe-SF parameters populating The One scaled to Planck Cosmology (test sample used to compute the contribution of the scaling to the covariance matrix).

\item Planck: Same as Planck Scaled, but for a simulation run with Planck cosmology (also used to compute the contribution of the scaling to the covariance matrix), and in the lower panel of Figure~\ref{fig:emulatorerror}.

\item Planck emulator: emulator instance with fixed Planck cosmology using the same SHAMe-SF parameters as the two previous samples.

\item Cosmology: test sample from the upper panel of Figure~\ref{fig:emulatorerror}, composed of combinations of SHAMe-SF parameters and cosmological parameters (same building process as the points to train the emulator, see Section~\ref{sec:emulator}).

\item Cosmology emulator: emulator instance for the same parameters as the Cosmology sample.

\end{itemize}

Following the procedure described in Section~\ref{sec:PSO}, and using the same priors from Section~\ref{sec:priors} (centring in the appropriate values for the sample with cosmologies different from Planck), we obtain the best-fit parameters for the points of each of the five samples using the covariance matrix of the MTNG-DESI sample (rescaled for each of the data points).

We show the $\chi^2$ of the fits and differences between the true value of the cosmological parameters and the inferences from the best fit for $\sig$, $\OmM$,$h$, $\OmMh$ and $S_8$ in Figure~\ref{fig:PSOerror}. As expected, here we also find a degeneracy between $\OmM$ and $h$. We do not find significant biases in any of the parameters (in all cases, “no deviation” falls within 1$\sigma$), but we tend to underestimate $h$ (and thus overestimate $\OmM$). Considering the width of the posteriors of the MTNG sample for each of the cosmological parameters, deviations are of the order of one-third of the width of the $1\sigma$ region for each parameter.

As a second test, we analyse whether the emulator is systematically biased in some region of the parameter space. For the MultiCosmology test set, we represent the values of the cosmological parameters coloured by the difference between their true value and the predicted value for each pair of parameters in Figure~\ref{fig:CosmoparamPSOerror}. Given the colour scheme, white regions indicate no bias for both parameters; purple, emerald, orange, and lime indicate bias in only one parameter; and blue, red, yellow, and green indicates bias in both parameters (and thus a degeneracy). As expected from Figure~\ref{fig:fullposterior}, we observe degeneracies between $\OmM$ and $h$. In the region close to the Planck cosmology, we infer for the best-fit higher values of $\OmM$ and lower values of $h$. The direction of this degeneracy was expected given the shape of the posteriors, but not necessarily the division of the biases in the parameter space (i.e., systematically underpredicting/overpredicting each parameter when looking for the best fit). A similar division into two regions occurs for parameters that are correlated by definition, e.g., $\OmMh$ and $h$. In the case of $\sig-\OmMh$, we do not find systematic degeneracies for the best fits in the central regions of the parameter space, but for the region with high $\sig$ we predict lower values of $\OmMh$ than the true value (green-yellow region).

\section{Removing scales}
\label{app:scaleproblems}

When analysing the scale-dependence of the constraints, we found that progressively removing small scales led to lower values of $\sig$ for DESI-ELGs and MTNG-DESI (biased for the latter, since we know the true values of the cosmological parameters in the simulation). As pointed out in Section~\ref{sec:MTNGscales}, the change in values of $\sig$ starts after removing points with $r_{\rm min} < 0.6~\hMpc$, roughly coinciding with the sizes of halos hosting satellites in the MTNG-DESI sample ($R_{200,c} =  0.37~\ihMpc$, for average halo mass of $\langle M_{\rm host, sat}\rangle= 10^{12.6}~\hMsun$). 

\begin{figure}
                \centering
                \includegraphics[width=0.42\textwidth]{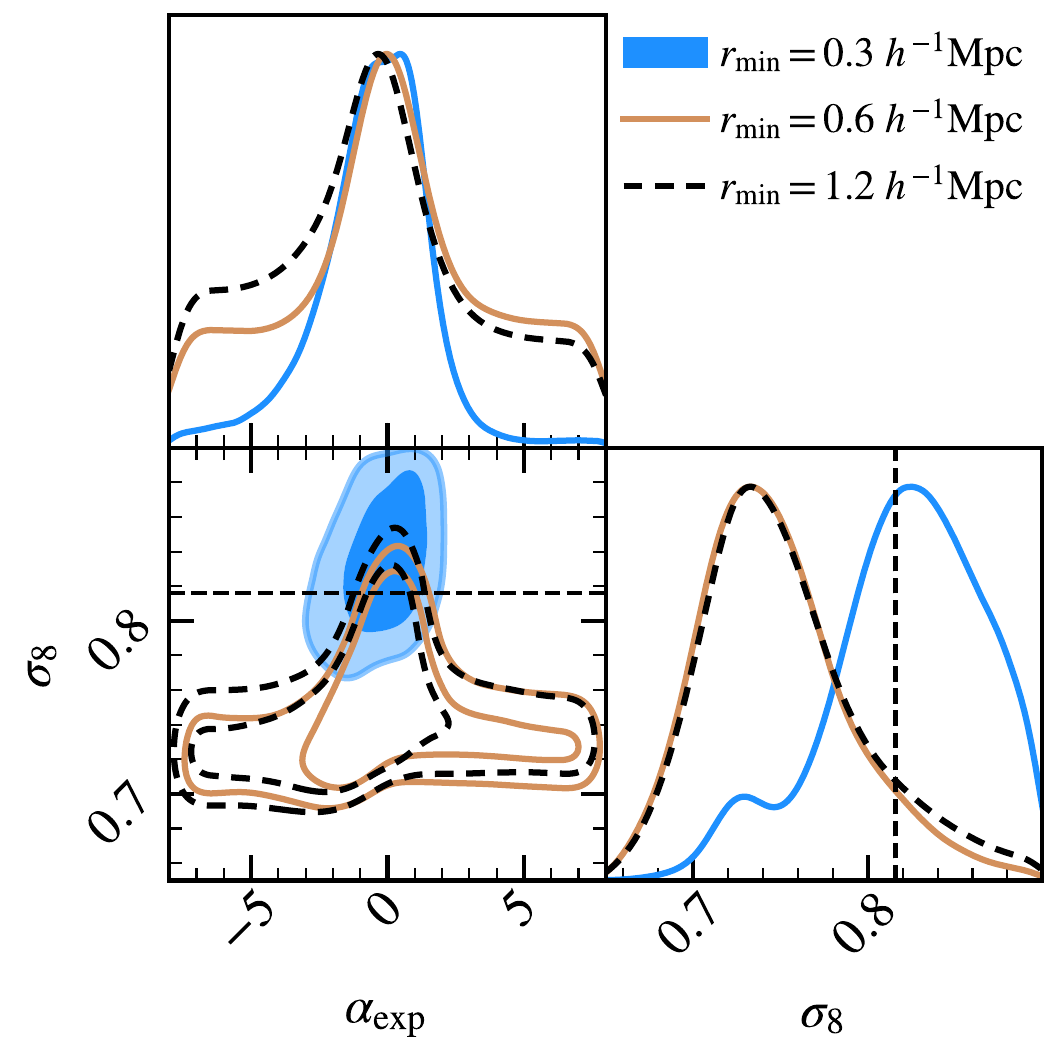}

        \caption{Projected constraints on $\sig$, and $\alpha_{\rm exp}$  for the MTNG-DESI sample for three different minimum scales. The black dashed lines on the histograms mark the true cosmology of the MTNG simulation. The best-fit of each sample is marked using coloured circles for the 2D distributions and dashed lines on the histograms.}
        \label{fig:ScaledegeneracyMTNGDESI}
\end{figure}

We analysed the posterior distributions for $\sig$ and the parameters more related to satellites for the fiducial scale cut ($r_{\rm min} = 0.3~\ihMpc$) and two runs with $r_{\rm min} = 0.6$ and $1.2~\ihMpc$. We show the 2D contours and 1D projections in Figure~\ref{fig:ScaledegeneracyMTNGDESI}. We find a degeneracy between $\sig$ and $\alpha_{\rm exp}$, the SHAMe-SF parameters that controls the difference in quenching strength with halo mass, creating projection effects on $\sig$. To compare the goodness of the fits for different minimum scales, we compute their $p$-values. To estimate the effective number of parameters, we follow \cite{Raveri:2019}. In the case of MTNG-DESI, we find a $ p$-value of almost 1 for all cuts, as expected from the low $\chi^2$ compared to the number of points, as shown in Figure~\ref{fig:clusteringMTNG}. The effective number of parameters is $\sim 13.8$ for $r_{\rm min}\leq 0.4~\ihMpc$,  $\sim 10$ for $0.6\leq r_{\rm min}\leq 1.2~\ihMpc$ and  $\sim 8$ for larger minimum scales. These scales coincide with the use of satellites to constrain SHAMe-SF parameters. For DESI-ELG with $r_{\min} = 0.3~\ihMpc$, we measure $p = 0.42$. For the scales where we obtain lower values of $\sig$, we find slightly larger p-values, and a change in the effective number of parameters when excluding scales within halos, from $N_{\rm eff}\sim 13$ for all samples with $r_{\rm min}<0.8~\hMpc$ to $N_{\rm eff} \sim 9$ for larger minimum scales.

In the case of MTNG-DESI, the MTNG value of $\sig$ is not excluded in the 2D contours. This underscores the importance of including small scales, especially when model parameters are mostly constrained by satellites, and of investigating degeneracies between model parameters and cosmological parameters, as well as jointly considering the analysis of the galaxy-halo connection (i.e., predicting average halo masses and satellite fractions) and cosmological constraints. 
For low values of $\sig$, one finds fewer satellite subhalos of a given mass, so it less matters how you quench them (thus the freedom of the value of $\alpha_{\rm exp}$). This explains the shape of the posterior: constrained for high $\sig$, almost flat for lower values.

\onecolumn

\section{Full posteriors}
\begin{figure*}[h!]
                \centering
                \includegraphics[width=0.95\textwidth]{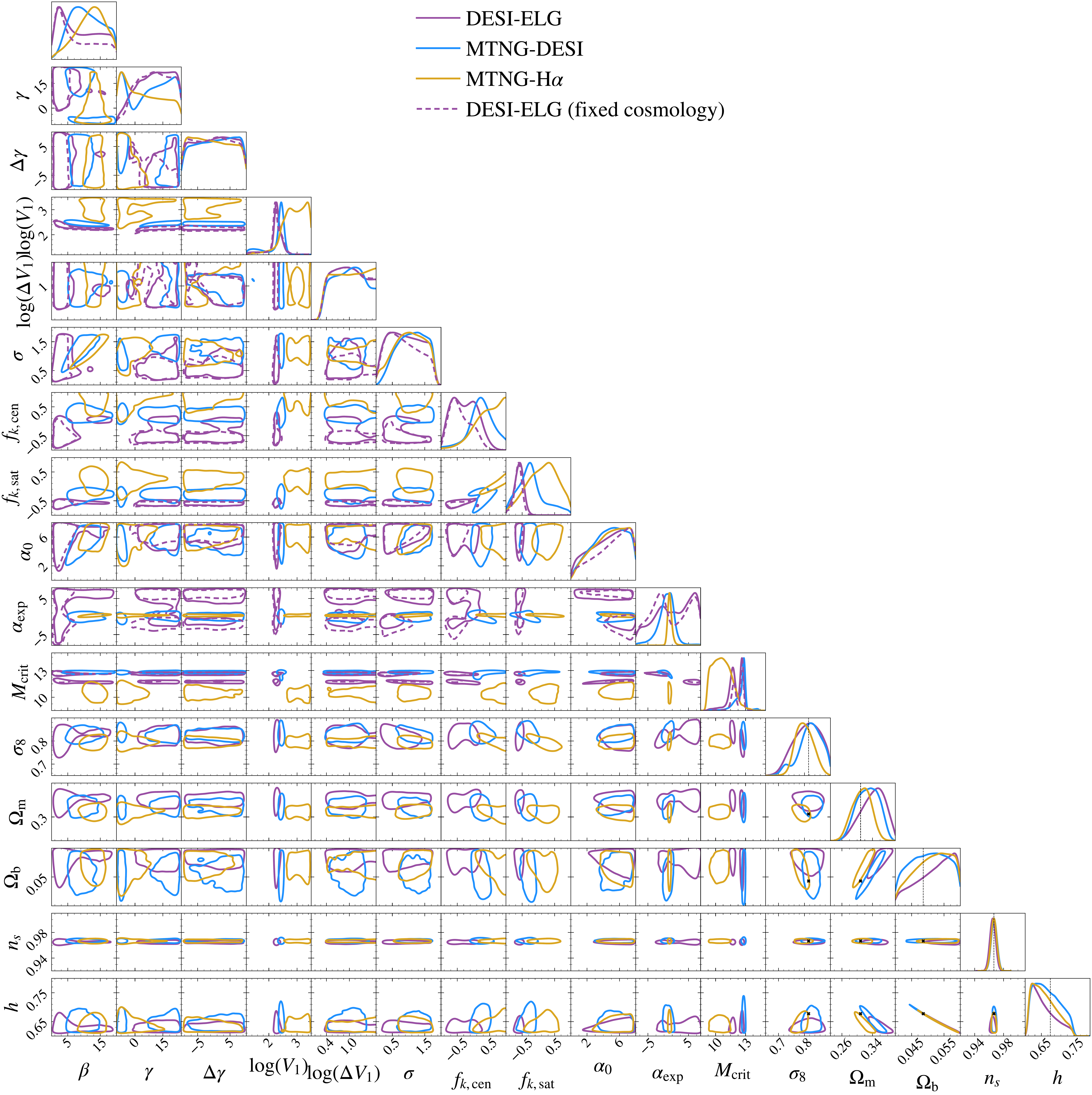}

        \caption{Projected posterior distribution ($1\sigma$) for all the parameters varied in the fiducial analysis (including the priors discussed in Section~\ref{sec:priors}) for the validation samples MTNG-DESI (blue) and MTNG-Halpha (yellow) and DESI-ELG (low-$z$, in purple). For the latter, we also include the posteriors in the SHAMe-SF parameters when fixing cosmology (dashed lines). 
}
        \label{fig:MTNGfull}
\end{figure*}

In Figure~\ref{fig:MTNGfull} we show the posterior distributions of the SHAMe-SF and cosmological parameters for the fiducial analysis of MTNG-DESI, MTNG-H$\alpha$ and DESI-ELG. In general, the SHAMe-SF parameters for DESI-ELG and MTNG-DESI are very similar, as expected given MTNG-DESI's selection criteria. The values of $V_1$ are similar to those found in \cite{SOM:2025DESI} when fitting the same sample. For MTNG-H$\alpha$, $V_1$ only has a constraint on the minimum value, which was expected since the sample contains galaxies with high stellar masses, populating the highest mass subhalos, which are not completely quenched yet at this redshift. As seen in previous works, we are prior-dominated in parameters that control the presence of subhalos with higher $\vpeak$ in the final sample ($\Delta V_1$ and $\Delta \gamma$), given their low effect on the clustering signal due to the shape of the mass and bias functions.

\label{app:fullposteriors}

\begin{figure*}[h!]
                \centering
                \includegraphics[height=0.40\textheight]{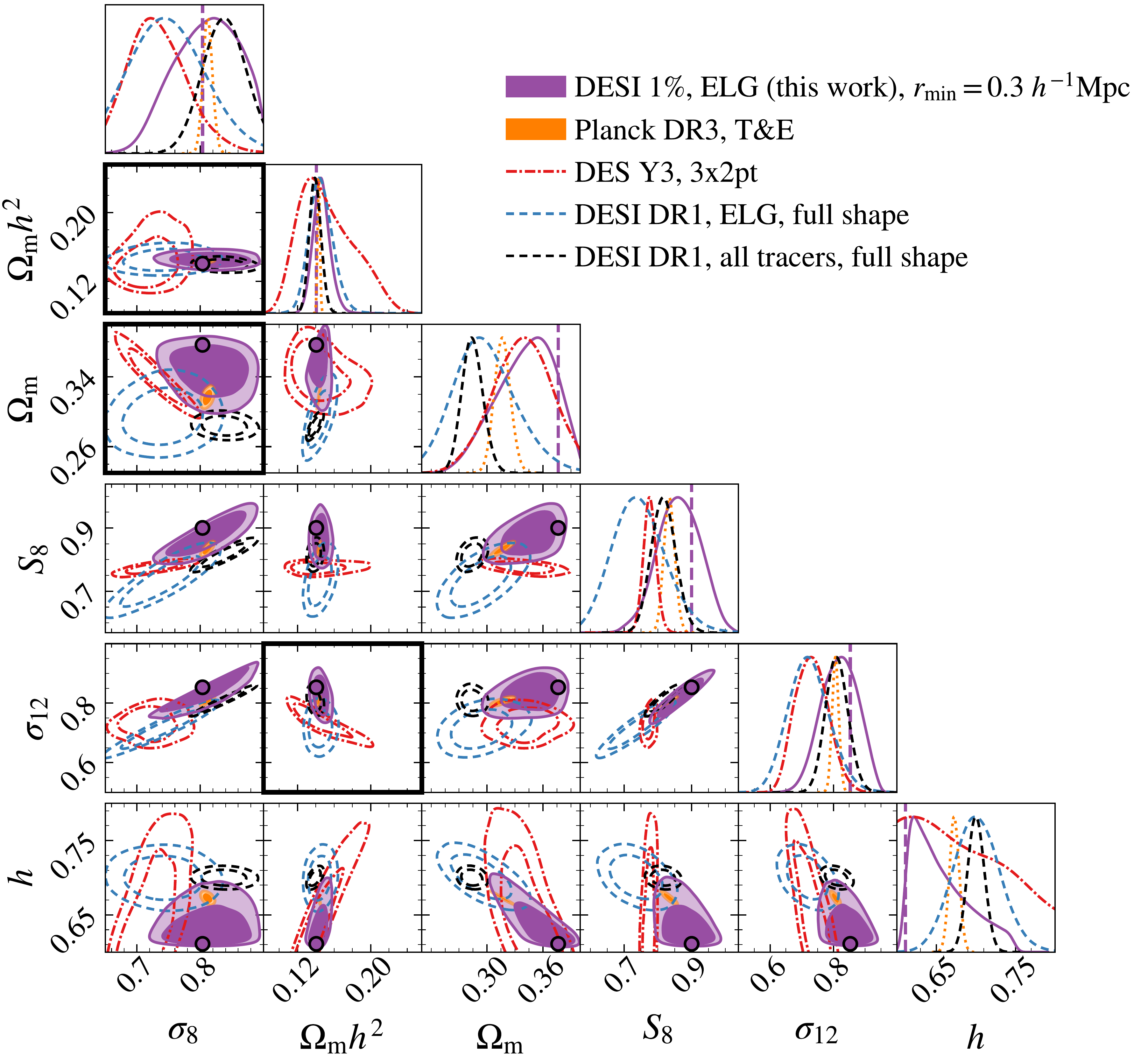}

        \caption{Projected posterior distribution for three cosmological parameters varied in the emulator ($\OmM$, $h$ and $\sig$) and the derived cosmological parameters $\OmMh$, $S_8$ and $\sigma_{12}$ for the DESI-ELG sample and the surveys discussed in Sect.~\ref{sec:DESIfidutial}. We show the best fit for the DESI-ELG sample using a dashed line (1D) and a point (2D).}
        \label{fig:DESIallcosmo}
\end{figure*}

\begin{figure*}[h!]
                \centering
                \includegraphics[height=0.40\textheight]{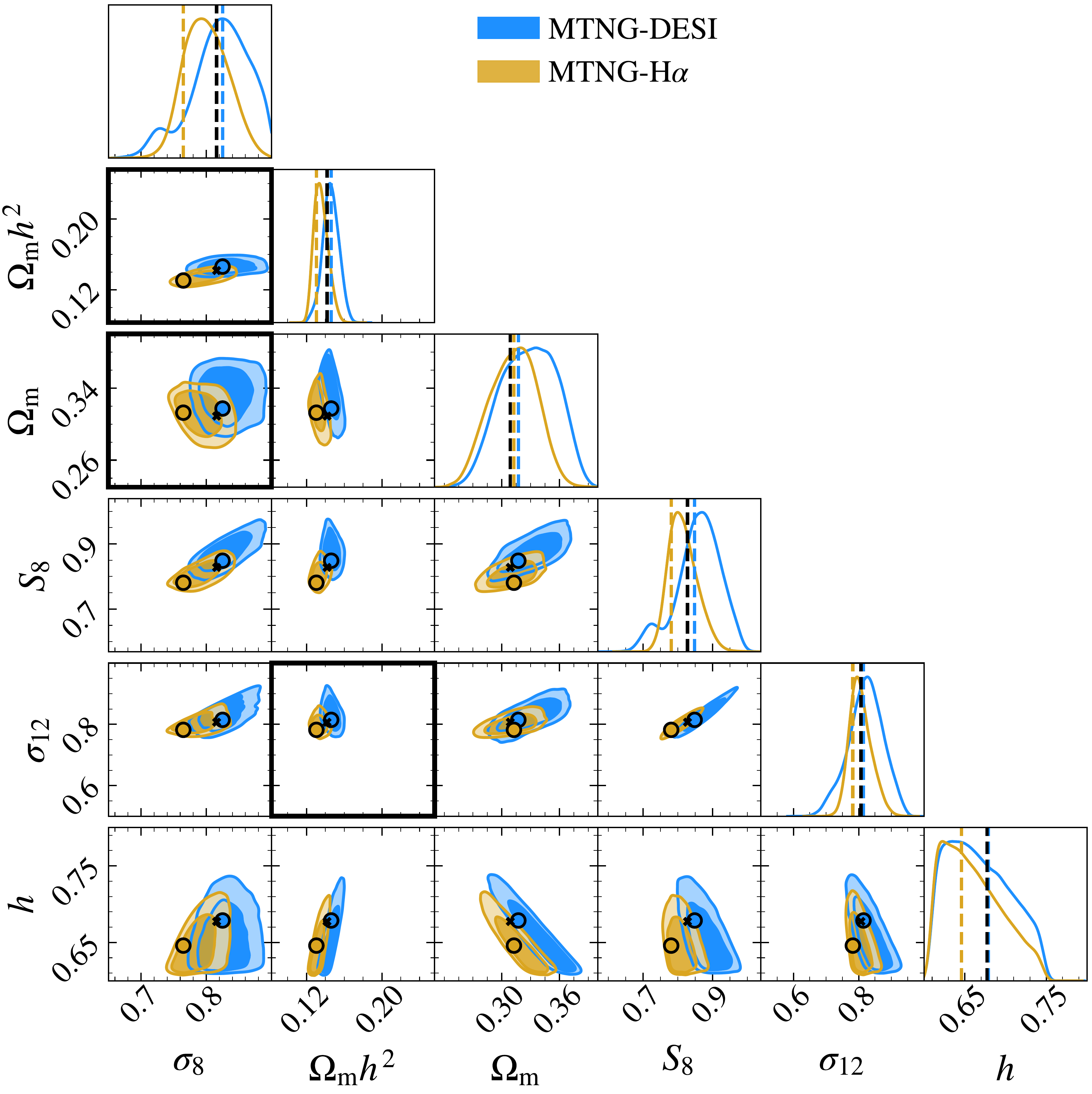}

        \caption{Same as Figure~\ref{fig:DESIallcosmo}, for the two validation samples defined for MTNG.}
        \label{fig:MTNGallapp}
\end{figure*}

In Figures~\ref{fig:DESIallcosmo} and~\ref{fig:MTNGallapp} we show the constraints on three cosmological parameters varied on the emulator ($\OmM$, $h$ and $\sig$) and the derived cosmological parameters $\OmMh$, $S_8$ and $\sigma_{12}$ for the two samples defined on the MTNG simulation and for DESI-ELGs (compared with the different surveys from Sect.~\ref{sec:DESIfidutial}). We highlight three panels in both figures: $\sig - \OmM$, $\sig - \OmMh$ and $\sigma_{12} - \OmMh$. As discussed in Section~\ref{sec:priors}, we chose $\sig - \OmMh$ as our fiducial parameter configuration, given the degeneracy between $\OmM$ and $h$. For our three samples, we measure low values of $h$, compatible with Planck within 1$\sigma$ for MTNG, and thus obtain higher values of $\OmM$ along the degeneracy line. If we compare our inferences in the $\sig$-$\OmM$ projection, they are compatible within $1\sigma$ with Planck, but only within $2\sigma$ with DES and DESI-full shape. Something similar happens when we compute $S_8$, given that its definition does not take into account the degeneracy between $\OmM$ and $h$. We do not find significant differences between using $\sig$ or $\sigma_{12}$ when combined with $\OmMh$. 
 
\label{app:fullposteriors}

\FloatBarrier

\section{Analysis of the DESI high-$z$ sample}

\label{app:DESIhigh}

We also analyse the high-$z$ bin of DESI-ELGs from \cite{Rocher2023:DESI} (see Sect.~\ref{sec:DESIobs}). For this sample, the $\sig$ values reached the upper limit of the prior. This behaviour persists regardless of the minimum scale considered, and different combinations of statistics are less constraining but still prefer higher values of $\sig$ near the prior limit. To test whether this is a systematic limitation of the model at high redshifts, we also built two auxiliary samples on MTNG equivalent to the low-$z$ case, one using a cut on SFR (H$\alpha$-like) and another using the same procedure as in MTNG-DESI for low-$z$, but changing the limit on the SFR until reaching a number density with the same incompleteness as the  low-$z$ (computed as $\bar{n}_{\rm MTNG-DESI}/\bar{n}_{\rm DESI-ELG}$ using the cuts from COSMOS). Note that this is approximate, without a crossmatch between DESI and COSMOS at this redshift.

\begin{figure*}[h!]
                \centering
                \includegraphics[width=0.95\textwidth]{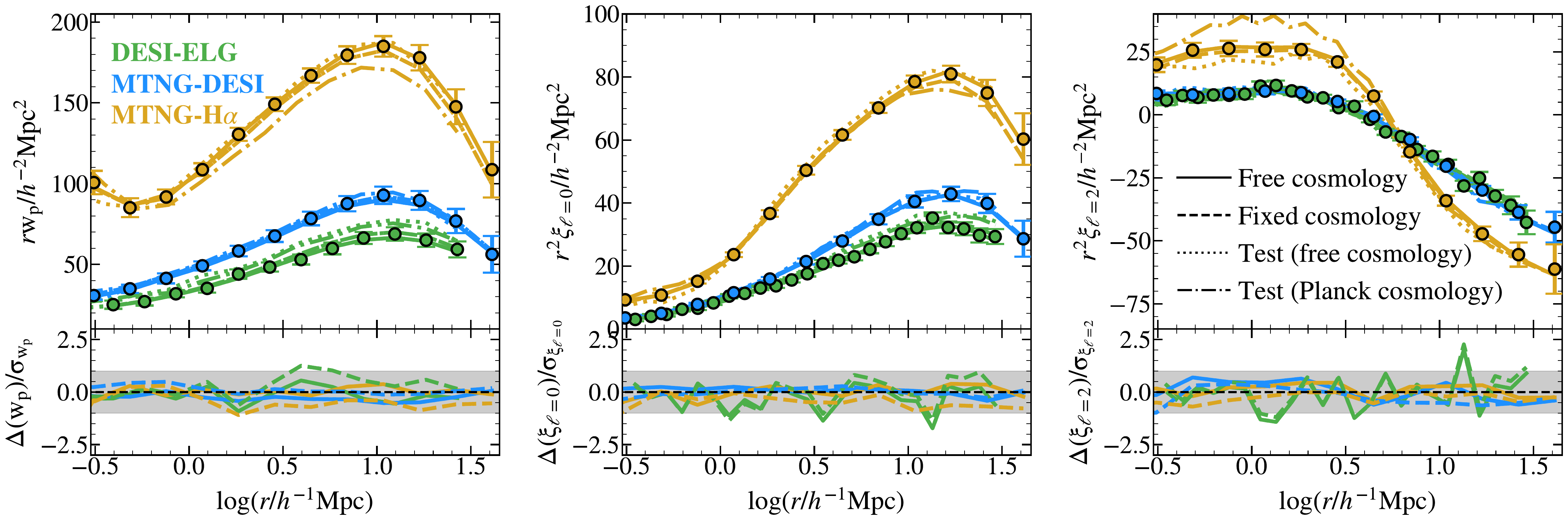}
                        
        \caption{Projected correlation function ($w_p$) and the monopole ($\xi_{\ell=0}$) and quadrupole ($\xi_{\ell=2}$) of the redshift--space correlation function of ELGs in DESI-ELGs for $1.1<z<1.6$ (green), and MTNG-DESI and MTNG-H$\alpha$ at $z = 1.327$ (blue and yellow, respectively). The corresponding best fits with the SHAMe-SF model are shown with solid lines (freeing the cosmological parameters) and dashed lines (fixing the cosmology). We add the evaluation of the best fits (fixed and free cosmology) for each sample as dash-dotted lines. Bottom panels: Difference between the data and the fit with SHAMe-SF in units of the diagonal elements of the respective covariance matrix.}
        \label{fig:DESIclusteringhighz}
\end{figure*}

\begin{figure*}[h!]
                \centering
                \includegraphics[width=0.85\textwidth]{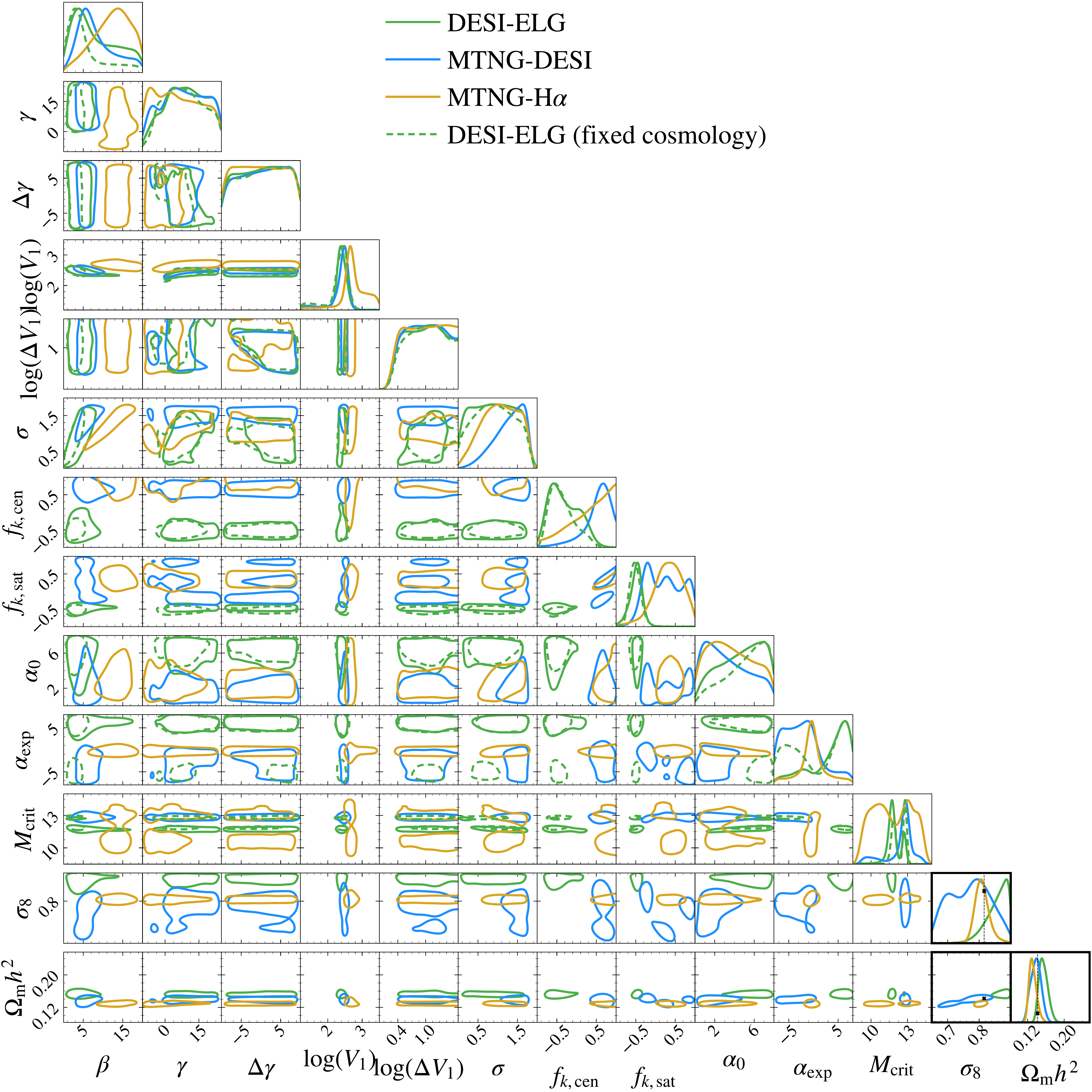}
        \caption{Similar to Figure~\ref{fig:MTNGfull}, for the samples defined at high-$z$. We show all the SHAMe-SF parameters and the cosmological parameters constrained in the analysis ($\sig$ and $\OmM$, panels with thicker lines).}
        \label{fig:DESIcosmohighz}
\end{figure*}

We show the clustering fits for the three samples in Figure~\ref{fig:DESIclusteringhighz}, and the cosmological constraints for the validation samples and DESI-ELG in Figure~\ref{fig:DESIcosmohighz}. MTNG-SFR behaves similarly to the low-$z$ case, but we cannot constrain $\sig$ for MTNG-DESI since the degeneracy with $\OmMh$ increases. Compared to DESI-ELG, none of the validation samples exhibits a similar behaviour regarding $\sig$. We report our findings here and wait for further data releases from DESI to analyse whether the results hold. We also note that the emulator errors are considerably higher for this number density. When comparing the posteriors on the SHAMe-SF parameters in Figure~\ref{fig:DESIcosmohighz} for DESI-ELG when freeing the cosmology and imposing Planck cosmology, we find similar differences in $\beta$ and $f_{k,\rm cen}$ as in the low$-z$ sample. For 
$\alpha_{\rm exp}$, we find a bi-modality also similar to the low$-z$ sample when the cosmology is fixed, but it disappears when freeing the cosmological parameters, keeping only the region hitting the upper limit of the prior.  

We include in Figure~\ref{fig:DESIclusteringhighz} the result of populating a simulation again using the best-fit parameters for the three samples. For both best-fits of DESI-ELG (free and fixed cosmology), the emulator underestimates the value of $\wpp$. In the case of MTNG-$H\alpha$, we also find differences for beyond $1\sigma$ for $\wpp$ and the smallest scales of the quadrupole. As the emulator is a key component of the analysis, we aim to investigate alternative techniques in future work, such as a likelihood emulator rather than a clustering emulator, and leave the door open to potential modifications of the SHAMe-SF model at higher redshifts. 

\end{appendix}

\end{document}